\documentclass[12pt]{article}
\usepackage{hyperref}
\usepackage{amsfonts}
\usepackage[dvips]{graphicx}
\usepackage{latexsym,amsmath,amssymb,graphics,stmaryrd}
\usepackage{array}
\usepackage{subfigure}
\usepackage{multirow}
\usepackage[dvips]{graphicx}
\usepackage[dvips]{color}
\usepackage{pstricks}
\usepackage{pst-node}
\usepackage{pst-coil}

\usepackage{dsfont}
\usepackage{amsthm}
\usepackage{graphics}
\usepackage{subfigure}
\usepackage{fancyhdr}
\usepackage[dvips]{color}

\makeatletter
\DeclareRobustCommand*{\bfseries}{%
\not@math@alphabet\bfseries\mathbf
\fontseries\bfdefault\selectfont
\boldmath
}
\makeatother

\usepackage{amsfonts}
\usepackage{bm}
\usepackage[dvips]{graphicx}
\usepackage{latexsym,amsmath,amssymb,graphics,stmaryrd}
\usepackage{cite}
\usepackage{array}
\usepackage{subfigure}
\usepackage{multirow}
\usepackage{epsfig}
\usepackage[dvips]{graphicx}
\usepackage[dvips]{color}
\usepackage{pstricks}
\usepackage{pst-node}
\usepackage{pst-plot}
\usepackage{amsthm}
\usepackage{graphics}
\usepackage{subfigure}
\usepackage{fancyhdr}

\usepackage{rotating}
\newlength{\arlength}
\newlength{\arheight}

\usepackage{feynmp}

\usepackage{feynmp}

\fancypagestyle{plain}{
\fancyhf{}
\newcommand{\fpage}{\iffloatpage{}{\thepage}}
\fancyfoot[C]{\fpage}

}

\newcommand{\col}{~,}
\newcommand{\pnt}{~.}
\newcommand{\AdS}{\text{AdS}}
\newcommand{\CFT}{\text{CFT}}

\newcommand{\YM}{{\scriptscriptstyle\text{YM}}}

\newcommand{\deriv}[2][]{\frac{\de #1}{\de #2}}
\newcommand{\parderiv}[2][]{\frac{\partial #1}{\partial #2}}

\newcommand{\de}{\operatorname{d}\!}

\newcommand{\e}{\operatorname{e}}

\newlength{\neglength}
\newlength{\diameter}

\newcommand{\svertex}[2]{%
\fmfiequ{#1}{point length(#2)/2 of (#2)}
}

\newcommand{\chionetwo}[1][black]{%
\fmftop{v1}
\fmfbottom{v4}
\fmfforce{(0.125w,h)}{v1}
\fmfforce{(0.125w,0)}{v4}
\fmffixed{(0.25w,0)}{v1,v2}
\fmffixed{(0.25w,0)}{v2,v3}
\fmffixed{(0.25w,0)}{v4,v5}
\fmffixed{(0.25w,0)}{v5,v6}
\fmffixed{(0,whatever)}{vc1,vc3}
\fmffixed{(0,whatever)}{vc2,vc4}
\fmf{plain,tension=0.5,right=0.25}{v1,vc1}
\fmf{plain,tension=0.5,left=0.25}{v2,vc1}
\fmf{phantom,tension=0.5,right=0.25}{v2,vc2}
\fmf{plain,tension=0.5,left=0.25}{v3,vc2}
\fmf{plain,tension=0.5,left=0.25}{v4,vc3}
\fmf{phantom,tension=0.5,right=0.25}{v5,vc3}
\fmf{plain,tension=0.5,left=0.25}{v5,vc4}
\fmf{plain,tension=0.5,right=0.25}{v6,vc4}
\fmf{plain,tension=1.25,left=0}{vc1,vc3}
\fmf{plain,tension=1.25,left=0}{vc2,vc4}
\fmffreeze
\fmf{plain,tension=1,left=0}{vc2,vc3}
\fmf{plain,tension=0.5,right=0,width=1mm}{v4,v6}
\fmffreeze
\fmfposition
\fmfipath{p[]}
\fmfipair{vd[],vm[],vu[]}
\fmfiset{p1}{vpath(__v1,__vc1)}
\fmfiset{p2}{vpath(__v2,__vc1)}
\fmfiset{p6}{vpath(__v3,__vc2)}
\fmfiset{p4}{vpath(__v4,__vc3)}
\fmfiset{p8}{vpath(__v5,__vc4)}
\fmfiset{p9}{vpath(__v6,__vc4)}
\fmfiset{p3}{vpath(__vc1,__vc3)}
\fmfiset{p7}{vpath(__vc2,__vc4)}
\fmfiset{p5}{vpath(__vc2,__vc3)}
\svertex{vm1}{p1}
\svertex{vm2}{p2}
\svertex{vm3}{p3}
\svertex{vm4}{p4}
\svertex{vm5}{p5}
\svertex{vm6}{p6}
\svertex{vm7}{p7}
\svertex{vm8}{p8}
\svertex{vm9}{p9}
}

\newcommand{\chionetwoone}[1][black]{%
\fmftop{v1}
\fmfbottom{v4}
\fmfforce{(0.125w,h)}{v1}
\fmfforce{(0.125w,0)}{v4}
\fmffixed{(0.25w,0)}{v1,v2}
\fmffixed{(0.25w,0)}{v2,v3}
\fmffixed{(0.25w,0)}{v4,v5}
\fmffixed{(0.25w,0)}{v5,v6}
\fmffixed{(0,whatever)}{vc1,vc3}
\fmffixed{(0,whatever)}{vb2,vb4}
\fmffixed{(0,whatever)}{vc1,vb1}
\fmffixed{(0,whatever)}{vc1,vb3}
\fmffixed{(whatever,0)}{vb1,vb2}
\fmffixed{(whatever,0)}{vb3,vb4}
\fmf{plain,tension=0.5,right=0.25}{v1,vc1}
\fmf{plain,tension=0.5,left=0.25}{v2,vc1}
\fmf{phantom,tension=0.5,right=0.25}{v2,vb2}
\fmf{plain,tension=0.5,left=0.25}{v3,vb2}
\fmf{plain,tension=0.5,left=0.25}{v4,vc3}
\fmf{plain,tension=0.5,right=0.25}{v5,vc3}
\fmf{phantom,tension=0.5,left=0.25}{v5,vb4}
\fmf{plain,tension=0.5,right=0.25}{v6,vb4}
\fmf{plain,tension=2,left=0}{vc1,vb1}
\fmf{plain,tension=2,left=0}{vb1,vb3}
\fmf{plain,tension=2,left=0}{vb3,vc3}
\fmf{plain,tension=2,left=0}{vb2,vb4}
\fmffreeze
\fmf{plain,tension=2,left=0}{vb1,vb2}
\fmf{plain,tension=2,left=0}{vb3,vb4}
\fmf{plain,tension=0.5,right=0,width=1mm}{v4,v6}
\fmffreeze
\fmfposition
\fmfipath{p[]}
\fmfipair{vd[],vm[],vu[]}
\fmfiset{p1}{vpath(__v1,__vc1)}
\fmfiset{p2}{vpath(__v2,__vc1)}
\fmfiset{p3}{vpath(__vc1,__vb1)}
\fmfiset{p4}{vpath(__vb1,__vb3)}
\fmfiset{p5}{vpath(__vb1,__vb2)}
\fmfiset{p6}{vpath(__v3,__vb2)}
\fmfiset{p7}{vpath(__vb2,__vb4)}
\fmfiset{p8}{vpath(__vb3,__vb4)}
\fmfiset{p9}{vpath(__v6,__vb4)}
\fmfiset{p10}{vpath(__vb3,__vc3)}
\fmfiset{p11}{vpath(__v4,__vc3)}
\fmfiset{p12}{vpath(__v5,__vc3)}
\svertex{vm1}{p1}
\svertex{vm2}{p2}
\svertex{vm3}{p3}
\svertex{vm4}{p4}
\svertex{vm5}{p5}
\svertex{vm6}{p6}
\svertex{vm7}{p7}
\svertex{vm8}{p8}
\svertex{vm9}{p9}
\svertex{vm10}{p10}
\svertex{vm11}{p11}
\svertex{vm12}{p12}
}

\newcommand{\chionetwothree}[1][black]{%
\fmftop{v1}
\fmfbottom{v5}
\fmfforce{(0.125w,h)}{v1}
\fmfforce{(0.125w,0)}{v5}
\fmffixed{(0.25w,0)}{v1,v2}
\fmffixed{(0.25w,0)}{v2,v3}
\fmffixed{(0.25w,0)}{v3,v4}
\fmffixed{(0.25w,0)}{v5,v6}
\fmffixed{(0.25w,0)}{v6,v7}
\fmffixed{(0.25w,0)}{v7,v8}
\fmffixed{(0,whatever)}{vc1,vc4}
\fmffixed{(0,whatever)}{vc2,vc5}
\fmffixed{(0,whatever)}{vc3,vc6}

\fmf{plain,tension=0.5,right=0.25}{v1,vc1}
\fmf{plain,tension=0.5,left=0.25}{v2,vc1}
\fmf{phantom,tension=0.5,right=0.25}{v2,vc2}
\fmf{plain,tension=0.5,left=0.25}{v3,vc2}
\fmf{phantom,tension=0.5,right=0.25}{v3,vc3}
\fmf{plain,tension=0.5,left=0.25}{v4,vc3}
\fmf{plain,tension=0.5,left=0.25}{v5,vc4}
\fmf{phantom,tension=0.5,right=0.25}{v6,vc4}
\fmf{plain,tension=0.5,left=0.25}{v6,vc5}
\fmf{phantom,tension=0.5,right=0.25}{v7,vc5}
\fmf{plain,tension=0.5,left=0.25}{v7,vc6}
\fmf{plain,tension=0.5,right=0.25}{v8,vc6}
\fmf{plain,tension=1.25,left=0}{vc1,vc4}
\fmf{plain,tension=1.25,left=0}{vc2,vc5}
\fmf{plain,tension=1.25,left=0}{vc3,vc6}
\fmffreeze
\fmf{plain,tension=1,left=0}{vc4,vc2}
\fmf{plain,tension=1,left=0}{vc5,vc3}
\fmf{plain,tension=0.5,right=0,width=1mm}{v5,v8}
\fmffreeze
\fmfposition
}

\newcommand{\chitwoonethree}[1][black]{%
\fmftop{v1}
\fmfbottom{v5}
\fmfforce{(0.125w,h)}{v1}
\fmfforce{(0.125w,0)}{v5}
\fmffixed{(0.25w,0)}{v1,v2}
\fmffixed{(0.25w,0)}{v2,v3}
\fmffixed{(0.25w,0)}{v3,v4}
\fmffixed{(0.25w,0)}{v5,v6}
\fmffixed{(0.25w,0)}{v6,v7}
\fmffixed{(0.25w,0)}{v7,v8}
\fmffixed{(whatever,0.5h)}{v5,vc1}
\fmffixed{(0,whatever)}{vc1,vc4}
\fmffixed{(0,whatever)}{vc2,vc5}
\fmffixed{(0,whatever)}{vc3,vc6}
\fmffixed{(whatever,0)}{vc1,vc3}
\fmffixed{(whatever,0)}{vc3,vc5}

\fmf{plain,tension=0.5,right=0.125}{v1,vc1}
\fmf{phantom,tension=0.5,left=0.25}{v2,vc1}
\fmf{plain,tension=0.5,right=0.25}{v2,vc2}
\fmf{plain,tension=0.5,left=0.25}{v3,vc2}
\fmf{phantom,tension=0.5,right=0.25}{v3,vc3}
\fmf{plain,tension=0.5,left=0.125}{v4,vc3}
\fmf{plain,tension=0.5,left=0.25}{v5,vc4}
\fmf{plain,tension=0.5,right=0.25}{v6,vc4}
\fmf{phantom,tension=0.5,left=0.25}{v6,vc5}
\fmf{phantom,tension=0.5,right=0.25}{v7,vc5}
\fmf{plain,tension=0.5,left=0.25}{v7,vc6}
\fmf{plain,tension=0.5,right=0.25}{v8,vc6}
\fmf{plain,tension=1.25,left=0}{vc1,vc4}
\fmf{plain,tension=1.25,left=0}{vc2,vc5}
\fmf{plain,tension=1.25,left=0}{vc3,vc6}
\fmffreeze
\fmf{plain,tension=1,left=0}{vc1,vc5}
\fmf{plain,tension=1,left=0}{vc5,vc3}
\fmf{plain,tension=0.5,right=0,width=1mm}{v5,v8}
\fmffreeze
\fmfposition
}

\newcommand{\chionethreetwo}[1][black]{%
\fmftop{v1}
\fmfbottom{v5}
\fmfforce{(0.125w,h)}{v1}
\fmfforce{(0.125w,0)}{v5}
\fmffixed{(0.25w,0)}{v1,v2}
\fmffixed{(0.25w,0)}{v2,v3}
\fmffixed{(0.25w,0)}{v3,v4}
\fmffixed{(0.25w,0)}{v5,v6}
\fmffixed{(0.25w,0)}{v6,v7}
\fmffixed{(0.25w,0)}{v7,v8}
\fmffixed{(whatever,0.5h)}{v5,vc2}
\fmffixed{(0,whatever)}{vc1,vc4}
\fmffixed{(0,whatever)}{vc2,vc5}
\fmffixed{(0,whatever)}{vc3,vc6}
\fmffixed{(whatever,0)}{vc2,vc4}
\fmffixed{(whatever,0)}{vc4,vc6}
\fmf{plain,tension=0.5,right=0.25}{v1,vc1}
\fmf{plain,tension=0.5,left=0.25}{v2,vc1}
\fmf{phantom,tension=0.5,right=0.25}{v2,vc2}
\fmf{phantom,tension=0.5,left=0.25}{v3,vc2}
\fmf{plain,tension=0.5,right=0.25}{v3,vc3}
\fmf{plain,tension=0.5,left=0.25}{v4,vc3}
\fmf{plain,tension=0.5,left=0.125}{v5,vc4}
\fmf{phantom,tension=0.5,right=0.25}{v6,vc4}
\fmf{plain,tension=0.5,left=0.25}{v6,vc5}
\fmf{plain,tension=0.5,right=0.25}{v7,vc5}
\fmf{phantom,tension=0.5,left=0.25}{v7,vc6}
\fmf{plain,tension=0.5,right=0.125}{v8,vc6}
\fmf{plain,tension=1.25,left=0}{vc1,vc4}
\fmf{plain,tension=1.25,left=0}{vc2,vc5}
\fmf{plain,tension=1.25,left=0}{vc3,vc6}
\fmffreeze
\fmf{plain,tension=1,left=0}{vc2,vc4}
\fmf{plain,tension=1,left=0}{vc2,vc6}
\fmf{plain,tension=0.5,right=0,width=1mm}{v5,v8}
\fmffreeze
\fmfposition
}

\newcommand{\chionetwothreetwo}[1][black]{%
\fmftop{v1}
\fmfbottom{v5}
\fmfforce{(0.125w,h)}{v1}
\fmfforce{(0.125w,0)}{v5}
\fmffixed{(0.25w,0)}{v1,v2}
\fmffixed{(0.25w,0)}{v2,v3}
\fmffixed{(0.25w,0)}{v3,v4}
\fmffixed{(0.25w,0)}{v5,v6}
\fmffixed{(0.25w,0)}{v6,v7}
\fmffixed{(0.25w,0)}{v7,v8}
\fmffixed{(0,whatever)}{va1,va2}
\fmffixed{(whatever,0)}{va1,vc1}
\fmffixed{(whatever,0)}{va2,vb3}
\fmffixed{(0,whatever)}{vc1,vc3}
\fmffixed{(0,whatever)}{vb2,vb4}
\fmffixed{(0,whatever)}{vc1,vb1}
\fmffixed{(0,whatever)}{vc1,vb3}
\fmffixed{(whatever,0)}{vb1,vb2}
\fmffixed{(whatever,0)}{vb3,vb4}
\fmf{phantom,tension=0.5,right=0.25}{v2,vc1}
\fmf{plain,tension=0.5,left=0.25}{v3,vc1}
\fmf{phantom,tension=0.5,right=0.25}{v3,vb2}
\fmf{plain,tension=0.5,left=0.25}{v4,vb2}
\fmf{plain,tension=0.5,left=0.25}{v6,vc3}
\fmf{plain,tension=0.5,right=0.25}{v7,vc3}
\fmf{phantom,tension=0.5,left=0.25}{v7,vb4}
\fmf{plain,tension=0.5,right=0.25}{v8,vb4}
\fmf{plain,tension=2,left=0}{vc1,vb1}
\fmf{plain,tension=2,left=0}{vb1,vb3}
\fmf{plain,tension=2,left=0}{vb3,vc3}
\fmf{plain,tension=2,left=0}{vb2,vb4}
\fmffreeze
\fmf{plain,tension=0.5,right=0.25}{v1,va1}
\fmf{plain,tension=0.5,left=0.25}{v2,va1}
\fmf{plain,tension=1}{va1,va2}
\fmf{plain,tension=0}{va2,vc1}
\fmf{plain,tension=0.5,left=0.25}{v5,va2}
\fmf{phantom,tension=0.5,right=0.25}{v6,va2}
\fmf{plain,tension=1,left=0}{vb1,vb2}
\fmf{plain,tension=1,left=0}{vb3,vb4}
\fmf{plain,tension=0.5,right=0,width=1mm}{v5,v8}
}

\newcommand{\chitwothreetwoone}[1][black]{%
\fmftop{v1}
\fmfbottom{v5}
\fmfforce{(0.125w,h)}{v1}
\fmfforce{(0.125w,0)}{v5}
\fmffixed{(0.25w,0)}{v1,v2}
\fmffixed{(0.25w,0)}{v2,v3}
\fmffixed{(0.25w,0)}{v3,v4}
\fmffixed{(0.25w,0)}{v5,v6}
\fmffixed{(0.25w,0)}{v6,v7}
\fmffixed{(0.25w,0)}{v7,v8}
\fmffixed{(0,whatever)}{va1,va2}
\fmffixed{(whatever,0)}{va2,vc3}
\fmffixed{(whatever,0)}{va1,vb1}
\fmffixed{(0,whatever)}{vc1,vc3}
\fmffixed{(0,whatever)}{vb2,vb4}
\fmffixed{(0,whatever)}{vc1,vb1}
\fmffixed{(0,whatever)}{vc1,vb3}
\fmffixed{(whatever,0)}{vb1,vb2}
\fmffixed{(whatever,0)}{vb3,vb4}
\fmf{plain,tension=0.5,right=0.25}{v2,vc1}
\fmf{plain,tension=0.5,left=0.25}{v3,vc1}
\fmf{phantom,tension=0.5,right=0.25}{v3,vb2}
\fmf{plain,tension=0.5,left=0.25}{v4,vb2}
\fmf{phantom,tension=0.5,left=0.25}{v6,vc3}
\fmf{plain,tension=0.5,right=0.25}{v7,vc3}
\fmf{phantom,tension=0.5,left=0.25}{v7,vb4}
\fmf{plain,tension=0.5,right=0.25}{v8,vb4}
\fmf{plain,tension=2,left=0}{vc1,vb1}
\fmf{plain,tension=2,left=0}{vb1,vb3}
\fmf{plain,tension=2,left=0}{vb3,vc3}
\fmf{plain,tension=2,left=0}{vb2,vb4}
\fmffreeze
\fmf{plain,tension=0.5,right=0.25}{v1,va1}
\fmf{phantom,tension=0.5,left=0.25}{v2,va1}
\fmf{plain,tension=1}{va1,va2}
\fmf{plain,tension=0}{va1,vc3}
\fmf{plain,tension=0.5,left=0.25}{v5,va2}
\fmf{plain,tension=0.5,right=0.25}{v6,va2}
\fmf{plain,tension=1,left=0}{vb1,vb2}
\fmf{plain,tension=1,left=0}{vb3,vb4}
\fmf{plain,tension=0.5,right=0,width=1mm}{v5,v8}
}

\newcommand{\chitwoonetwothree}[1][black]{%
\fmftop{v1}
\fmfbottom{v5}
\fmfforce{(0.125w,h)}{v1}
\fmfforce{(0.125w,0)}{v5}
\fmffixed{(0.25w,0)}{v1,v2}
\fmffixed{(0.25w,0)}{v2,v3}
\fmffixed{(0.25w,0)}{v3,v4}
\fmffixed{(0.25w,0)}{v5,v6}
\fmffixed{(0.25w,0)}{v6,v7}
\fmffixed{(0.25w,0)}{v7,v8}
\fmffixed{(0,whatever)}{va1,va2}
\fmffixed{(whatever,0)}{va2,vc3}
\fmffixed{(whatever,0)}{va1,vb1}
\fmffixed{(0,whatever)}{vc1,vc3}
\fmffixed{(0,whatever)}{vb2,vb4}
\fmffixed{(0,whatever)}{vc1,vb1}
\fmffixed{(0,whatever)}{vc1,vb3}
\fmffixed{(whatever,0)}{vb1,vb2}
\fmffixed{(whatever,0)}{vb3,vb4}
\fmf{plain,tension=0.5,right=0.25}{v2,vc1}
\fmf{plain,tension=0.5,left=0.25}{v3,vc1}
\fmf{plain,tension=0.5,right=0.25}{v1,vb2}
\fmf{phantom,tension=0.5,left=0.25}{v2,vb2}
\fmf{plain,tension=0.5,left=0.25}{v6,vc3}
\fmf{phantom,tension=0.5,right=0.25}{v7,vc3}
\fmf{plain,tension=0.5,left=0.25}{v5,vb4}
\fmf{phantom,tension=0.5,right=0.25}{v6,vb4}
\fmf{plain,tension=2,left=0}{vc1,vb1}
\fmf{plain,tension=2,left=0}{vb1,vb3}
\fmf{plain,tension=2,left=0}{vb3,vc3}
\fmf{plain,tension=2,left=0}{vb2,vb4}
\fmffreeze
\fmf{phantom,tension=0.5,right=0.25}{v3,va1}
\fmf{plain,tension=0.5,left=0.25}{v4,va1}
\fmf{plain,tension=1}{va1,va2}
\fmf{plain,tension=0}{va1,vc3}
\fmf{plain,tension=0.5,left=0.25}{v7,va2}
\fmf{plain,tension=0.5,right=0.25}{v8,va2}
\fmf{plain,tension=1,left=0}{vb1,vb2}
\fmf{plain,tension=1,left=0}{vb3,vb4}
\fmf{plain,tension=0.5,right=0,width=1mm}{v5,v8}
}

\newcommand{\chithreeonetwoone}[1][black]{%
\fmftop{v1}
\fmfbottom{v5}
\fmfforce{(0.125w,h)}{v1}
\fmfforce{(0.125w,0)}{v5}
\fmffixed{(0.25w,0)}{v1,v2}
\fmffixed{(0.25w,0)}{v2,v3}
\fmffixed{(0.25w,0)}{v3,v4}
\fmffixed{(0.25w,0)}{v5,v6}
\fmffixed{(0.25w,0)}{v6,v7}
\fmffixed{(0.25w,0)}{v7,v8}
\fmffixed{(0,whatever)}{va1,va2}
\fmffixed{(whatever,0)}{va1,vc1}
\fmffixed{(whatever,0)}{va2,vb1}
\fmffixed{(0,whatever)}{vc1,vc3}
\fmffixed{(0,whatever)}{vb2,vb4}
\fmffixed{(0,whatever)}{vc1,vb1}
\fmffixed{(0,whatever)}{vc1,vb3}
\fmffixed{(whatever,0)}{vb1,vb2}
\fmffixed{(whatever,0)}{vb3,vb4}
\fmf{plain,tension=0.5,right=0.25}{v1,vc1}
\fmf{plain,tension=0.5,left=0.25}{v2,vc1}
\fmf{phantom,tension=0.5,right=0.25}{v2,vb2}
\fmf{phantom,tension=0.5,left=0.25}{v3,vb2}
\fmf{plain,tension=0.5,left=0.25}{v5,vc3}
\fmf{plain,tension=0.5,right=0.25}{v6,vc3}
\fmf{phantom,tension=0.5,left=0.125}{v6,vb4}
\fmf{plain,tension=0.5,right=0.125}{v7,vb4}
\fmf{plain,tension=2,left=0}{vc1,vb1}
\fmf{plain,tension=2,left=0}{vb1,vb3}
\fmf{plain,tension=2,left=0}{vb3,vc3}
\fmf{plain,tension=2,left=0}{vb2,vb4}
\fmffreeze
\fmf{plain,tension=0.5,right=0.25}{v3,va1}
\fmf{plain,tension=0.5,left=0.25}{v4,va1}
\fmf{plain,tension=1}{va1,va2}
\fmf{plain,tension=0}{va2,vb2}
\fmf{phantom,tension=0.5,left=0.125}{v7,va2}
\fmf{plain,tension=0.5,right=0.125}{v8,va2}
\fmf{plain,tension=1,left=0}{vb1,vb2}
\fmf{plain,tension=1,left=0}{vb3,vb4}
\fmf{plain,tension=0.5,right=0,width=1mm}{v5,v8}
}

\newcommand{\chionethreetwothree}[1][black]{%
\fmftop{v1}
\fmfbottom{v5}
\fmfforce{(0.125w,h)}{v1}
\fmfforce{(0.125w,0)}{v5}
\fmffixed{(0.25w,0)}{v1,v2}
\fmffixed{(0.25w,0)}{v2,v3}
\fmffixed{(0.25w,0)}{v3,v4}
\fmffixed{(0.25w,0)}{v5,v6}
\fmffixed{(0.25w,0)}{v6,v7}
\fmffixed{(0.25w,0)}{v7,v8}
\fmffixed{(0,whatever)}{va1,va2}
\fmffixed{(whatever,0)}{va1,vc1}
\fmffixed{(whatever,0)}{va2,vb1}
\fmffixed{(0,whatever)}{vc1,vc3}
\fmffixed{(0,whatever)}{vb2,vb4}
\fmffixed{(0,whatever)}{vc1,vb1}
\fmffixed{(0,whatever)}{vc1,vb3}
\fmffixed{(whatever,0)}{vb1,vb2}
\fmffixed{(whatever,0)}{vb3,vb4}
\fmf{plain,tension=0.5,right=0.25}{v3,vc1}
\fmf{plain,tension=0.5,left=0.25}{v4,vc1}
\fmf{phantom,tension=0.5,right=0.25}{v2,vb2}
\fmf{phantom,tension=0.5,left=0.25}{v3,vb2}
\fmf{plain,tension=0.5,left=0.25}{v7,vc3}
\fmf{plain,tension=0.5,right=0.25}{v8,vc3}
\fmf{plain,tension=0.5,left=0.125}{v6,vb4}
\fmf{phantom,tension=0.5,right=0.125}{v7,vb4}
\fmf{plain,tension=2,left=0}{vc1,vb1}
\fmf{plain,tension=2,left=0}{vb1,vb3}
\fmf{plain,tension=2,left=0}{vb3,vc3}
\fmf{plain,tension=2,left=0}{vb2,vb4}
\fmffreeze
\fmf{plain,tension=0.5,right=0.25}{v1,va1}
\fmf{plain,tension=0.5,left=0.25}{v2,va1}
\fmf{plain,tension=1}{va1,va2}
\fmf{plain,tension=0}{va2,vb2}
\fmf{plain,tension=0.5,left=0.125}{v5,va2}
\fmf{phantom,tension=0.5,right=0.125}{v6,va2}
\fmf{plain,tension=1,left=0}{vb1,vb2}
\fmf{plain,tension=1,left=0}{vb3,vb4}
\fmf{plain,tension=0.5,right=0,width=1mm}{v5,v8}
}

\newcommand{\chionetwoonethree}[1][black]{%
\fmftop{v1}
\fmfbottom{v5}
\fmfforce{(0.125w,h)}{v1}
\fmfforce{(0.125w,0)}{v5}
\fmffixed{(0.25w,0)}{v1,v2}
\fmffixed{(0.25w,0)}{v2,v3}
\fmffixed{(0.25w,0)}{v3,v4}
\fmffixed{(0.25w,0)}{v5,v6}
\fmffixed{(0.25w,0)}{v6,v7}
\fmffixed{(0.25w,0)}{v7,v8}
\fmffixed{(0,whatever)}{va1,va2}
\fmffixed{(whatever,0)}{va1,vb3}
\fmffixed{(whatever,0)}{va2,vc3}
\fmffixed{(0,whatever)}{vc1,vc3}
\fmffixed{(0,whatever)}{vb2,vb4}
\fmffixed{(0,whatever)}{vc1,vb1}
\fmffixed{(0,whatever)}{vc1,vb3}
\fmffixed{(whatever,0)}{vb1,vb2}
\fmffixed{(whatever,0)}{vb3,vb4}
\fmf{plain,tension=0.5,right=0.25}{v1,vc1}
\fmf{plain,tension=0.5,left=0.25}{v2,vc1}
\fmf{phantom,tension=0.5,right=0.125}{v2,vb2}
\fmf{plain,tension=0.5,left=0.125}{v3,vb2}
\fmf{plain,tension=0.5,left=0.25}{v5,vc3}
\fmf{plain,tension=0.5,right=0.25}{v6,vc3}
\fmf{phantom,tension=0.5,left=0.25}{v6,vb4}
\fmf{phantom,tension=0.5,right=0.25}{v7,vb4}
\fmf{plain,tension=2,left=0}{vc1,vb1}
\fmf{plain,tension=2,left=0}{vb1,vb3}
\fmf{plain,tension=2,left=0}{vb3,vc3}
\fmf{plain,tension=2,left=0}{vb2,vb4}
\fmffreeze
\fmf{phantom,tension=0.5,right=0.125}{v3,va1}
\fmf{plain,tension=0.5,left=0.125}{v4,va1}
\fmf{plain,tension=1}{va1,va2}
\fmf{plain,tension=0}{va1,vb4}
\fmf{plain,tension=0.5,left=0.25}{v7,va2}
\fmf{plain,tension=0.5,right=0.25}{v8,va2}
\fmf{plain,tension=1,left=0}{vb1,vb2}
\fmf{plain,tension=1,left=0}{vb3,vb4}
\fmf{plain,tension=0.5,right=0,width=1mm}{v5,v8}
}

\newcommand{\chionetwoonetwoone}[1][black]{%
\fmftop{v1}
\fmfbottom{v4}
\fmfforce{(0.125w,h)}{v1}
\fmfforce{(0.125w,0)}{v4}
\fmffixed{(0.25w,0)}{v1,v2}
\fmffixed{(0.25w,0)}{v2,v3}
\fmffixed{(0.25w,0)}{v4,v5}
\fmffixed{(0.25w,0)}{v5,v6}
\fmffixed{(0,whatever)}{vc1,vc3}
\fmffixed{(0,whatever)}{vb2,vb4}
\fmffixed{(0,whatever)}{vc2,vb2}
\fmffixed{(0,whatever)}{vc4,vb4}
\fmffixed{(0,whatever)}{vc1,vb1}
\fmffixed{(0,whatever)}{vc1,vb3}
\fmffixed{(whatever,0)}{vc1,vc2}
\fmffixed{(whatever,0)}{vb1,vb2}
\fmffixed{(whatever,0)}{vb3,vb4}
\fmffixed{(whatever,0)}{vc3,vc4}
\fmf{plain,tension=0.5,right=0.25}{v1,vc1}
\fmf{plain,tension=0.5,left=0.25}{v2,vc1}
\fmf{phantom,tension=0.5,right=0.25}{v2,vc2}
\fmf{plain,tension=0.5,left=0.25}{v3,vc2}
\fmf{plain,tension=0.5,left=0.25}{v4,vc3}
\fmf{phantom,tension=0.5,right=0.25}{v5,vc3}
\fmf{plain,tension=0.5,left=0.25}{v5,vc4}
\fmf{plain,tension=0.5,right=0.25}{v6,vc4}
\fmf{plain,tension=2,left=0}{vc1,vb1}
\fmf{plain,tension=2,left=0}{vb1,vb3}
\fmf{plain,tension=2,left=0}{vb3,vc3}
\fmf{plain,tension=2,left=0}{vc2,vb2}
\fmf{plain,tension=2,left=0}{vb2,vb4}
\fmf{plain,tension=2,left=0}{vb4,vc4}
\fmffreeze
\fmf{plain,tension=2,left=0}{vb1,vc2}
\fmf{plain,tension=2,left=0}{vb3,vb2}
\fmf{plain,tension=2,left=0}{vc3,vb4}
\fmf{plain,tension=0.5,right=0,width=1mm}{v4,v6}
}

\DeclareMathOperator{\tr}{tr}

\DeclareMathOperator{\T}{T}

\DeclareMathOperator{\Top}{T}
\DeclareMathOperator{\Kop}{K}
\DeclareMathOperator{\Rop}{R}

\DeclareMathOperator{\D}{D}

\newlength{\eqoff}

\newlength{\unit}
\newlength{\linew}
\usepackage[a4paper,top=2.6cm,bottom=2.6cm,left=2.7cm,right=2.7cm,
marginparsep=0.25cm, 
marginparwidth=2.25cm 
]{geometry}

\usepackage[textsize=tiny]{todonotes}

\newcommand{\Calternative}[2]{#1}

\numberwithin{equation}{section}

\newgray{ogray}{0.85}
\newgray{hatchgray}{1}
\newgray{sgray}{0.8}

\newcommand{\sfillstyle}{solid}

\newlength{\rad}
\newlength{\roff}
\newlength{\ri}
\psset{xunit=\unit,yunit=\unit,runit=\unit}
\newlength{\dlinewidth}
\newlength{\doublesep}
\psset{doublesep=\doublesep}
\psset{linewidth=\linew}
\newlength{\auxlen}
\newlength{\linearc}
\newlength{\xa}
\newlength{\ya}
\newlength{\xb}
\newlength{\yb}
\newlength{\xc}
\newlength{\yc}
\newlength{\xd}
\newlength{\yd}
\newlength{\xe}
\newlength{\ye}
\newlength{\xf}
\newlength{\yf}

\newlength{\yg}

\newlength{\yh}

\newcommand{\recthreevertex}[4][white]{%
\setlength{\xa}{-0.7071\auxlen}
\addtolength{\xa}{-0.7071\dlinewidth}
\setlength{\xb}{0\unit}
\addtolength{\xb}{-0.7071\auxlen}
\setlength{\xc}{-0.7071\auxlen}
\addtolength{\xc}{0.7071\unit}
\setlength{\xd}{-0.7071\auxlen}
\addtolength{\xd}{1.2071\unit}

\setlength{\ya}{-0.7071\auxlen}
\addtolength{\ya}{-0.7071\dlinewidth}
\setlength{\yb}{0\unit}
\addtolength{\yb}{-0.7071\auxlen}
\setlength{\yc}{-0.7071\auxlen}
\addtolength{\yc}{-0.7071\dlinewidth}
\addtolength{\yc}{0.7071\unit}
\setlength{\yd}{0.7071\auxlen}
\addtolength{\yd}{0.7071\dlinewidth}
\addtolength{\yd}{-0.7071\unit}
\setlength{\ye}{0\unit}
\addtolength{\ye}{0.7071\auxlen}
\setlength{\yf}{0.7071\auxlen}
\addtolength{\yf}{0.7071\dlinewidth}
\psset{doubleline=false}
\rput{0}(#2\unit,#3\unit){%
\pscustom[fillstyle=\sfillstyle,fillcolor=#1,linecolor=#1,linewidth=0pt]{%
\rotate{#4}
\psline[liftpen=1,linearc=\linearc](\xa,\yb)(-0.7071\dlinewidth,0)(\xa,\ye)
\psline[liftpen=1,linearc=\linearc](\xb,\yf)(\xc,\yd)(\xd,\yd)
\psline[liftpen=1,linearc=\linearc](\xd,\yc)(\xc,\yc)(\xb,\ya)}
\pscustom{%
\rotate{#4}
\psline[liftpen=1,linearc=\linearc](\xa,\yb)(-0.7071\dlinewidth,0)(\xa,\ye)
\psline[liftpen=2,linearc=\linearc](\xb,\yf)(\xc,\yd)(\xd,\yd)
\psline[liftpen=2,linearc=\linearc](\xd,\yc)(\xc,\yc)(\xb,\ya)
}
}
}

\newcommand{\fourvertex}[4][white]{%
\setlength{\xa}{0\unit}
\addtolength{\xa}{-1\unit}
\setlength{\xb}{0\unit}
\addtolength{\xb}{-0.5\dlinewidth}
\setlength{\xc}{0\unit}
\addtolength{\xc}{0.5\dlinewidth}
\setlength{\xd}{0\unit}
\addtolength{\xd}{1\unit}
\setlength{\ya}{0\unit}
\addtolength{\ya}{-1\unit}
\setlength{\yb}{0\unit}
\addtolength{\yb}{-0.5\dlinewidth}
\setlength{\yc}{0\unit}
\addtolength{\yc}{0.5\dlinewidth}
\setlength{\yd}{0\unit}
\addtolength{\yd}{1\unit}
\psset{doubleline=false}
\rput{0}(#2\unit,#3\unit){%
\pscustom[fillstyle=\sfillstyle,fillcolor=#1,linecolor=#1,linewidth=0pt]{%
\rotate{#4}
\psline[liftpen=1,linearc=\linearc](\xc,\ya)(\xc,\yb)(\xd,\yb)
\psline[liftpen=1,linearc=\linearc](\xd,\yc)(\xc,\yc)(\xc,\yd)
\psline[liftpen=1,linearc=\linearc](\xb,\yd)(\xb,\yc)(\xa,\yc)
\psline[liftpen=1,linearc=\linearc](\xa,\yb)(\xb,\yb)(\xb,\ya)}
\pscustom{%
\rotate{#4}
\psline[liftpen=1,linearc=\linearc](\xc,\ya)(\xc,\yb)(\xd,\yb)
\psline[liftpen=2,linearc=\linearc](\xd,\yc)(\xc,\yc)(\xc,\yd)
\psline[liftpen=2,linearc=\linearc](\xb,\yd)(\xb,\yc)(\xa,\yc)
\psline[liftpen=2,linearc=\linearc](\xa,\yb)(\xb,\yb)(\xb,\ya)
}
}
}

\newcommand{\uoutex}[3][white]{%
\setlength{\xb}{#2\unit}
\addtolength{\xb}{-1.5\unit}
\setlength{\xc}{#2\unit}
\setlength{\ya}{#3\unit}
\addtolength{\ya}{0.5\dlinewidth}
\setlength{\yb}{#3\unit}
\addtolength{\yb}{-0.5\dlinewidth}
\setlength{\yc}{#3\unit}
\addtolength{\yc}{-1.5\unit}
\psset{doubleline=false}
\pscustom[fillstyle=\sfillstyle,fillcolor=#1,linecolor=#1]{%
\psline[liftpen=1,linearc=\linearc](\xc,\yb)(\xb,\yb)(\xb,\yc)
\psline[liftpen=1](\xb,\ya)(\xc,\ya)}
\psline[linearc=\linearc](\xb,\yc)(\xb,\yb)(\xc,\yb)
\psline(\xb,\ya)(\xc,\ya)
}
\newcommand{\doutex}[3][white]{%
\setlength{\xb}{#2\unit}
\addtolength{\xb}{-1.5\unit}
\setlength{\xc}{#2\unit}
\setlength{\ya}{#3\unit}
\addtolength{\ya}{-0.5\dlinewidth}
\setlength{\yb}{#3\unit}
\addtolength{\yb}{0.5\dlinewidth}
\setlength{\yc}{#3\unit}
\addtolength{\yc}{1.5\unit}
\psset{doubleline=false}
\pscustom[fillstyle=\sfillstyle,fillcolor=#1,linecolor=#1]{%
\psline[linearc=\linearc](\xc,\yb)(\xb,\yb)(\xb,\yc)
\psline(\xb,\ya)(\xc,\ya)}
\psline[linearc=\linearc](\xb,\yc)(\xb,\yb)(\xc,\yb)
\psline(\xb,\ya)(\xc,\ya)
}
\newcommand{\dinex}[3][white]{%
\setlength{\xb}{#2\unit}
\addtolength{\xb}{1.5\unit}
\setlength{\xc}{#2\unit}
\setlength{\ya}{#3\unit}
\addtolength{\ya}{-0.5\dlinewidth}
\setlength{\yb}{#3\unit}
\addtolength{\yb}{0.5\dlinewidth}
\setlength{\yc}{#3\unit}
\addtolength{\yc}{1.5\unit}
\psset{doubleline=false}
\pscustom[fillstyle=\sfillstyle,fillcolor=#1,linecolor=#1]{%
\psline[linearc=\linearc](\xc,\yb)(\xb,\yb)(\xb,\yc)
\psline(\xb,\ya)(\xc,\ya)}
\psline[linearc=\linearc](\xb,\yc)(\xb,\yb)(\xc,\yb)
\psline(\xb,\ya)(\xc,\ya)
}
\newcommand{\uinex}[3][white]{%
\setlength{\xb}{#2\unit}
\addtolength{\xb}{1.5\unit}
\setlength{\xc}{#2\unit}
\setlength{\ya}{#3\unit}
\addtolength{\ya}{0.5\dlinewidth}
\setlength{\yb}{#3\unit}
\addtolength{\yb}{-0.5\dlinewidth}
\setlength{\yc}{#3\unit}
\addtolength{\yc}{-1.5\unit}
\psset{doubleline=false}
\pscustom[fillstyle=\sfillstyle,fillcolor=#1,linecolor=#1]{%
\psline[linearc=\linearc](\xc,\yb)(\xb,\yb)(\xb,\yc)
\psline[liftpen=1](\xb,\ya)(\xc,\ya)}
\psline[linearc=\linearc](\xb,\yc)(\xb,\yb)(\xc,\yb)
\psline(\xb,\ya)(\xc,\ya)
}
\newcommand{\ioutex}[3][white]{%
\setlength{\xb}{#2\unit}
\addtolength{\xb}{-1.5\unit}
\setlength{\xc}{#2\unit}
\setlength{\ya}{#3\unit}
\addtolength{\ya}{1.5\unit}
\setlength{\yb}{#3\unit}
\addtolength{\yb}{0.5\dlinewidth}
\setlength{\yc}{#3\unit}
\addtolength{\yc}{-0.5\dlinewidth}
\setlength{\yd}{#3\unit}
\addtolength{\yd}{-1.5\unit}
\psset{doubleline=false}
\pscustom[fillstyle=\sfillstyle,fillcolor=#1,linecolor=#1]{%
\psline[linearc=\linearc](\xc,\yb)(\xb,\yb)(\xb,\ya)
\psline[liftpen=1,linearc=\linearc](\xb,\yd)(\xb,\yc)(\xc,\yc)}
\psline[linecolor=#1,linewidth=\linew](\xb,\ya)(\xb,\yd)
\psline[linearc=\linearc]{-C}(\xc,\yb)(\xb,\yb)(\xb,\ya)
\psline[liftpen=1,linearc=\linearc](\xb,\yd)(\xb,\yc)(\xc,\yc)
}
\newcommand{\iinex}[3][white]{%
\setlength{\xb}{#2\unit}
\addtolength{\xb}{1.5\unit}
\setlength{\xc}{#2\unit}
\setlength{\ya}{#3\unit}
\addtolength{\ya}{1.5\unit}
\setlength{\yb}{#3\unit}
\addtolength{\yb}{0.5\dlinewidth}
\setlength{\yc}{#3\unit}
\addtolength{\yc}{-0.5\dlinewidth}
\setlength{\yd}{#3\unit}
\addtolength{\yd}{-1.5\unit}
\psset{doubleline=false}
\pscustom[fillstyle=\sfillstyle,fillcolor=#1,linecolor=#1,linewidth=0pt]{%
\psline[linearc=\linearc](\xc,\yb)(\xb,\yb)(\xb,\ya)
\psline[liftpen=1,linearc=\linearc](\xb,\yd)(\xb,\yc)(\xc,\yc)}
\psline[linecolor=#1,linewidth=\linew](\xb,\ya)(\xb,\yd)
\psline[linearc=\linearc]{-C}(\xc,\yb)(\xb,\yb)(\xb,\ya)
\psline[liftpen=1,linearc=\linearc](\xb,\yd)(\xb,\yc)(\xc,\yc)
}

\newlength{\armlen}
\newcounter{nnodenum}
\newcounter{mnodenum}
\numberwithin{equation}{section}
\newcommand{\vacpolp}[2][0.5]{%
\fmfcmd{
begingroup;
save t,v,tv,do,di,ppol,pstr,dia;
path ppol,pstr;
pair v[],tv[],do[],di[];
ppol=#2;
t3=arctime (0.5*arclength ppol) of ppol;
v3=point t3 of ppol;
dia=#1 arclength ppol; 
fill(fullcircle scaled dia shifted v3) withcolor 0.2black;
endgroup;
}
}

\newcommand{\ltwoopren}[1]{%
\settoheight{\eqoff}{$\times$}%
\setlength{\eqoff}{0.5\eqoff}%
\addtolength{\eqoff}{-7.5\unitlength}%
\raisebox{\eqoff}{%
\fmfframe(1,-1.25)(1,1.25){%
\begin{fmfchar*}(15,15)
\fmfleft{vl}
\fmfright{vr}
\fmfforce{(0.5w,0)}{vo}
\fmfforce{(0,0.866025h)}{vl}
\fmfforce{(w,0.866025h)}{vr}
\fmffreeze
\fmfposition
{#1}
\fmfiv{decor.shape=circle,decor.filled=full,decor.size=3}{vloc(__vo)}
\end{fmfchar*}}}}

\begin{document}
\begin{fmffile}{graphs1}
\fmfcmd{%
thin := 1pt; 
thick := 2thin;
arrow_len := 4mm;
arrow_ang := 15;
curly_len := 3mm;
dash_len := 1.5mm; 
dot_len := 1mm; 
wiggly_len := 2mm; 
wiggly_slope := 60;
zigzag_len := 2mm;
zigzag_width := 2thick;
decor_size := 5mm;
dot_size := 2thick;
}


\fmfcmd{%
marksize=2mm;
def draw_mark(expr p,a) =
  begingroup
    save t,tip,dma,dmb; pair tip,dma,dmb;
    t=arctime a of p;
    tip =marksize*unitvector direction t of p;
    dma =marksize*unitvector direction t of p rotated -45;
    dmb =marksize*unitvector direction t of p rotated 45;
    linejoin:=beveled;
    draw (-.5dma.. .5tip-- -.5dmb) shifted point t of p;
  endgroup
enddef;
style_def derplain expr p =
    save amid;
    amid=.5*arclength p;
    draw_mark(p, amid);
    draw p;
enddef;
def draw_marks(expr p,a) =
  begingroup
    save t,tip,dma,dmb,dmo; pair tip,dma,dmb,dmo;
    t=arctime a of p;
    tip =marksize*unitvector direction t of p;
    dma =marksize*unitvector direction t of p rotated -45;
    dmb =marksize*unitvector direction t of p rotated 45;
    dmo =marksize*unitvector direction t of p rotated 90;
    linejoin:=beveled;
    draw (-.5dma.. .5tip-- -.5dmb) shifted point t of p withcolor 0white;
    draw (-.5dmo.. .5dmo) shifted point t of p;
  endgroup
enddef;
style_def derplains expr p =
    save amid;
    amid=.5*arclength p;
    draw_marks(p, amid);
    draw p;
enddef;
def draw_markss(expr p,a) =
  begingroup
    save t,tip,dma,dmb,dmo; pair tip,dma,dmb,dmo;
    t=arctime a of p;
    tip =marksize*unitvector direction t of p;
    dma =marksize*unitvector direction t of p rotated -45;
    dmb =marksize*unitvector direction t of p rotated 45;
    dmo =marksize*unitvector direction t of p rotated 90;
    linejoin:=beveled;
    draw (-.5dma.. .5tip-- -.5dmb) shifted point t of p withcolor 0white;
    draw (-.5dmo.. .5dmo) shifted point arctime a+0.25 mm of p of p;
    draw (-.5dmo.. .5dmo) shifted point arctime a-0.25 mm of p of p;
  endgroup
enddef;
style_def derplainss expr p =
    save amid;
    amid=.5*arclength p;
    draw_markss(p, amid);
    draw p;
enddef;
style_def dblderplain expr p =
    save amidm;
    save amidp;
    amidm=.5*arclength p-0.75mm;
    amidp=.5*arclength p+0.75mm;
    draw_mark(p, amidm);
    draw_mark(p, amidp);
    draw p;
enddef;
style_def dblderplains expr p =
    save amidm;
    save amidp;
    amidm=.5*arclength p-0.75mm;
    amidp=.5*arclength p+0.75mm;
    draw_mark(p, amidm);
    draw_marks(p, amidp);
    draw p;
enddef;
style_def dblderplainss expr p =
    save amidm;
    save amidp;
    amidm=.5*arclength p-0.75mm;
    amidp=.5*arclength p+0.75mm;
    draw_mark(p, amidm);
    draw_markss(p, amidp);
    draw p;
enddef;
style_def dblderplainsss expr p =
    save amidm;
    save amidp;
    amidm=.5*arclength p-0.75mm;
    amidp=.5*arclength p+0.75mm;
    draw_marks(p, amidm);
    draw_markss(p, amidp);
    draw p;
enddef;
}

\fmfcmd{%
style_def plain_ar expr p =
  cdraw p;
  shrink (0.6);
  cfill (arrow p);
  endshrink;
enddef;
style_def plain_rar expr p =
  cdraw p; 
  shrink (0.6);
  cfill (arrow reverse(p));
  endshrink;
enddef;
style_def dashes_ar expr p =
  draw_dashes p;
  shrink (0.6);
  cfill (arrow p);
  endshrink;
enddef;
style_def dashes_rar expr p =
  draw_dashes p;
  shrink (0.6);
  cfill (arrow reverse(p));
  endshrink;
enddef;
style_def dots_ar expr p =
  draw_dots p;
  shrink (0.6);
  cfill (arrow p);
  endshrink;
enddef;
style_def dots_rar expr p =
  draw_dots p;
  shrink (0.6);
  cfill (arrow reverse(p));
  endshrink;
enddef;
}

\fmfcmd{%
style_def phantom_cross expr p =
    save amid,ang;
    amid=.5*length p;
    ang= angle direction amid of p;
    draw ((polycross 4) scaled 8 rotated ang) shifted point amid of p;
enddef;
}


\begingroup\parindent0pt
\begin{flushright}\footnotesize
\texttt{HU-MATH-2014-07}\\
\texttt{HU-EP-14/15}
\end{flushright}
\vspace*{2em}
\centering
\begingroup\LARGE
\bf
A piece of cake:
the ground-state energies in
$\gamma_i$-deformed $\mathcal{N}=4$ SYM theory 
\\ 
at leading wrapping order %
\par\endgroup
\vspace{1.5em}
\begingroup\large\bf
Jan Fokken,
Christoph Sieg,
Matthias Wilhelm
\par\endgroup
\thispagestyle{empty}
\vspace{1em}
\begingroup\itshape
Institut f\"ur Mathematik und Institut f\"ur Physik\\
Humboldt-Universit\"at zu Berlin\\
IRIS Geb\"aude \\
Zum Grossen Windkanal 6 \\
12489 Berlin
\par\endgroup
\vspace{1em}
\begingroup\ttfamily
fokken, csieg, mwilhelm @physik.hu-berlin.de \\
\par\endgroup
\vspace{1.5em}
\endgroup

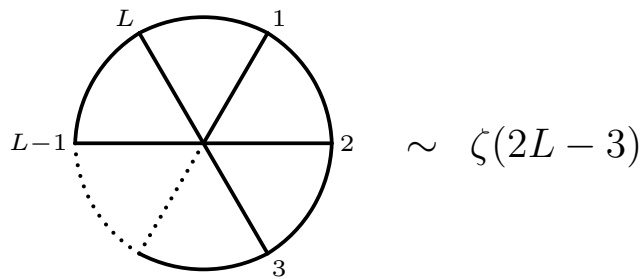
\vspace{0.5cm}
\begin{equation*}
\scalebox{1.5}{
\settoheight{\eqoff}{$\times$}%
\setlength{\eqoff}{0.5\eqoff}%
\addtolength{\eqoff}{-13\unitlength}%
\raisebox{\eqoff}{%
\fmfframe(3,-2)(0,-2){%
\begin{fmfchar*}(30,30)
  \fmfleft{in}
  \fmfright{out1}
\fmf{phantom}{in,v1}
\fmf{phantom}{out,v2}
\fmfforce{(0,0.5h)}{in}
\fmfforce{(w,0.5h)}{out}
\fmfpoly{phantom}{v1,va4,va3,v2,va2,va1}
\fmffixed{(0.75w,0)}{v1,v2}
\fmf{phantom}{vc,v1}
\fmf{plain}{vc,v2}
\fmffreeze
\fmf{plain,left=0.25}{v1,va1}
\fmf{plain,left=0.25}{va1,va2}
\fmf{plain,left=0.25}{va2,v2}
\fmf{plain,left=0.25}{v2,va3}
\fmf{plain,left=0.25}{va3,va4}
\fmf{dots,left=0.25}{va4,v1}
\fmf{plain}{vc,va1}
\fmf{plain}{vc,va2}
\fmf{plain}{vc,va3}
\fmf{dots}{vc,va4}
\fmf{plain}{vc,v1}
\fmffreeze
\fmfv{l=$\scriptscriptstyle L$,l.dist=2}{va1}
\fmfv{l=$\scriptscriptstyle 1$,l.dist=2}{va2}
\fmfv{l=$\scriptscriptstyle 2$,l.dist=2}{v2}
\fmfv{l=$\scriptscriptstyle 3$,l.dist=2}{va3}
\fmfv{l=$\scriptscriptstyle L-1$,l.dist=2}{v1}
\end{fmfchar*}}}
}
\quad
\scalebox{1.3}{\text{$\sim$}}\quad
\scalebox{1.3}{\text{$\zeta(2L-3)$}}
\end{equation*}
\vspace{0.5cm}

\paragraph{Abstract.}
In the non-supersymmetric $\gamma_i$-deformed $\mathcal{N}=4$ SYM theory,
the scaling dimensions of the operators $\tr[Z^L]$ composed of $L$ scalar fields $Z$ receive finite-size wrapping and prewrapping corrections in the 't Hooft limit.  
In this paper, we calculate these scaling dimensions to leading wrapping order directly from Feynman diagrams. For  $L\ge3$, the result is 
proportional to the  maximally transcendental `cake' integral. 
It matches with an earlier result obtained from the integrability-based 
L\"uscher corrections, TBA and Y-system equations. 
At $L=2$, where the integrability-based equations yield infinity, we find a finite rational result. This result is renormalization-scheme dependent 
due to the non-vanishing $\beta$-function of an induced quartic scalar double-trace 
coupling, on which we have reported earlier. This explicitly shows that conformal invariance is broken -- even in the 't Hooft limit.


\newpage


\section{Introduction and summary}

\label{sec:introduction}

In this paper, we provide a field-theoretic test of integrability in the $\gamma_i$-deformed $\mathcal{N}=4$ SYM theory ($\gamma_i$-deformation). This theory was proposed as the field-theory part of a non-supersymmetric example of the $\AdS/\CFT$ correspondence \cite{Frolov:2005dj}, which is obtained by applying a three-parameter deformation to both sides of the original correspondence \cite{Maldacena:1997re,Gubser:1998bc,Witten:1998qj}. On the string theory side, three consecutive T-duality, shift, T-duality (TsT) transformations -- each depending on one of the real parameters $\gamma_i$, $i=1,2,3$ -- are applied to the $\text{S}^5$ factor of the $\AdS_5\times\text{S}^5$ background. This breaks the $SO(6)$ isometry group to its $U(1)_{q^1}\times U(1)_{q^2}\times U(1)_{q^3}$ Cartan subgroup. On the gauge-theory side, phase factors deform the Yukawa-type (fermion-fermion-scalar) and F-term (four-scalar) couplings of the $\mathcal{N}=4$ SYM theory with gauge group $SU(N)$. They depend on the $\gamma_i$ and the three Cartan charges $(q^1,q^2,q^3)$. In the limit of equal deformation parameters $\gamma_1=\gamma_2=\gamma_3=-\pi\beta$, a simple ($\mathcal{N}=1$) supersymmetry is restored, and one obtains the setup of Lunin and Maldacena \cite{Lunin:2005jy}. The gauge theory becomes the (real) $\beta$-deformation, which is a special case of the exactly marginal deformations of $\mathcal{N}=4$ SYM theory classified by Leigh and Strassler \cite{Leigh:1995ep}. Like the undeformed $\mathcal{N}=4$ SYM theory, also these deformations are most  accessible in the 't Hooft (planar) limit \cite{'tHooft:1973jz}, where $N\to\infty$ and the Yang-Mills coupling constant $g_\YM\to0$ such that the 't Hooft  coupling $\lambda=g_\YM^2N$ is kept fixed. In this limit, the string theory becomes free and in the gauge theory non-planar vacuum diagrams are suppressed.\footnote{Non-planar non-vacuum diagrams may, however, become planar when connected to external states, and thus may contribute in the 't Hooft limit \cite{Sieg:2005kd}. They give rise to finite-size effects, which are the main subject of this work.}

The $\beta$- and $\gamma_i$-deformation share certain important properties with their parent $\mathcal{N}=4$ SYM theory. One of these is claimed to be integrability in the planar limit. In the asymptotic regime, i.e.\ in the absence of finite-size effects, the dilatation operator can be obtained directly from its undeformed counterpart via a relation\footnote{This relation follows when a theorem formulated by Filk for spacetime non-commutative field theories in \cite{Filk:1996dm} is adapted to the deformed theories, implementing the deformations via noncommutative Moyal-like $\ast$-products \cite{Lunin:2005jy}.} between planar single-trace Feynman diagrams of elementary interactions: in the deformed theory such a diagram is given by its undeformed counterpart multiplied by a phase factor which is determined from the order and $(q^1,q^2,q^3)$-charge of the external fields alone. This relation was used in \cite{Beisert:2005if} to determine the one-loop dilatation operator in terms of the $\mathcal{N}=4$ SYM theory expression \cite{Beisert:2003jj}.
The obtained result, as well as the deformed gravity background \cite{Frolov:2005dj}, is compatible \cite{Roiban:2003dw,Beisert:2005if} with the integrability found in the original $\AdS/\CFT$ correspondence, see the review collection \cite{Beisert:2010jr} and in particular chapter \cite{Zoubos:2010kh} therein. In the integrability-based approach, the deformation can be incorporated by introducing twists into the boundary conditions of the asymptotic Bethe ansatz \cite{Beisert:2005if}.\footnote{The twisted Bethe ansatz can be derived from a twisted transfer matrix \cite{Gromov:2010dy} corresponding to operational twisted boundary conditions \cite{Arutyunov:2010gu} or, alternatively, a twisted S-matrix \cite{Ahn:2010ws}.}

A simple test of the claimed integrability, also beyond the asymptotic regime, can be performed by analyzing the spectrum of composite operators that are protected in the $\mathcal{N}=4$ SYM theory but acquire anomalous dimensions in the $\beta$- and $\gamma_i$-deformation. If, in addition, such operators are determined uniquely by their global charges, operator mixing cannot occur. Thus, the calculations become relatively simple but still yield  highly non-trivial results. In the $\beta$-deformation, the single-impurity operators of the $SU(2)$ subsectors are of this type. For generic lengths $L\ge2$, they are given by
\begin{equation}\label{firstexstate}
\mathcal{O}_{L,1}=\tr[(\phi^i)^{L-1}\phi^j]
\col\qquad i,j=1,2,3
\col\qquad j\neq i
\col
\end{equation}
and they correspond to single-magnon states in the  spin-chain picture. In the asymptotic regime, their anomalous dimensions (energies in the spin-chain picture) are determined by the dispersion relation of the twisted Bethe ansatz \cite{Beisert:2005if}. These findings can directly be verified in the Feynman diagram approach, where the modifications that capture the deformation \cite{Fiamberti:2008sm} can easily be incorporated into the explicit three-loop calculation of \cite{Sieg:2010tz} and the all-loop argument of \cite{Gross:2002su}. Beyond the asymptotic regime, finite-size corrections have to be taken into account. For the operators \eqref{firstexstate} with $L\ge 3$, these are the wrapping corrections\footnote{Their general properties were first analyzed in the Feynman diagram approach in \cite{Sieg:2005kd} and then in the context of $\AdS/\CFT$-integrability in  \cite{Ambjorn:2005wa}.}, which start at loop order $K=L$ called critical wrapping order. By a direct Feynman diagram calculation at this order, explicit results were obtained up to eleven loops \cite{Fiamberti:2008sn}. These results were successfully reproduced in \cite{Gunnesson:2009nn} for $\beta=\frac{1}{2}$ and in \cite{Gromov:2010dy} and \cite{Arutyunov:2010gu} for generic $\beta$, based on the L\"uscher corrections, Y-system and TBA equations, respectively.

There are, however, important properties of the deformations that are not rooted in the undeformed $\mathcal{N}=4$ SYM theory; in particular, they affect the anomalous dimensions and hence integrability, forcing us to discuss the $L=2$ case of the operators \eqref{firstexstate} separately. While the $\mathcal{N}=4$ SYM theory is essentially the same if the gauge group $SU(N)$ is replaced by $U(N)$, this is no longer the case after the the theory is deformed. In the $\beta$-deformation with gauge group $U(N)$, quantum corrections induce the running of a quartic scalar double-trace coupling, which breaks conformal invariance \cite{Freedman:2005cg}. In the $SU(N)$ theory, this coupling is at its non-vanishing IR fix point value \cite{Hollowood:2004ek}, such that this theory is conformal. As explained in detail in our recent works \cite{Fokken:2013aea,Fokken:2013mza}, the role of this double-trace coupling in the $SU(N)$ theory can be understood in terms of the finite-size effect of prewrapping, which is caused by the absence of the $U(1)$ mode in the $SU(N)$ theory.
Although this double-trace coupling has a prefactor of $\frac{1}{N}$, it can contribute at the leading (planar) order in the large-$N$ expansion: the underlying mechanism is the same as in the case of wrapping, but the contributions start one loop order earlier, i.e.\ it can affect length-$L$ operators already at $K=L-1$ loops. For the operators \eqref{firstexstate}, this occurs only at $L=2$. The anomalous dimension of $\mathcal{O}_{2,1}$ is vanishing to all loop orders in the $SU(N)$ theory \cite{Fokken:2013mza},\footnote{See \cite{Freedman:2005cg} and \cite{Penati:2005hp} for explicit one- and two-loop calculations, respectively.} while it is non-vanishing already at one loop in the $U(N)$ theory without tree-level double-trace coupling \cite{Freedman:2005cg}. At one loop, the dilatation operator and twisted asymptotic Bethe ansatz of \cite{Beisert:2005if} reproduce the latter result for $\mathcal{O}_{2,1}$. In \cite{Fokken:2013mza}, we have incorporated the prewrapping effect into the one-loop dilatation operator of \cite{Beisert:2005if}, which then captures the complete one-loop spectrum of the $\beta$-deformation with $SU(N)$ gauge group. It is an open problem how to incorporate prewrapping in addition to wrapping into the integrability-based approach of L\"uscher corrections, TBA and Y-system equations. In fact, the present TBA result of \cite{Arutyunov:2010gu} for the operators \eqref{firstexstate} is logarithmically divergent when evaluated at $L=2$.\footnote{Such a divergence was encountered earlier in the expressions for the ground-state energy of the TBA \cite{Frolov:2009in}. In \cite{deLeeuw:2012hp}, it was found that the divergent ground-state energy vanishes in the undeformed theory when a regulating twist is introduced in the $\text{AdS}_5$ directions. This regularization extends to the ground state of the supersymmetric deformations \cite{FrolovPC}.%
}

In the $\gamma_i$-deformation with either gauge group $SU(N)$ or $U(N)$, a further type of double-trace coupling occurs in the action \cite{Fokken:2013aea}. It reads 
\begin{equation}\label{Qiiii}
-\frac{g_\YM^2}{N}\sum_{i=1}^3(Q^{ii}_{\text{F}\,ii}+\delta Q^{ii}_{\text{F}\,ii})\tr[\bar\phi_i\bar\phi_i]\tr[\phi^i\phi^i]
\col
\end{equation}
where throughout this paper Einstein's convention of implicit summation never applies. In this expression, $Q^{ii}_{\text{F}\,ii}$ denotes the (undetermined) tree-level coupling constant, which has to be included in the action since one-loop corrections induce a counter-term contribution $\delta Q^{ii}_{\text{F}\,ii}$. In combination with the self-energy counter term of the scalar fields, it generates the $\beta$-function 
\begin{equation}
\begin{aligned}\label{betaQ}
\beta_{Q_{\text{F}\,ii}^{ii}}
&=4g^2\big(4\sin^2\gamma_i^+\sin^2\gamma_i^-+(Q^{ii}_{\text{F}\,ii})^2\big)+\mathcal{O}(g^4)
\col
\end{aligned}
\end{equation}
where $g=\frac{\sqrt{\lambda}}{4\pi}$ is the effective planar coupling constant and \Calternative{}{the $\gamma_i^\pm$ for $i=1,2,3$ read}
\begin{equation}\label{gammapmdef}
\gamma_1^\pm=\mp\frac{1}{2}(\gamma_2\pm\gamma_3)
\col\qquad
\gamma_2^\pm=\mp\frac{1}{2}(\gamma_3\pm\gamma_1)
\col\qquad
\gamma_3^\pm=\mp\frac{1}{2}(\gamma_1\pm\gamma_2)\pnt
\end{equation}
The function $\beta_{Q_{\text{F}\,ii}^{ii}}$ has no fix-points in the perturbative regime of small $g$. Hence, this type of double-trace coupling is running, and conformal invariance is broken in the $\gamma_i$-deformation; see our paper \cite{Fokken:2013aea} for a detailed discussion in the context of the AdS/CFT correspondence. In analogy to the double-trace coupling in the $\beta$-deformation, also the coupling \eqref{Qiiii} has a prefactor of $\tfrac{1}{N}$ and enters the planar spectrum of the theory via a finite-size effect one-loop order earlier than the critical wrapping order.\footnote{Hence, as already explained in our paper \cite{Fokken:2013aea}, even in the planar limit conformal invariance is broken by the running of the double-trace coupling \eqref{Qiiii}. In this paper, we demonstrate this at an explicit example. In the later work \cite{Jin:2013baa}, the running of the double-trace coupling \eqref{Qiiii} was confirmed. Note, however, that the author of \cite{Jin:2013baa} nevertheless claims that the $\gamma_i$-deformation is `conformally invariant in the planar limit'.} By a slight generalization of the notion, we also associate it with prewrapping.

Operators even simpler than those in \eqref{firstexstate} allow for a test of the claimed integrability in the $\gamma_i$-deformation. For generic lengths $L\ge2$, these operators are given by
\begin{equation}\label{groundstate}
\mathcal{O}_L=\tr[(\phi^i)^L]\col
\end{equation}
and they correspond to ground states in the  spin-chain picture. They have the properties mentioned above \eqref{firstexstate}, i.e.\ they are protected in the $\mathcal{N}=4$ SYM theory and are uniquely determined by their global charges. In contrast to the single-impurity operators $\mathcal{O}_{L,1}$, they are even protected in the $\beta$-deformation. In the $\gamma_i$-deformation, they do not receive corrections from the twisted Bethe ansatz at the asymptotic level \cite{Ahn:2011xq}, but solely from finite-size effects. For $L\ge3$, their anomalous dimensions were determined in \cite{Ahn:2011xq} in the integrability-based approach as L\"uscher corrections and from the TBA and Y-system equations up to next-to-leading wrapping order. At $L=2$, the equations of \cite{Ahn:2011xq} diverge in a similar fashion as those of the $\beta$-deformation mentioned above.

In this paper, we determine the planar anomalous dimensions of the operators \eqref{groundstate} at leading wrapping order directly from Feynman diagrams. For $L\ge 3$, the calculation can be reduced to only four Feynman diagrams. They are proportional to the maximally transcendental `cake' integral of \cite{Broadhurst:1985vq} and hence to the Riemann $\zeta$-function $\zeta(2L-3)$, such that we find 
\begin{equation}
\begin{aligned}
\gamma_{\mathcal{O}_L}&=-64g^{2L}\sin^2\frac{L\gamma_i^+}{2}\sin^2\frac{L\gamma_i^-}{2}
\binom{2L-3}{L-1}\zeta(2L-3)
\col
\end{aligned}
\end{equation}
where $\gamma_i^\pm$ are defined in \eqref{gammapmdef}.
Our result for $\gamma_{\mathcal{O}_L}$ matches the leading-order expression obtained in \cite{Ahn:2011xq} from integrability.\footnote{Note that one has to absorb a factor of $2$ into $g$ and a factor $L$ into $\gamma_i^\pm$ in order to match the definitions of \cite{Ahn:2011xq}.}  For $L=2$, we obtain the following result for the planar anomalous dimension:
\begin{equation}\label{gammaO2res}
\begin{aligned}
\gamma_{\mathcal{O}_2}&=4g^2Q_{\text{F}\,ii}^{ii}-32g^4\sin^2\gamma_i^+\sin^2\gamma_i^- -2g^2\varrho\,\beta_{Q_{\text{F}\,ii}^{ii}}
\pnt
\end{aligned}
\end{equation}
Already at one loop, it receives a contribution which is proportional to the tree-level coupling $Q_{\text{F}\,ii}^{ii}$ in \eqref{Qiiii} and entirely originates from prewrapping. For $Q_{\text{F}\,ii}^{ii}=0$, the remaining two-loop term can be traced back to wrapping diagrams only and a counter-term contribution involving $\delta Q_{\text{F}\,ii}^{ii}$ of \eqref{Qiiii}. Since $Q_{\text{F}\,ii}^{ii}$ is running, the two-loop term of \eqref{gammaO2res} depends on the chosen renormalization scheme. This scheme dependence is indicated by the parameter $\varrho$, and it is proportional to the $\beta$-function for $Q_{\text{F}\,ii}^{ii}$ given in \eqref{betaQ}.\footnote{It is a well-known fact that in a conformal field theory the (anomalous) scaling dimensions of gauge-invariant composite operators are observables and are hence renormalization-scheme independent. The presence of $\varrho$ in the planar anomalous dimension \eqref{gammaO2res} therefore explicitly shows that the $\gamma_i$-deformation is not conformally invariant in the planar limit -- in contrast to the claim of \cite{Jin:2013baa}.} In the dimensional reduction (DR) scheme used in the main part of this paper, we have $\varrho=0$, while in the modified dimensional reduction ($\overline{\text{DR}}$) scheme we have $\varrho=-\gamma_{\text{E}}+\ln4\pi$.\footnote{The DR and $\overline{\text{DR}}$ schemes are the supersymmetric analoga of the widely used minimal subtraction (MS) scheme of \cite{'tHooft:1973mm} and the modified minimal subtraction ($\overline{\text{MS}}$) scheme of \cite{Bardeen:1978yd}, respectively.}

In \cite{Fokken:2013aea}, we have proposed the following test of the integrability-based approach, which involves the $L=2$ result \eqref{gammaO2res}. The first step is to find a finite and correct integrability-based description for the $L=2$ single-impurity operator \eqref{firstexstate} in the conformal $\beta$-deformation (with gauge group $SU(N)$). Such a description must exist if this theory is integrable as claimed. The second step is to apply the resulting modified description to the $L=2$ ground-state operator \eqref{groundstate} in the $\gamma_i$-deformation. 
If the corresponding equations still yield an infinite result for the anomalous dimension, we can associate the previously encountered divergence in the $L=2$ states with the running of a contributing double-trace coupling and hence the breakdown of conformal invariance. If the resulting value is, however, finite, chances are high that it coincides with the expression \eqref{gammaO2res} for a particular choice of the tree-level coupling $Q_{\text{F}\,ii}^{ii}$ and the scheme, i.e.\ the parameter $\varrho$. In particular, the two-loop contribution in \eqref{gammaO2res} has the same functional dependence on the deformation parameters $\gamma_i$ as the one found in \cite{Ahn:2011xq} from the integrability-based equations in the following cases: $Q_{\text{F}\,ii}^{ii}=0$, or  $Q_{\text{F}\,ii}^{ii}\propto\sin^2\gamma_i^+\sin^2\gamma_i^-$, or $\varrho=0$ and $Q_{\text{F}\,ii}^{ii}$ arbitrary.
The integrability-based description might then capture also the non-conformal theory in a fixed scheme. Further tests of prewrapping-affected states sensitive to the non-conformality would be required to check if this is indeed the case.

This paper is organized as follows. 
In Section~\ref{sec:wrapdefdep}, we analyze the diagrams which determine the planar anomalous dimensions of the composite operators \eqref{groundstate} and formulate restrictive criteria for them to have a non-trivial deformation dependence. Since the contributions from the deformation-independent diagrams can be reconstructed from the deformation-dependent ones, this drastically reduces the calculational effort. Section~\ref{sec:calculation} contains the main part of the calculation, which treats the $L\ge3$ case and the $L=2$ case in Subsections \ref{sec:Lge3} and \ref{sec:Le2}, respectively. In Appendix~\ref{app:action}, we present the action of the $\gamma_i$-deformation as well as our notation and conventions. Some auxiliary identities for the calculation in Section~\ref{sec:calculation} are derived in Appendix~\ref{app:tensorid}. We refer the reader to Appendix~\ref{app:renormalization} for a short review of the renormalization theory of composite operators. In Appendix~\ref{app:schemedep}, we discuss the renormalization-scheme dependence emerging at $L=2$.

\section{Deformation-dependence of diagrams}
\label{sec:wrapdefdep}
In this section, we analyze the diagrams which contribute to the renormalization of the composite operators \eqref{groundstate} at any loop order $K$. We identify a subclass of them which contains all diagrams with  a non-trivial deformation dependence. Only these diagrams have to be evaluated explicitly. The contribution from the deformation-independent ones can be reconstructed using the fact that the operators \eqref{groundstate} are protected in the undeformed theory as well as in the $\beta$-deformation.

As reviewed in Appendix \ref{app:renormalization}, the renormalization constant $\mathcal{Z}_{\mathcal{O}_L}$ of the composite operators $\mathcal{O}_L$ is determined by the same diagrams that yield the UV divergence of the connected Green function $\langle\mathcal{O}_L(x)\bar\phi_i(x_1)\dots\bar\phi_i(x_L)\rangle_{\text{c}}$ (albeit occurring in different linear combinations). Each diagram contributing to the connected Green function consists of the operator $\mathcal{O}_L$ and a subdiagram of the elementary interactions, which contains all information on the deformation dependence. If we remove $\mathcal{O}_L$, the resulting subdiagram is a direct product of $c$ connected pieces, which we label by $\xi=1,\dots,c$. In each piece, $R_\xi$ external fields $\phi^i$ and $R_\xi$ external fields $\bar\phi_i$ interact, where the $R_\xi$ obey the condition $\sum_\xi R_\xi=L$.
For $K\le L-2$ loops, each such piece is a planar single-trace diagram with color structure $\tr[(\phi^i)^{R_\xi}(\bar\phi_i)^{R_\xi}]$. For $K\ge L-1$, all fields of the operator can also interact in a single non-planar piece ($c=1$ and $R_1=L$) such that the respective subdiagram has the double-trace color structure $\tr[(\phi^i)^L]\tr[(\bar\phi_i)^L]$.

As mentioned in the introduction, a planar single-trace diagram of elementary interactions in the $\gamma_i$- and $\beta$-deformation is given by its counterpart in the undeformed parent theory times a phase factor which is determined from the order and $(q^1,q^2,q^3)$-charge of the external fields alone. This relation is based on the adaption of Filk's theorem for spacetime-noncommutative field theories \cite{Filk:1996dm}, and in the formulation of \cite{Khoze:2005nd} it reads
\begin{equation}\label{diagrel}
\begin{aligned}
\settoheight{\eqoff}{$+$}%
\setlength{\eqoff}{0.5\eqoff}%
\addtolength{\eqoff}{-5.5\unit}%
\raisebox{\eqoff}{%
\begin{pspicture}(-2.5,-1)(18.5,10)
\rput[r](1.5,9){$\scriptstyle A_{R}$}
\rput[r](1.5,6){$\scriptstyle A_{R-1}$}
\rput[r](1.5,0){$\scriptstyle A_{1}$}
\rput[l](14.5,0){$\scriptstyle A_{2R}$}
\rput[l](14.5,6){$\scriptstyle A_{R+2}$}
\rput[l](14.5,9){$\scriptstyle A_{R+1}$}
\uinex{2}{9}
\iinex{2}{6}
\dinex{2}{0}
\setlength{\ya}{9\unit}
\addtolength{\ya}{0.5\dlinewidth}
\setlength{\yb}{0\unit}
\addtolength{\yb}{-0.5\dlinewidth}
\setlength{\xc}{7.5\unit}
\setlength{\yc}{4.5\unit}
\addtolength{\yc}{-0.5\dlinewidth}
\setlength{\xd}{8.5\unit}
\setlength{\yd}{4.5\unit}
\addtolength{\yd}{0.5\dlinewidth}
\psline(3.5,\ya)(12.5,\ya)
\psline[linestyle=dotted](3.5,4.5)(3.5,1.5)
\psline(12.5,\yb)(3.5,\yb)
\doutex{14}{0}
\ioutex{14}{6}
\uoutex{14}{9}
\psline[linestyle=dotted](12.5,4.5)(12.5,1.5)
\setlength{\xa}{3.5\unit}
\addtolength{\xa}{\dlinewidth}
\setlength{\xb}{12.5\unit}
\addtolength{\xb}{-\dlinewidth}
\setlength{\ya}{9\unit}
\addtolength{\ya}{-0.5\dlinewidth}
\setlength{\yb}{0\unit}
\addtolength{\yb}{0.5\dlinewidth}
\pscustom[linecolor=gray,fillstyle=solid,fillcolor=gray,linearc=\linearc]{%
\psline(\xa,4.5)(\xa,\ya)(\xb,\ya)(\xb,4.5)
\psline[liftpen=2](\xb,4.5)(\xb,\yb)(\xa,\yb)(\xa,4.5)
}
\rput(8,6){planar}
\rput(8,3){$\gamma_i$-def.}
\end{pspicture}}
&=
\settoheight{\eqoff}{$+$}%
\setlength{\eqoff}{0.5\eqoff}%
\addtolength{\eqoff}{-5.5\unit}%
\raisebox{\eqoff}{%
\begin{pspicture}(-2.5,-1)(18.5,10)
\rput[r](1.5,9){$\scriptstyle A_{R}$}
\rput[r](1.5,6){$\scriptstyle A_{R-1}$}
\rput[r](1.5,0){$\scriptstyle A_{1}$}
\rput[l](14.5,0){$\scriptstyle A_{2R}$}
\rput[l](14.5,6){$\scriptstyle A_{R+2}$}
\rput[l](14.5,9){$\scriptstyle A_{R+1}$}
\uinex{2}{9}
\iinex{2}{6}
\dinex{2}{0}
\setlength{\ya}{9\unit}
\addtolength{\ya}{0.5\dlinewidth}
\setlength{\yb}{0\unit}
\addtolength{\yb}{-0.5\dlinewidth}
\setlength{\xc}{7.5\unit}
\setlength{\yc}{4.5\unit}
\addtolength{\yc}{-0.5\dlinewidth}
\setlength{\xd}{8.5\unit}
\setlength{\yd}{4.5\unit}
\addtolength{\yd}{0.5\dlinewidth}
\psline(3.5,\ya)(12.5,\ya)
\psline[linestyle=dotted](3.5,4.5)(3.5,1.5)
\psline(12.5,\yb)(3.5,\yb)
\doutex{14}{0}
\ioutex{14}{6}
\uoutex{14}{9}
\psline[linestyle=dotted](12.5,4.5)(12.5,1.5)
\setlength{\xa}{3.5\unit}
\addtolength{\xa}{\dlinewidth}
\setlength{\xb}{12.5\unit}
\addtolength{\xb}{-\dlinewidth}
\setlength{\ya}{9\unit}
\addtolength{\ya}{-0.5\dlinewidth}
\setlength{\yb}{0\unit}
\addtolength{\yb}{0.5\dlinewidth}
\pscustom[linecolor=gray,fillstyle=solid,fillcolor=gray,linearc=\linearc]{%
\psline(\xa,4.5)(\xa,\ya)(\xb,\ya)(\xb,4.5)
\psline[liftpen=2](\xb,4.5)(\xb,\yb)(\xa,\yb)(\xa,4.5)
}
\rput(8,6){planar}
\rput(8,3){$\mathcal{N}=4$}
\end{pspicture}}
\,\Phi(A_{1}\ast A_{2}\ast\dots
\ast A_{2R})
\pnt
\end{aligned}
\end{equation}
where the arbitrary planar elementary interactions between the external fields $A_n$, $n=1,\dots,2R$ are depicted as gray-shaded regions. The operator $\Phi$ extracts the phase factor of its argument, which is determined by the non-commutative $\ast$-product defined in \eqref{astprod}. Relation \eqref{diagrel} directly applies to each of the $c$ connected single-trace pieces of the subdiagram of elementary interactions. In this case, $A_1,\dots,A_R$ become identical scalar fields $\phi^i$ and $A_{R+1},\dots,A_{2R}$ become the respective anti-scalar fields $\bar\phi_i$, where $R\in\{R_1,\dots,R_c\}$. The $\ast$-products then reduce to ordinary products yielding $\Phi=1$, and correspondingly each piece individually and the subdiagram as a whole is deformation-independent. In the asymptotic regime, i.e.\ for loop orders $K\le L-2$, these deformation-independent diagrams are the only contributions to the renormalization constant $\mathcal{Z}_{\mathcal{O}_L}$. At least for $K\le L-2$ loops, the composite operators \eqref{groundstate} are thus protected as in the parent $\mathcal{N}=4$ SYM theory.

At $K\ge L-1$ loops, also diagrams containing connected subdiagrams with double-trace structure $\tr[(\phi^i)^L]\tr[(\bar\phi_i)^L]$ can contribute. They are associated with finite-size effects, i.e with the prewrapping and wrapping corrections at $K\ge L-1$ and $K\ge L$ loops, respectively. These diagrams are not captured by relation \eqref{diagrel}. Moreover, their deformation-dependence cannot be determined from the extension of relation \eqref{diagrel} to multi-trace diagrams formulated in \cite{Fokken:2013mza}, since their individual trace factors carry net $(q^1,q^2,q^3)$-charge.

Subdiagrams associated with prewrapping contributions contain couplings or contributions to the propagator that are of double-trace type. The prewrapping effect already present in the $\beta$-deformation cannot affect the operators \eqref{groundstate}, as discussed in \cite{Fokken:2013mza}. Hence, the coupling \eqref{Qiiii} is the only source of prewrapping contributions to $\mathcal{Z}_{\mathcal{O}_L}$. According to the criteria developed in \cite{Fokken:2013mza}, this coupling can only contribute if one of its trace factors carries the same $(q^1,q^2,q^3)$-charge as the operator $\mathcal{O}_L$. This restricts prewrapping contributions to $L=2$. Since the coupling \eqref{Qiiii} is deformation-dependent, so are these prewrapping contributions.

Subdiagrams associated with wrapping contributions contain loops that wrap around the $L$ external fields $\bar\phi_i$ thereby generating the double-trace structure. By imposing conditions on the wrapping loops, the sum of all wrapping-type subdiagrams can be decomposed into two classes one of which contains only deformation-independent diagrams.
This decomposition reads
\begin{equation}\label{wrapdiagdecomp}
\settoheight{\eqoff}{$+$}%
\setlength{\eqoff}{0.5\eqoff}%
\addtolength{\eqoff}{-10.5\unit}%
\raisebox{\eqoff}{%
\begin{pspicture}(-0.5,-6)(21.5,15)
%
\rput[r](1.5,9){$\scriptstyle \phi^i$}
\rput[r](1.5,0){$\scriptstyle \phi^i$}
\rput[l](14.5,0){$\scriptstyle \bar\phi_i$}
\rput[l](14.5,9){$\scriptstyle \bar\phi_i$}
%
\uinex{2}{9}
\iinex{2}{6}
\dinex{2}{0}
\setlength{\xa}{4\unit}
\addtolength{\xa}{-0.5\dlinewidth}
\setlength{\ya}{9\unit}
\addtolength{\ya}{0.5\dlinewidth}
\setlength{\xb}{6.5\unit}
\addtolength{\xb}{-0.5\dlinewidth}
\setlength{\yb}{14\unit}
\addtolength{\yb}{0.5\dlinewidth}
\setlength{\xc}{9.5\unit}
\addtolength{\xc}{0.5\dlinewidth}
\setlength{\yc}{11\unit}
\addtolength{\yc}{-0.5\dlinewidth}
\setlength{\xd}{12\unit}
\addtolength{\xd}{0.5\dlinewidth}
\setlength{\yd}{0\unit}
\addtolength{\yd}{-0.5\dlinewidth}
\setlength{\xe}{17.5\unit}
\addtolength{\xe}{-0.5\dlinewidth}
\setlength{\ye}{-2\unit}
\addtolength{\ye}{0.5\dlinewidth}
\setlength{\xf}{20.5\unit}
\addtolength{\xf}{0.5\dlinewidth}
\setlength{\yf}{-5\unit}
\addtolength{\yf}{-0.5\dlinewidth}
\setlength{\yg}{8\unit}
\addtolength{\yg}{-0.5\dlinewidth}
\setlength{\yh}{1\unit}
\addtolength{\yh}{0.5\dlinewidth}
\psline[liftpen=1,linearc=\linearc](\xa,\ya)(\xb,\ya)(\xb,\yb)(\xf,\yb)(\xf,\yg)
\psline[liftpen=1,linearc=\linearc](\xf,\yg)(\xf,\yh)
\psline[liftpen=1,linearc=\linearc](\xf,\yh)(\xf,\yf)(\xb,\yf)(\xb,\yd)(\xa,\yd)
\psline[liftpen=1,linearc=\linearc](\xd,\ya)(\xc,\ya)(\xc,\yc)(\xe,\yc)(\xe,\yg)
\psline[liftpen=1,linearc=\linearc](\xe,\yg)(\xe,\yh)
\psline[liftpen=1,linearc=\linearc](\xe,\yh)(\xe,\ye)(\xc,\ye)(\xc,\yd)(\xd,\yd)
%
%
\psline[linestyle=dotted](3.5,4.5)(3.5,1.5)
\doutex{14}{0}
\ioutex{14}{6}
\uoutex{14}{9}
\psline[linestyle=dotted](12.5,4.5)(12.5,1.5)
\setlength{\xa}{4\unit}
\addtolength{\xa}{0.5\dlinewidth}
\setlength{\ya}{9\unit}
\addtolength{\ya}{-0.5\dlinewidth}
\setlength{\xb}{6.5\unit}
\addtolength{\xb}{0.5\dlinewidth}
\setlength{\yb}{14\unit}
\addtolength{\yb}{-0.5\dlinewidth}
\setlength{\xc}{9.5\unit}
\addtolength{\xc}{-0.5\dlinewidth}
\setlength{\yc}{11\unit}
\addtolength{\yc}{0.5\dlinewidth}
\setlength{\xd}{12\unit}
\addtolength{\xd}{-0.5\dlinewidth}
\setlength{\yd}{0\unit}
\addtolength{\yd}{0.5\dlinewidth}
\setlength{\xe}{17.5\unit}
\addtolength{\xe}{0.5\dlinewidth}
\setlength{\ye}{-2\unit}
\addtolength{\ye}{-0.5\dlinewidth}
\setlength{\xf}{20.5\unit}
\addtolength{\xf}{-0.5\dlinewidth}
\setlength{\yf}{-5\unit}
\addtolength{\yf}{0.5\dlinewidth}
\setlength{\yg}{8\unit}
\addtolength{\yg}{0.5\dlinewidth}
\setlength{\yh}{1\unit}
\addtolength{\yh}{-0.5\dlinewidth}
%
%
\pscustom[linecolor=gray,fillstyle=solid,fillcolor=gray,linearc=\linearc]{%
\psline[liftpen=1,linearc=\linearc](\xb,\ya)(\xd,\ya)(\xd,\yd)(\xa,\yd)(\xa,\ya)(\xb,\ya)
\psline[liftpen=2,linearc=\linearc](\xb,\ya)(\xb,\yb)(\xf,\yb)(\xf,\yg)
\psline[liftpen=1](\xf,\yg)(\xe,\yg)
\psline[liftpen=1,linearc=\linearc](\xe,\yg)(\xe,\yc)(\xc,\yc)(\xc,\ya)
\psline[liftpen=2,linearc=\linearc](\xb,\yd)(\xb,\yf)(\xf,\yf)(\xf,\yh)
\psline[liftpen=1](\xf,\yh)(\xe,\yh)
\psline[liftpen=1,linearc=\linearc](\xe,\yh)(\xe,\ye)(\xc,\ye)(\xc,\yd)
}
\pscustom[linecolor=gray,fillstyle=solid,fillcolor=gray]{%
\psline[liftpen=1](\xe,\yg)(\xf,\yg)(\xf,\yh)
\psline[liftpen=2](\xf,\yh)(\xe,\yh)(\xe,\yg)
}
\end{pspicture}}
\,=
\settoheight{\eqoff}{$+$}%
\setlength{\eqoff}{0.5\eqoff}%
\addtolength{\eqoff}{-10.5\unit}%
\raisebox{\eqoff}{%
\begin{pspicture}(-0.5,-6)(21.5,15)
%
\rput[r](1.5,9){$\scriptstyle \phi^i$}
\rput[r](1.5,0){$\scriptstyle \phi^i$}
\rput[l](14.5,0){$\scriptstyle \bar\phi_i$}
\rput[l](14.5,9){$\scriptstyle \bar\phi_i$}
%
\uinex{2}{9}
\iinex{2}{6}
\dinex{2}{0}
\setlength{\xa}{4\unit}
\addtolength{\xa}{-0.5\dlinewidth}
\setlength{\ya}{9\unit}
\addtolength{\ya}{0.5\dlinewidth}
\setlength{\xb}{6.5\unit}
\addtolength{\xb}{-0.5\dlinewidth}
\setlength{\yb}{14\unit}
\addtolength{\yb}{0.5\dlinewidth}
\setlength{\xc}{9.5\unit}
\addtolength{\xc}{0.5\dlinewidth}
\setlength{\yc}{11\unit}
\addtolength{\yc}{-0.5\dlinewidth}
\setlength{\xd}{12\unit}
\addtolength{\xd}{0.5\dlinewidth}
\setlength{\yd}{0\unit}
\addtolength{\yd}{-0.5\dlinewidth}
\setlength{\xe}{17.5\unit}
\addtolength{\xe}{-0.5\dlinewidth}
\setlength{\ye}{-2\unit}
\addtolength{\ye}{0.5\dlinewidth}
\setlength{\xf}{20.5\unit}
\addtolength{\xf}{0.5\dlinewidth}
\setlength{\yf}{-5\unit}
\addtolength{\yf}{-0.5\dlinewidth}
\setlength{\yg}{8\unit}
\addtolength{\yg}{-0.5\dlinewidth}
\setlength{\yh}{1\unit}
\addtolength{\yh}{0.5\dlinewidth}
\psline[liftpen=1,linearc=\linearc](\xa,\ya)(\xb,\ya)(\xb,\yb)(\xf,\yb)(\xf,\yg)
\psline[liftpen=1,linearc=\linearc](\xf,\yg)(\xf,\yh)
\psline[liftpen=1,linearc=\linearc](\xf,\yh)(\xf,\yf)(\xb,\yf)(\xb,\yd)(\xa,\yd)
\psline[liftpen=1,linearc=\linearc](\xd,\ya)(\xc,\ya)(\xc,\yc)(\xe,\yc)(\xe,\yg)
\psline[liftpen=1,linearc=\linearc](\xe,\yg)(\xe,\yh)
\psline[liftpen=1,linearc=\linearc](\xe,\yh)(\xe,\ye)(\xc,\ye)(\xc,\yd)(\xd,\yd)
%
%
\psline[linestyle=dotted](3.5,4.5)(3.5,1.5)
\doutex{14}{0}
\ioutex{14}{6}
\uoutex{14}{9}
\psline[linestyle=dotted](12.5,4.5)(12.5,1.5)
\setlength{\xa}{4\unit}
\addtolength{\xa}{0.5\dlinewidth}
\setlength{\ya}{9\unit}
\addtolength{\ya}{-0.5\dlinewidth}
\setlength{\xb}{6.5\unit}
\addtolength{\xb}{0.5\dlinewidth}
\setlength{\yb}{14\unit}
\addtolength{\yb}{-0.5\dlinewidth}
\setlength{\xc}{9.5\unit}
\addtolength{\xc}{-0.5\dlinewidth}
\setlength{\yc}{11\unit}
\addtolength{\yc}{0.5\dlinewidth}
\setlength{\xd}{12\unit}
\addtolength{\xd}{-0.5\dlinewidth}
\setlength{\yd}{0\unit}
\addtolength{\yd}{0.5\dlinewidth}
\setlength{\xe}{17.5\unit}
\addtolength{\xe}{0.5\dlinewidth}
\setlength{\ye}{-2\unit}
\addtolength{\ye}{-0.5\dlinewidth}
\setlength{\xf}{20.5\unit}
\addtolength{\xf}{-0.5\dlinewidth}
\setlength{\yf}{-5\unit}
\addtolength{\yf}{0.5\dlinewidth}
\setlength{\yg}{8\unit}
\addtolength{\yg}{0.5\dlinewidth}
\setlength{\yh}{1\unit}
\addtolength{\yh}{-0.5\dlinewidth}
%
%
\pscustom[linecolor=gray,fillstyle=solid,fillcolor=gray,linearc=\linearc]{%
\psline[liftpen=1,linearc=\linearc](\xb,\ya)(\xd,\ya)(\xd,\yd)(\xa,\yd)(\xa,\ya)(\xb,\ya)
\psline[liftpen=2,linearc=\linearc](\xb,\ya)(\xb,\yb)(\xf,\yb)(\xf,\yg)
\psline[liftpen=1](\xf,\yg)(\xe,\yg)
\psline[liftpen=1,linearc=\linearc](\xe,\yg)(\xe,\yc)(\xc,\yc)(\xc,\ya)
\psline[liftpen=2,linearc=\linearc](\xb,\yd)(\xb,\yf)(\xf,\yf)(\xf,\yh)
\psline[liftpen=1](\xf,\yh)(\xe,\yh)
\psline[liftpen=1,linearc=\linearc](\xe,\yh)(\xe,\ye)(\xc,\ye)(\xc,\yd)
}

\pscustom[linecolor=gray,fillstyle=solid,fillcolor=gray]{%
\psline[liftpen=1](\xe,\yg)(\xf,\yg)(\xf,\yh)
\psline[liftpen=2](\xf,\yh)(\xe,\yh)(\xe,\yg)
}
\setlength{\xb}{8\unit}
\setlength{\yb}{12.5\unit}
\setlength{\xc}{19\unit}
\setlength{\yc}{-3.5\unit}
\setlength{\yg}{8\unit}
\addtolength{\yg}{0.5\dlinewidth}
\setlength{\yh}{1\unit}
\addtolength{\yh}{-0.5\dlinewidth}
\psline[liftpen=2,linearc=\linearc](\xc,\yh)(\xc,\yc)(\xb,\yc)(\xb,\yb)(\xc,\yb)
(\xc,\yg)
\psline[liftpen=2,linearc=\linearc](\xc,\yg)(\xc,\yh)
\end{pspicture}}
\,+
\settoheight{\eqoff}{$+$}%
\setlength{\eqoff}{0.5\eqoff}%
\addtolength{\eqoff}{-10.5\unit}%
\raisebox{\eqoff}{%
\begin{pspicture}(-0.5,-6)(21.5,15)
%
\rput[r](1.5,9){$\scriptstyle \phi^i$}
\rput[r](1.5,0){$\scriptstyle \phi^i$}
\rput[l](14.5,0){$\scriptstyle \bar\phi_i$}
\rput[l](14.5,9){$\scriptstyle \bar\phi_i$}
%
\uinex{2}{9}
\iinex{2}{6}
\dinex{2}{0}
\setlength{\xa}{4\unit}
\addtolength{\xa}{-0.5\dlinewidth}
\setlength{\ya}{9\unit}
\addtolength{\ya}{0.5\dlinewidth}
\setlength{\xb}{6.5\unit}
\addtolength{\xb}{-0.5\dlinewidth}
\setlength{\yb}{14\unit}
\addtolength{\yb}{0.5\dlinewidth}
\setlength{\xc}{9.5\unit}
\addtolength{\xc}{0.5\dlinewidth}
\setlength{\yc}{11\unit}
\addtolength{\yc}{-0.5\dlinewidth}
\setlength{\xd}{12\unit}
\addtolength{\xd}{0.5\dlinewidth}
\setlength{\yd}{0\unit}
\addtolength{\yd}{-0.5\dlinewidth}
\setlength{\xe}{17.5\unit}
\addtolength{\xe}{-0.5\dlinewidth}
\setlength{\ye}{-2\unit}
\addtolength{\ye}{0.5\dlinewidth}
\setlength{\xf}{20.5\unit}
\addtolength{\xf}{0.5\dlinewidth}
\setlength{\yf}{-5\unit}
\addtolength{\yf}{-0.5\dlinewidth}
\setlength{\yg}{8\unit}
\addtolength{\yg}{-0.5\dlinewidth}
\setlength{\yh}{1\unit}
\addtolength{\yh}{0.5\dlinewidth}
\psline[liftpen=1,linearc=\linearc](\xa,\ya)(\xb,\ya)(\xb,\yb)(\xf,\yb)(\xf,\yg)
\psline[liftpen=1,linearc=\linearc](\xf,\yg)(\xf,\yh)
\psline[liftpen=1,linearc=\linearc](\xf,\yh)(\xf,\yf)(\xb,\yf)(\xb,\yd)(\xa,\yd)
\psline[liftpen=1,linearc=\linearc](\xd,\ya)(\xc,\ya)(\xc,\yc)(\xe,\yc)(\xe,\yg)
\psline[liftpen=1,linearc=\linearc](\xe,\yg)(\xe,\yh)
\psline[liftpen=1,linearc=\linearc](\xe,\yh)(\xe,\ye)(\xc,\ye)(\xc,\yd)(\xd,\yd)
%
%
\psline[linestyle=dotted](3.5,4.5)(3.5,1.5)
\doutex{14}{0}
\ioutex{14}{6}
\uoutex{14}{9}
\psline[linestyle=dotted](12.5,4.5)(12.5,1.5)
\setlength{\xa}{4\unit}
\addtolength{\xa}{0.5\dlinewidth}
\setlength{\ya}{9\unit}
\addtolength{\ya}{-0.5\dlinewidth}
\setlength{\xb}{6.5\unit}
\addtolength{\xb}{0.5\dlinewidth}
\setlength{\yb}{14\unit}
\addtolength{\yb}{-0.5\dlinewidth}
\setlength{\xc}{9.5\unit}
\addtolength{\xc}{-0.5\dlinewidth}
\setlength{\yc}{11\unit}
\addtolength{\yc}{0.5\dlinewidth}
\setlength{\xd}{12\unit}
\addtolength{\xd}{-0.5\dlinewidth}
\setlength{\yd}{0\unit}
\addtolength{\yd}{0.5\dlinewidth}
\setlength{\xe}{17.5\unit}
\addtolength{\xe}{0.5\dlinewidth}
\setlength{\ye}{-2\unit}
\addtolength{\ye}{-0.5\dlinewidth}
\setlength{\xf}{20.5\unit}
\addtolength{\xf}{-0.5\dlinewidth}
\setlength{\yf}{-5\unit}
\addtolength{\yf}{0.5\dlinewidth}
\setlength{\yg}{8\unit}
\addtolength{\yg}{0.5\dlinewidth}
\setlength{\yh}{1\unit}
\addtolength{\yh}{-0.5\dlinewidth}
%
%
\pscustom[linecolor=gray,fillstyle=solid,fillcolor=gray,linearc=\linearc]{%
\psline[liftpen=1,linearc=\linearc](\xb,\ya)(\xd,\ya)(\xd,\yd)(\xa,\yd)(\xa,\ya)(\xb,\ya)
\psline[liftpen=2,linearc=\linearc](\xb,\ya)(\xb,\yb)(\xf,\yb)(\xf,\yg)
\psline[liftpen=1](\xf,\yg)(\xe,\yg)
\psline[liftpen=1,linearc=\linearc](\xe,\yg)(\xe,\yc)(\xc,\yc)(\xc,\ya)
\psline[liftpen=2,linearc=\linearc](\xb,\yd)(\xb,\yf)(\xf,\yf)(\xf,\yh)
\psline[liftpen=1](\xf,\yh)(\xe,\yh)
\psline[liftpen=1,linearc=\linearc](\xe,\yh)(\xe,\ye)(\xc,\ye)(\xc,\yd)
}
%
%
\addtolength{\yh}{0.5\linew}
\addtolength{\yg}{-0.5\linew}
\pssin[periods=8,coilarm=0.1,amplitude=0.2](\xe,\yg)(\xe,\yh)
\pssin[periods=8,coilarm=0.1,amplitude=0.2](\xf,\yg)(\xf,\yh)
\newlength{\yaaa}
\setlength{\yaaa}{0.5\yh}
\addtolength{\yaaa}{0.5\yg}
\addtolength{\xe}{0.7\dlinewidth}
\addtolength{\xf}{-0.7\dlinewidth}
\psline[linestyle=dotted,dotsep=1.5pt](\xe,\yaaa)(\xf,\yaaa)
\newlength{\xaaa}
\setlength{\xaaa}{0.5\xe}
\addtolength{\xaaa}{0.5\xf}
%
\end{pspicture}}
\pnt
\end{equation}
The diagrams in the first class on the rhs.\ contain at least one wrapping loop that is purely made out of matter-type fields, i.e.\ a closed path running around the wrapping loop can be built only from matter-type propagators joining in any type of vertices. Such a path is depicted as a solid cycle. Diagrams in the second class on the rhs.\ do not contain such a closed loop, i.e.\ in all closed paths along the wrapping loops at least one gauge-field propagator occurs. This is represented by the wiggly lines.

We can now prove that the diagrams of the second class are undeformed. Given such a diagram, we remove all gauge-field propagators and replace the vertices at their ends according to 
\begin{equation}
\label{vertexreplacement} 
\setlength{\unit}{0.4cm}
\psset{xunit=\unit,yunit=\unit,runit=\unit}
\setlength{\dlinewidth}{0.75\unit}
\setlength{\linew}{1pt}
\setlength{\doublesep}{\dlinewidth}
\addtolength{\doublesep}{-\linew}
\psset{doublesep=\doublesep}
\psset{linewidth=\linew}
\setlength{\auxlen}{-0.2929\dlinewidth}
\addtolength{\auxlen}{\unit}
\setlength{\linearc}{0.75\unit}
\settoheight{\eqoff}{$+$}%
\setlength{\eqoff}{0.5\eqoff}%
\addtolength{\eqoff}{-2\unit}%
\raisebox{\eqoff}{%
\begin{pspicture}(-2,-2)(2,2)
\recthreevertex{0}{0}{0}
\psset{doubleline=true}
\psline(-1.4142,1.4142)(-0.7071,0.7071)
\psline(-0.7071,-0.7071)(-1.4142,-1.4142)
\psline(0.7071,0)(1.7071,0)
\psset{doubleline=false}
\psline(-1.4142,1.4142)(0,0)
\psline(0,0)(-1.4142,-1.4142)
\pscoil[coilwidth=0.1666,coilheight=3,coilarm=0,coilaspect=0]{-}(0,0)(1.7071,0)
\end{pspicture}}
\col\,
\settoheight{\eqoff}{$+$}%
\setlength{\eqoff}{0.5\eqoff}%
\addtolength{\eqoff}{-2\unit}%
\raisebox{\eqoff}{%
\begin{pspicture}(-2,-2)(2,2)
\fourvertex{0}{0}{45}
\psset{doubleline=true}
\psline(-1.4142,1.4142)(-0.7071,0.7071)
\psline(1.4142,1.4142)(0.7071,0.7071)
\psline(1.4142,-1.4142)(0.7071,-0.7071)
\psline(-0.7071,-0.7071)(-1.4142,-1.4142)
\psset{doubleline=false}
\psline(-1.4142,1.4142)(0,0)
\psline(0,0)(-1.4142,-1.4142)
\pscoil[coilwidth=0.1666,coilheight=3,coilarm=0,coilaspect=0]{-}(0,0)(1.4142,1.4142)
\pscoil[coilwidth=0.1666,coilheight=3,coilarm=0,coilaspect=0]{-}(0,0)(1.4142,-1.4142)
\end{pspicture}}
\quad\longrightarrow\quad
\settoheight{\eqoff}{$+$}%
\setlength{\eqoff}{0.5\eqoff}%
\addtolength{\eqoff}{-2\unit}%
\raisebox{\eqoff}{%
\begin{pspicture}(-2,-2)(0.2921,2)
\psset{doubleline=true}
\psline[linearc=\linearc](-1.4142,1.4142)(0,0)(-1.4142,-1.4142)
\psset{doubleline=false}
\psline[linearc=\linearc](-1.4142,1.4142)(0,0)(-1.4142,-1.4142)
\end{pspicture}}
\pnt
\end{equation}
As in \eqref{wrapdiagdecomp}, the central solid lines in \eqref{vertexreplacement} stand for matter-type fields. Since the gauge-boson interactions are undeformed, the resulting diagram has the same dependence on the deformation parameters as the original one. 
Furthermore, from the definition of the second class it follows immediately that the above procedure cuts each closed path along the wrapping loops at least once. The resulting diagram hence does no longer have a wrapping loop; instead, it is a planar single-trace diagram (or a product thereof). Thus, relation \eqref{diagrel} can be applied to it (or each of its factors) showing that the diagram is undeformed. All deformation-dependent wrapping diagrams must hence be contained in the first class.

\section{Finite-size corrections to the ground state energies}
\label{sec:calculation}

In the following, we determine the anomalous dimensions of the composite operators \eqref{groundstate} to leading wrapping order $K=L$ from Feynman diagrams. Specializing the previous discussion to $K=L$, the only diagrams which can be affected by the deformation are wrapping diagrams with a single matter-type wrapping loop and all prewrapping diagrams, i.e.\ diagrams in which the double-trace coupling \eqref{Qiiii} occurs. These diagrams have to be evaluated explicitly. The contributions from all other (deformation-independent) diagrams can be reconstructed from the condition that the operators \eqref{groundstate} are protected in the undeformed theory.

\subsection{\texorpdfstring{Generic case $L\ge3$}{Generic case L>=3} }
\label{sec:Lge3}

At $L\ge3$, prewrapping is absent and all deformation-dependent diagrams are of wrapping type with a matter-type wrapping loop. Only four of these diagrams are non-vanishing. Using the conventions in Appendix \ref{app:action} and the identities \eqref{eq:Qidentity}, \eqref{eq:rhoidentity} given in Appendix~\ref{app:tensorid}, they evaluate to
\begin{equation}
\begin{aligned}\label{wrapLdiags}
S(L)=
\settoheight{\eqoff}{$\times$}%
\setlength{\eqoff}{0.5\eqoff}%
\addtolength{\eqoff}{-16\unitlength}%
\raisebox{\eqoff}{%
\fmfframe(7,1)(3,1){%
\begin{fmfchar*}(30,30)
  \fmfleft{in}
  \fmfright{out1}
\fmf{phantom}{in,v5}
\fmf{phantom}{out,v2}
\fmf{phantom}{in,va5}
\fmf{phantom}{out,va2}
\fmfforce{(0,0.5h)}{in}
\fmfforce{(w,0.5h)}{out}
\fmfpoly{phantom}{v1,v6,v5,v4,v3,v2}
\fmffixed{(0.75w,0)}{v5,v2}
\fmfpoly{phantom}{va1,va6,va5,va4,va3,va2}
\fmffixed{(w,0)}{va5,va2}
\fmf{phantom}{vc,v1}
\fmf{phantom}{vc,v4}
\fmffreeze
\fmf{plain_ar,left=0}{v1,v2}
\fmf{plain_ar,left=0}{v2,v3}
\fmf{plain_ar,left=0}{v3,v4}
\fmf{dots,left=0}{v4,v5}
\fmf{plain_ar,left=0}{v5,v6}
\fmf{plain_ar,left=0}{v6,v1}
\fmf{plain_rar}{vc,v1}
\fmf{plain_rar}{vc,v2}
\fmf{plain_rar}{vc,v3}
\fmf{dots}{vc,v4}
\fmf{plain_rar}{vc,v5}
\fmf{plain_rar}{vc,v6}
\fmf{plain_ar}{va1,v1}
\fmf{plain_ar}{va2,v2}
\fmf{plain_ar}{va3,v3}
\fmf{dots}{va4,v4}
\fmf{plain_ar}{va5,v5}
\fmf{plain_ar}{va6,v6}
\fmffreeze
%
\fmfv{l=$\scriptscriptstyle 1$,l.dist=2}{va1}
\fmfv{l=$\scriptscriptstyle 2$,l.dist=2}{va2}
\fmfv{l=$\scriptscriptstyle 3$,l.dist=2}{va3}
\fmfv{l=$\scriptscriptstyle {L-1}$,l.dist=2}{va5}
\fmfv{l=$\scriptscriptstyle L$,l.dist=2}{va6}
\fmfiv{decor.shape=circle,decor.filled=full,decor.size=3}{vloc(__vc)}
\end{fmfchar*}}}
&=g_\YM^{2L}N^L \sum_{j=1}^3 (\hat Q_{ij}^{ji})^L P_L\\[-1.5\baselineskip]
&=g_\YM^{2L}N^L
\Big(2\e^{iL\gamma_i^-}\cos{L\gamma_i^+}+\frac{1}{2^L}\Big)
P_L
\col\\
\bar{S}(L)=
\settoheight{\eqoff}{$\times$}%
\setlength{\eqoff}{0.5\eqoff}%
\addtolength{\eqoff}{-16\unitlength}%
\raisebox{\eqoff}{%
\fmfframe(7,1)(3,1){%
\begin{fmfchar*}(30,30)
  \fmfleft{in}
  \fmfright{out1}
\fmf{phantom}{in,v5}
\fmf{phantom}{out,v2}
\fmf{phantom}{in,va5}
\fmf{phantom}{out,va2}
\fmfforce{(0,0.5h)}{in}
\fmfforce{(w,0.5h)}{out}
\fmfpoly{phantom}{v1,v6,v5,v4,v3,v2}
\fmffixed{(0.75w,0)}{v5,v2}
\fmfpoly{phantom}{va1,va6,va5,va4,va3,va2}
\fmffixed{(w,0)}{va5,va2}
\fmf{phantom}{vc,v1}
\fmf{phantom}{vc,v4}
\fmffreeze
\fmf{plain_rar,left=0}{v1,v2}
\fmf{plain_rar,left=0}{v2,v3}
\fmf{plain_rar,left=0}{v3,v4}
\fmf{dots,left=0}{v4,v5}
\fmf{plain_rar,left=0}{v5,v6}
\fmf{plain_rar,left=0}{v6,v1}
\fmf{plain_rar}{vc,v1}
\fmf{plain_rar}{vc,v2}
\fmf{plain_rar}{vc,v3}
\fmf{dots}{vc,v4}
\fmf{plain_rar}{vc,v5}
\fmf{plain_rar}{vc,v6}
\fmf{plain_ar}{va1,v1}
\fmf{plain_ar}{va2,v2}
\fmf{plain_ar}{va3,v3}
\fmf{dots}{va4,v4}
\fmf{plain_ar}{va5,v5}
\fmf{plain_ar}{va6,v6}
\fmffreeze
%
\fmfv{l=$\scriptscriptstyle 1$,l.dist=2}{va1}
\fmfv{l=$\scriptscriptstyle 2$,l.dist=2}{va2}
\fmfv{l=$\scriptscriptstyle 3$,l.dist=2}{va3}
\fmfv{l=$\scriptscriptstyle {L-1}$,l.dist=2}{va5}
\fmfv{l=$\scriptscriptstyle L$,l.dist=2}{va6}
\fmfiv{decor.shape=circle,decor.filled=full,decor.size=3}{vloc(__vc)}
\end{fmfchar*}}}
&=g_\YM^{2L}N^L \sum_{j=1}^3(\hat Q_{ji}^{ij})^L P_L\\[-1.5\baselineskip]
&=g_\YM^{2L}N^L
\Big(2\e^{-iL\gamma_i^-}\cos{L\gamma_i^+}+\frac{1}{2^L}\Big)P_L
\col\\
F(L)=
\settoheight{\eqoff}{$\times$}%
\setlength{\eqoff}{0.5\eqoff}%
\addtolength{\eqoff}{-16\unitlength}%
\raisebox{\eqoff}{%
\fmfframe(7,1)(3,1){%
\begin{fmfchar*}(30,30)
  \fmfleft{in}
  \fmfright{out1}
\fmf{phantom}{in,v10}
\fmf{phantom}{out,v4}
\fmf{phantom}{in,va5}
\fmf{phantom}{out,va2}
\fmfforce{(0,0.5h)}{in}
\fmfforce{(w,0.5h)}{out}
\fmfpoly{phantom}{v1,v12,v11,v10,v9,v8,v7,v6,v5,v4,v3,v2}
\fmffixed{(0.75w,0)}{v10,v4}
\fmfpoly{phantom}{va1,va6,va5,va4,va3,va2}
\fmffixed{(w,0)}{va5,va2}
\fmf{phantom}{vc,v1}
\fmf{phantom}{vc,v7}
\fmffreeze
\fmf{dashes_ar,left=0}{v1,v2}
\fmf{dashes_rar,left=0}{v2,v3}
\fmf{dashes_ar,left=0}{v3,v4}
\fmf{dashes_rar,left=0}{v4,v5}
\fmf{dashes_ar,left=0}{v5,v6}
\fmf{dots,left=0}{v6,v7}
\fmf{dots,left=0}{v7,v8}
\fmf{dashes_rar,left=0}{v8,v9}
\fmf{dashes_ar,left=0}{v9,v10}
\fmf{dashes_rar,left=0}{v10,v11}
\fmf{dashes_ar,left=0}{v11,v12}
\fmf{dashes_rar,left=0}{v12,v1}
\fmf{plain_rar}{vc,v1}
\fmf{plain_rar}{vc,v3}
\fmf{plain_rar}{vc,v5}
\fmf{dots}{vc,v7}
\fmf{plain_rar}{vc,v9}
\fmf{plain_rar}{vc,v11}
\fmf{plain_ar}{va1,v2}
\fmf{plain_ar}{va2,v4}
\fmf{plain_ar}{va3,v6}
\fmf{dots}{va4,v8}
\fmf{plain_ar}{va5,v10}
\fmf{plain_ar}{va6,v12}
\fmffreeze
%
\fmfv{l=$\scriptscriptstyle 1$,l.dist=2}{va1}
\fmfv{l=$\scriptscriptstyle 2$,l.dist=2}{va2}
\fmfv{l=$\scriptscriptstyle 3$,l.dist=2}{va3}
\fmfv{l=$\scriptscriptstyle {L-1}$,l.dist=2}{va5}
\fmfv{l=$\scriptscriptstyle L$,l.dist=2}{va6}
\fmfiv{decor.shape=circle,decor.filled=full,decor.size=3}{vloc(__vc)}
\end{fmfchar*}}}
&=g_\YM^{2L}N^L\tr\big[((\rho^{\dagger\,i})(\rho_{i})^{\T})^L\big]2(-1)^{L-1}P_L\vphantom{{} \sum_{j=1}^3{}}\\[-1.5\baselineskip]
&=-4g_\YM^{2L}N^L \cos L\gamma_i^+P_L\vphantom{\Big(\Big)}\col\\
\tilde F(L)
=
\settoheight{\eqoff}{$\times$}%
\setlength{\eqoff}{0.5\eqoff}%
\addtolength{\eqoff}{-16\unitlength}%
\raisebox{\eqoff}{%
\fmfframe(7,1)(3,1){%
\begin{fmfchar*}(30,30)
  \fmfleft{in}
  \fmfright{out1}
\fmf{phantom}{in,v10}
\fmf{phantom}{out,v4}
\fmf{phantom}{in,va5}
\fmf{phantom}{out,va2}
\fmfforce{(0,0.5h)}{in}
\fmfforce{(w,0.5h)}{out}
\fmfpoly{phantom}{v1,v12,v11,v10,v9,v8,v7,v6,v5,v4,v3,v2}
\fmffixed{(0.75w,0)}{v10,v4}
\fmfpoly{phantom}{va1,va6,va5,va4,va3,va2}
\fmffixed{(w,0)}{va5,va2}
\fmf{phantom}{vc,v1}
\fmf{phantom}{vc,v7}
\fmffreeze
\fmf{dashes_rar,left=0}{v1,v2}
\fmf{dashes_ar,left=0}{v2,v3}
\fmf{dashes_rar,left=0}{v3,v4}
\fmf{dashes_ar,left=0}{v4,v5}
\fmf{dashes_rar,left=0}{v5,v6}
\fmf{dots,left=0}{v6,v7}
\fmf{dots,left=0}{v7,v8}
\fmf{dashes_ar,left=0}{v8,v9}
\fmf{dashes_rar,left=0}{v9,v10}
\fmf{dashes_ar,left=0}{v10,v11}
\fmf{dashes_rar,left=0}{v11,v12}
\fmf{dashes_ar,left=0}{v12,v1}
\fmf{plain_rar}{vc,v1}
\fmf{plain_rar}{vc,v3}
\fmf{plain_rar}{vc,v5}
\fmf{dots}{vc,v7}
\fmf{plain_rar}{vc,v9}
\fmf{plain_rar}{vc,v11}
\fmf{plain_ar}{va1,v2}
\fmf{plain_ar}{va2,v4}
\fmf{plain_ar}{va3,v6}
\fmf{dots}{va4,v8}
\fmf{plain_ar}{va5,v10}
\fmf{plain_ar}{va6,v12}
\fmffreeze
%
\fmfv{l=$\scriptscriptstyle 1$,l.dist=2}{va1}
\fmfv{l=$\scriptscriptstyle 2$,l.dist=2}{va2}
\fmfv{l=$\scriptscriptstyle 3$,l.dist=2}{va3}
\fmfv{l=$\scriptscriptstyle {L-1}$,l.dist=2}{va5}
\fmfv{l=$\scriptscriptstyle L$,l.dist=2}{va6}
\fmfiv{decor.shape=circle,decor.filled=full,decor.size=3}{vloc(__vc)}
\end{fmfchar*}}}
&=g_\YM^{2L}N^L
\tr[((\tilde\rho^{\dagger\,i})(\tilde\rho_{i})^{\T})^L]2(-1)^{L-1}P_L\vphantom{{} \sum_{j=1}^3{}}\\[-1.5\baselineskip]
&=-4g_\YM^{2L}N^L \cos L\gamma_i^-P_L\vphantom{\Big(\Big)}
\col
\end{aligned}
\end{equation}
where scalar and fermionic fields are represented by solid and dashed lines, respectively. The composite operator \eqref{groundstate} is drawn as the central dot. All these diagrams depend on the scalar `cake' integral $P_L$. Its diagrammatic representation and its UV divergence 
$\mathcal{P}_L$ read \cite{Broadhurst:1985vq} 
\begin{equation}\label{PL}
P_L=
\settoheight{\eqoff}{$\times$}%
\setlength{\eqoff}{0.5\eqoff}%
\addtolength{\eqoff}{-8\unitlength}%
\raisebox{\eqoff}{%
\fmfframe(3,-2)(0,-2){%
\begin{fmfchar*}(20,20)
  \fmfleft{in}
  \fmfright{out1}
\fmf{phantom}{in,v1}
\fmf{phantom}{out,v2}
\fmfforce{(0,0.5h)}{in}
\fmfforce{(w,0.5h)}{out}
\fmfpoly{phantom}{v1,va4,va3,v2,va2,va1}
\fmffixed{(0.75w,0)}{v1,v2}
\fmf{phantom}{vc,v1}
\fmf{plain}{vc,v2}
\fmffreeze
\fmf{plain,left=0.25}{v1,va1}
\fmf{plain,left=0.25}{va1,va2}
\fmf{plain,left=0.25}{va2,v2}
\fmf{plain,left=0.25}{v2,va3}
\fmf{plain,left=0.25}{va3,va4}
\fmf{dots,left=0.25}{va4,v1}
\fmf{plain}{vc,va1}
\fmf{plain}{vc,va2}
\fmf{plain}{vc,va3}
\fmf{dots}{vc,va4}
\fmf{plain}{vc,v1}
\fmffreeze
\fmfv{l=$\scriptscriptstyle L$,l.dist=2}{va1}
\fmfv{l=$\scriptscriptstyle 1$,l.dist=2}{va2}
\fmfv{l=$\scriptscriptstyle 2$,l.dist=2}{v2}
\fmfv{l=$\scriptscriptstyle 3$,l.dist=2}{va3}
\fmfv{l=$\scriptscriptstyle L-1$,l.dist=2}{v1}
\end{fmfchar*}}}
\col\qquad
\mathcal{P}_L
=\Kop(P_L)
=\frac{1}{(4\pi)^{2L}}\frac{1}{\varepsilon}
\frac{2}{L}\binom{2L-3}{L-1}\zeta(2L-3)
\col
\end{equation}
where in $D=4-2\varepsilon$ dimensions the operator $\Kop$ extracts all poles in $\varepsilon$. The integral $P_L$ is free of subdivergences, and hence its overall UV divergence is given by a simple $\frac{1}{\varepsilon}$-pole.

The diagrams  $F(L)$ and $\tilde F(L)$ in \eqref{wrapLdiags} contain two particular configurations of the Yukawa vertices. The scalar fields of any two adjacent vertices on the fermionic wrapping loop always play a different role in the diagram: one is external while the other is contracted with the composite operator.
In all other possible diagrams with a fermionic wrapping loop, the scalar fields of at least two adjacent Yukawa vertices are both either external or contracted with the composite operator. Such diagrams vanish, as can be easily seen from the following contractions of the corresponding coupling tensors
\begin{equation}
\begin{aligned}
\settoheight{\eqoff}{$\times$}%
\setlength{\eqoff}{0.5\eqoff}%
\addtolength{\eqoff}{-5\unitlength}%
\raisebox{\eqoff}{%
\fmfframe(4,2)(4,2){%
\begin{fmfchar*}(24,6)
\fmfleft{vl}
\fmfright{vr}
\fmfforce{(0,0)}{vl}
\fmfforce{(w,0)}{vr}
\fmfforce{(0.333w,h)}{vi1}
\fmfforce{(0.666w,h)}{vi2}
\fmfforce{(0.333w,0)}{vc1}
\fmfforce{(0.666w,0)}{vc2}
\fmf{dashes_ar}{vl,vc1}
\fmf{dashes_rar}{vc1,vc2}
\fmf{dashes_ar}{vc2,vr}
\fmffreeze
\fmfposition
\fmf{plain_ar}{vi1,vc1}
\fmf{plain_ar}{vi2,vc2}
\fmfiv{label=$\scriptstyle A$,l.dist=2}{vloc(__vl)}
\fmfiv{label=$\scriptstyle i$,l.dist=2}{vloc(__vi1)}
\fmfiv{label=$\scriptstyle i$,l.dist=2}{vloc(__vi2)}
\fmfiv{label=$\scriptstyle B$,l.dist=2}{vloc(__vr)}
\end{fmfchar*}}}
&\propto
\sum_{C=1}^4\rho_{i\,CA}\tilde \rho_i^{BC}
=0\col\\
\settoheight{\eqoff}{$\times$}%
\setlength{\eqoff}{0.5\eqoff}%
\addtolength{\eqoff}{-5\unitlength}%
\raisebox{\eqoff}{%
\fmfframe(4,2)(4,2){%
\begin{fmfchar*}(24,6)
\fmfleft{vl}
\fmfright{vr}
\fmfforce{(0,0)}{vl}
\fmfforce{(w,0)}{vr}
\fmfforce{(0.333w,h)}{vi1}
\fmfforce{(0.666w,h)}{vi2}
\fmfforce{(0.333w,0)}{vc1}
\fmfforce{(0.666w,0)}{vc2}
\fmf{dashes_rar}{vl,vc1}
\fmf{dashes_ar}{vc1,vc2}
\fmf{dashes_rar}{vc2,vr}
\fmffreeze
\fmfposition
\fmf{plain_ar}{vi1,vc1}
\fmf{plain_ar}{vi2,vc2}
\fmfiv{label=$\scriptstyle A$,l.dist=2}{vloc(__vl)}
\fmfiv{label=$\scriptstyle i$,l.dist=2}{vloc(__vi1)}
\fmfiv{label=$\scriptstyle i$,l.dist=2}{vloc(__vi2)}
\fmfiv{label=$\scriptstyle B$,l.dist=2}{vloc(__vr)}
\end{fmfchar*}}}
&\propto
\sum_{C=1}^4\tilde \rho_i^{CA}\rho_{i\,BC}
=0\pnt
\end{aligned}
\end{equation}

The negative sum of the poles of \eqref{wrapLdiags} yields the contribution of all deformation-dependent diagrams to the renormalization constant 
$\mathcal{Z}_{\mathcal{O}_L}$. It is given by
\begin{equation}\label{deltaZOLdef}
\begin{aligned}
\delta\mathcal{Z}_{\mathcal{O}_L,\text{def}}
&=
-\Kop[S(L)+\bar{S}(L)+F(L)+\tilde F(L)]\\
&=4g_\YM^{2L}N^L\Big(\cos L\gamma_i^++\cos L\gamma_i^--\cos L\gamma_i^+\cos L\gamma_i^--\frac{1}{2^{L+1}}\Big)\mathcal{P}_L
\pnt
\end{aligned}
\end{equation}
Already at this point, the vanishing of the divergences for the operators \eqref{groundstate} in the $\beta$-deformation provides a non-trivial check: for $\gamma_1=\gamma_2=\gamma_3=-\pi\beta$, which corresponds to $\gamma_i^+=-\pi\beta$, $\gamma_i^-=0$, the above expression has to be independent of $\beta$, such that it cancels with the remaining deformation-independent diagrams. This is indeed the case. Then, \eqref{deltaZOLdef} directly determines the contribution from the deformation-independent diagrams as
\begin{equation}\label{deltaZOLnondef}
\begin{aligned}
\delta\mathcal{Z}_{\mathcal{O}_L,\text{non-def}}
&=-\delta\mathcal{Z}_{\mathcal{O}_L,\text{def}}\,\Big|_{\gamma_i^\pm=0}
=-4g_\YM^{2L}N^L\Big(1-\frac{1}{2^{L+1}}\Big)\mathcal{P}_L
\pnt
\end{aligned}
\end{equation}
The renormalization constant $\mathcal{Z}_{\mathcal{O}_L}$ to $L$-loop order then reads
\begin{equation}\label{ZL}
\begin{aligned}
\mathcal{Z}_{\mathcal{O}_L}
&=
1+\delta\mathcal{Z}_{\mathcal{O}_L,\text{def}}
+\delta\mathcal{Z}_{\mathcal{O}_L,\text{non-def}}
&=1-16g_\YM^{2L}N^L\sin^2\frac{L\gamma_i^+}{2}\sin^2\frac{L\gamma_i^-}{2}\mathcal{P}_L\pnt
\end{aligned}
\end{equation}

Inserting this expression into \eqref{gammaOLdef} and using the explicit result for $\mathcal{P}_L$ given in \eqref{PL}, we obtain the anomalous dimension 
\begin{equation}
\begin{aligned}\label{gammaOL}
\gamma_{\mathcal{O}_L}&=-64g^{2L}\sin^2\frac{L\gamma_i^+}{2}\sin^2\frac{L\gamma_i^-}{2}
\binom{2L-3}{L-1}\zeta(2L-3)\col
\end{aligned}
\end{equation}
where we have absorbed powers of $4\pi$ into the effective planar coupling constant $g=\frac{\sqrt{\lambda}}{4\pi}$. It exactly matches the expression found from the integrability-based equations in \cite{Ahn:2011xq}.\footnote{Note that one has to absorb a factor of a factor of $2$ into $g$ and a factor $L$ into $\gamma_i^\pm$ in order to match the definitions of \cite{Ahn:2011xq}.} 

\subsection{\texorpdfstring{Special case $L=2$}{Special case L=2}}
\label{sec:Le2}

At $L=2$, there are also deformation-dependent contributions from prewrapping diagrams in addition to those from wrapping diagrams. Since the prewrapping effect contributes already at one loop, we split the renormalization constant of the operator \eqref{groundstate} at $L=2$ as 
\begin{equation}\label{Z2}
\mathcal{Z}_{\mathcal{O}_2}
=1+\delta\mathcal{Z}^{(1)}_{\mathcal{O}_2}
+\delta\mathcal{Z}^{(2)}_{\mathcal{O}_2}
+\mathcal{O}(g^6)\col
\end{equation}
where the superscript in parenthesis denotes the loop order.

The only deformation-dependent one-loop diagram is a prewrapping diagram involving the double-trace coupling \eqref{Qiiii} as subdiagram. It reads
\begin{equation}
\label{prewrapdiag}
\ltwoopren{\fmfforce{(0.5w,0.5h)}{vc}
\fmf{plain_rar,left=0.75}{vo,vc}\fmf{plain_rar,right=0.75}{vo,vc}
\fmf{plain_rar}{vc,vl}\fmf{plain_rar}{vc,vr}
\fmfiv{label=$\scriptstyle Q_{\text{F}}$,l.a=0,l.dist=4}{vloc(__vc)}}
=-2g_\YM^2NQ_{\text{F}\,ii}^{ii}I_1
\col
\end{equation}
where the integral $I_1$ is specified below. All other planar one-loop diagrams are independent of the deformation according to the discussion in Section \ref{sec:wrapdefdep}. Their net contribution vanishes 
since the composite operator is protected in the undeformed theory and in the $\beta$-deformation where $Q_{\text{F}\,ii}^{ii}=0$. The one-loop contribution to the renormalization constant \eqref{Z2} is therefore given by
\begin{equation}\label{deltaZO2oneloop}
\delta\mathcal{Z}^{(1)}_{\mathcal{O}_2}=2g_\YM^2NQ_{\text{F}\,ii}^{ii}\Kop[I_1]\pnt
\end{equation}

The two-loop calculation requires the one-loop diagram \eqref{prewrapdiag}, the remaining de\-for\-ma\-tion-in\-de\-pen\-dent one-loop one-particle-irreducible (1PI) diagrams and their counter terms. They occur as subdiagrams and hence we have to evaluate them explicitly keeping also finite terms. The 1PI diagrams of operator renormalization and the self-energy correction of the scalar fields respectively read
\begin{equation}\label{subdiags}
\begin{gathered}
\ltwoopren{\fmfforce{(0.5w,0.5h)}{vc}
\fmf{plain_rar,left=0.75}{vo,vc}\fmf{plain_rar,right=0.75}{vo,vc}
\fmf{plain_rar}{vc,vl}\fmf{plain_rar}{vc,vr}
\fmffreeze
\fmfposition
}
=g_\YM^2NI_1
\col\qquad
\ltwoopren{
\fmf{plain_rar}{vo,vcl}\fmf{plain_rar}{vcl,vl}
\fmf{plain_rar}{vo,vcr}\fmf{plain_rar}{vcr,vr}
\fmffreeze
\fmf{photon}{vcl,vcr}
}
=g_\YM^2N\alpha I_1
\col\\
\settoheight{\eqoff}{$\times$}%
\setlength{\eqoff}{0.5\eqoff}%
\addtolength{\eqoff}{-5.625\unitlength}%
\raisebox{\eqoff}{%
\fmfframe(1,0)(1,0){%
\begin{fmfchar*}(15,11.25)
\fmfleft{v1}
\fmfright{v2}
\fmffixed{(0.5w,0)}{vc1,vc2}
\fmf{plain_ar}{v1,vc1}
\fmf{plain_ar}{vc2,v2}
\fmf{phantom,left=1}{vc1,vc2}
\fmf{phantom,left=1}{vc2,vc1}
\fmffreeze
\fmfposition
\fmfipath{p[]}
\fmfiset{p1}{vpath(__vc1,__vc2)}
\fmfiset{p2}{vpath(__vc2,__vc1)}
\fmfcmd{fill(p1--p2--cycle) withcolor 0.2black;}
\end{fmfchar*}}}
=g_\YM^2Np^{2(1-\varepsilon)}
(-(1+\alpha)I_1+2(\alpha-1)I'_1)\col
\end{gathered}
\end{equation}
where $\alpha$ is the gauge-fixing parameter and $p_\nu$ is the external momentum. Since the composite operators are gauge invariant, $\alpha$ has to drop out of the final result, and this serves as a check of our calculation. The above expressions depend on the integrals $I_1$ and $I_1'$, which are evaluated in terms of the $G$-functions \cite{Chetyrkin:1980pr,Vladimirov:1979zm} 
\begin{equation}\label{Gdef}
\begin{aligned}
G(\alpha,\beta)
&=\frac{1}{(4\pi)^\frac{D}{2}}
\frac{\Gamma(\alpha+\beta-\tfrac{D}{2})\Gamma(\tfrac{D}{2}-\alpha)\Gamma(\tfrac{D}{2}-\beta)}{\Gamma(\alpha)\Gamma(\beta)\Gamma(D-\alpha-\beta)}
\col\\
G_1(\alpha,\beta)
&=\frac{1}{2}(G(\alpha,\beta)-G(\alpha,\beta-1)+G(\alpha-1,\beta))
\end{aligned}
\end{equation}
and are explicitly given by
\begin{equation}
\begin{aligned}
I_1
&=
\settoheight{\eqoff}{$\times$}%
\setlength{\eqoff}{0.5\eqoff}%
\addtolength{\eqoff}{-4.5\unitlength}%
\raisebox{\eqoff}{%
\fmfframe(0,-3)(0,-3){%
\begin{fmfchar*}(15,15)
  \fmfleft{in}
  \fmfright{out1}
\fmf{plain}{in,v1}
\fmf{plain}{out,v2}
\fmfforce{(0,0.5h)}{in}
\fmfforce{(w,0.5h)}{out}
\fmffixed{(0.666w,0)}{v1,v2}
\fmf{plain,right=0.5,l.dist=2}{v1,v2}
\fmf{plain,left=0.5,l.dist=2}{v1,v2}
\end{fmfchar*}}}
=\frac{\mu^{4-D}}{p^{2(2-\frac{D}{2})}}G(1,1)=\frac{1}{(4\pi)^2}\Big(\frac{1}{\varepsilon}+2-\gamma_{\text{E}}+\ln\frac{4\pi\mu^2}{p^2}+\mathcal{O}(\varepsilon)\Big)
\col\\
I_1'
&=
\settoheight{\eqoff}{$\times$}%
\setlength{\eqoff}{0.5\eqoff}%
\addtolength{\eqoff}{-5.5\unitlength}%
\raisebox{\eqoff}{%
\fmfframe(0,2)(0,-6){%
\begin{fmfchar*}(15,15)
  \fmfleft{in}
  \fmfright{out1}
\fmf{plain}{in,v1}
\fmf{plain_rar}{out,v2}
\fmfforce{(0,0.5h)}{in}
\fmfforce{(w,0.5h)}{out}
\fmffixed{(0.666w,0)}{v1,v2}
\fmfpoly{phantom}{v1,v2,v3}
\fmf{plain}{v1,v2}
\fmf{plain}{v1,v3}
\fmf{plain_ar}{v3,v2}
\end{fmfchar*}}}
=\frac{\mu^{4-D}}{p^{2(2-\frac{D}{2})}}G_1(2,1)
=\frac{1}{(4\pi)^2}+\mathcal{O}(\varepsilon)
\col
\end{aligned}
\end{equation}
where $p_\nu$ is the external momentum and $\gamma_{\text{E}}$ is the Euler-Mascheroni constant. The arrows in the second diagram indicate that the respective momenta occur in a scalar product in the numerator. The integrals contain a power of the 't Hooft mass $\mu$, originating  from a rescaling of the Yang-Mills coupling constant in order to render it dimensionless in $D=4-2\varepsilon$ dimensions \cite{'tHooft:1973mm}. In the divergent integral $I_1$ the $\mu$-dependence starts in the finite terms, while in the finite integral $I_1'$ the $\mu$-dependence is postponed to the terms of order $\mathcal{O}(\varepsilon)$. Using these expansions, the counter terms for the diagrams \eqref{prewrapdiag} and \eqref{subdiags} read
\begin{equation}\label{cttwopointv}
\begin{gathered}
\ltwoopren{
\fmf{plain_rar}{vo,vl}
\fmf{plain_rar}{vo,vr}
\fmffreeze
\fmfiv{d.sh=cross,d.size=8}{vloc(__vo)}
\fmfiv{label=$\scriptstyle Q_{\text{F}}$,l.a=0,l.dist=6}{vloc(__vo)}}
=\delta J^{(1)}_{\mathcal{O}_2,\text{def}}
=2g^2Q_{\text{F}\,ii}^{ii}\frac{1}{\varepsilon}
\col\qquad
\ltwoopren{
\fmf{plain_rar}{vo,vl}
\fmf{plain_rar}{vo,vr}
\fmffreeze
\fmfiv{d.sh=cross,d.size=8}{vloc(__vo)}
}
=\delta J^{(1)}_{\mathcal{O}_2,\text{non-def}}
=-g^2(1+\alpha)\frac{1}{\varepsilon}
\col\\
\settoheight{\eqoff}{$\times$}%
\setlength{\eqoff}{0.5\eqoff}%
\addtolength{\eqoff}{-5.625\unitlength}%
\raisebox{\eqoff}{%
\fmfframe(1,0)(1,0){%
\begin{fmfchar*}(15,11.25)
\fmfleft{v1}
\fmfright{v2}
\fmf{plain_ar}{v1,vc1}
\fmf{plain_ar}{vc1,v2}
\fmffreeze
\fmfposition
\fmfv{d.sh=cross,d.size=8}{vc1}
\end{fmfchar*}}}
{}={}
-p^2
\delta^{(1)}_\phi
\col\qquad\delta^{(1)}_\phi=-g^2(1+\alpha)\frac{1}{\varepsilon}
\end{gathered}
\end{equation}
where we have split the contributions to the counter term for the composite operator into deformation-dependent and deformation-independent ones.\footnote{Note that the deformation-independent counter term of operator renormalization and the one of the self energy are equal, $\delta J^{(1)}_{\mathcal{O}_2,\text{non-def}}=\delta^{(1)}_\phi$, and hence their contributions cancel as expected when the expression for the operator renormalization constant \eqref{ZOL} is expanded to one loop.}

At two loops, two types of deformation-dependent diagrams contribute. First, there are the wrapping diagrams \eqref{wrapLdiags}, which have to be evaluated at $L=2$. Second, there are diagrams which are deformation-dependent since they contain at least one coupling $Q_{\text{F}\,ii}^{ii}$ or one of the counter terms $\delta Q_{\text{F}\,ii}^{ii}$, $\delta J_{\mathcal{O}_2,\text{def}}$. These prewrapping-generated contributions vanish in the $\beta$-deformation and in the undeformed theory. Hence, the contribution from the deformation-independent diagrams can be reconstructed from the one of the deformation-dependent wrapping diagrams alone. Their sum is the contribution from all diagrams that involve elementary single-trace couplings only. It is essentially given by setting $L=2$ in \eqref{ZL}.
We only have to be careful when extracting the divergence $\mathcal{P}_2$ of the respective cake integral. This integral contains an IR divergence which alters the $\frac{1}{\varepsilon}$-poles coming from the UV divergences. In order to avoid this IR divergence, we have to inject an external momentum $p_\nu$ into the composite operator. The resulting integral $I_2$ and its pole-part $\Kop[I_2]$ are given by
\begin{equation}\label{eq:I2}
\begin{aligned}
I_2
&=
\settoheight{\eqoff}{$\times$}%
\setlength{\eqoff}{0.5\eqoff}%
\addtolength{\eqoff}{-5.5\unitlength}%
\raisebox{\eqoff}{%
\fmfframe(0,2)(0,-6){%
\begin{fmfchar*}(15,15)
  \fmfleft{in}
  \fmfright{out1}
\fmf{plain}{in,v1}
\fmf{plain}{out,v2}
\fmfforce{(0,0.5h)}{in}
\fmfforce{(w,0.5h)}{out}
\fmffixed{(0.666w,0)}{v1,v2}
\fmfpoly{phantom}{v1,v2,v3}
\fmf{plain}{v1,v2}
\fmf{plain}{v1,v3}
\fmf{plain,right=0.25}{v3,v2}
\fmf{plain,left=0.25}{v3,v2}
\end{fmfchar*}}}
=\frac{\mu^{2(4-D)}}{p^{4(2-\frac{D}{2})}}G(1,1)G(3-\tfrac{D}{2},1)
\col\\
\Kop[I_2]&=\frac{1}{(4\pi)^4}\Big(\frac{1}{2\varepsilon^2}
+\frac{1}{\varepsilon}\Big(\frac{5}{2}-\gamma_{\text{E}}+\ln\frac{4\pi\mu^2}{p^2}\Big)\Big)\col
\end{aligned}
\end{equation}
where the latter replaces $\mathcal{P}_2$ in \eqref{ZL}. Accordingly, the two-loop contribution to $\mathcal{Z}_{\mathcal{O}_2}$ from all diagrams which only involve single-trace couplings is given by 
\begin{equation}\label{deltaZO2st}
\delta\mathcal{Z}_{\mathcal{O}_2,\text{st}}^{(2)}=-16g_\YM^4N^2\sin^2\gamma_i^+\sin^2\gamma_i^-\Kop[I_2]
\pnt
\end{equation}
This expression contains a $\tfrac{1}{\varepsilon}$-pole depending on $\ln p^2$, which cannot be absorbed into a local counter term for $\mathcal{O}_2$. It originates from a non-subtracted subdivergence of the integral $I_2$ given in \eqref{eq:I2}. Consistency requires that this subdivergence is subtracted by contributions from other Feynman diagrams, such that the result only contains the overall UV-divergence 
\begin{equation}
\begin{aligned}\label{KRI2}
\mathcal{I}_2&=\Kop\Rop[I_2]=\Kop[I_2-\Kop[I_1]I_1]=
\frac{1}{(4\pi)^4}\Big(-\frac{1}{2\varepsilon^2}+\frac{1}{2\varepsilon}\Big)
\col
\end{aligned}
\end{equation}
where the operation $\Rop$ subtracts the subdivergence.\footnote{A $\frac{1}{\varepsilon^2}$-pole persists, indicating that the contribution originated from a diagram with a one-loop subdivergence.}
This shows that truncating the action to only single-trace terms is inconsistent -- even in the planar limit.
Concretely, the subdivergence in \eqref{deltaZO2st} can be traced back to the one-loop renormalization of the quartic double-trace coupling \eqref{Qiiii}. The counter term of this coupling was determined in \cite{Fokken:2013aea}, and it reads
\begin{equation}
\begin{aligned}\label{deltaQ}
\delta Q^{ii}_{\text{F}\,ii}
&=2g_\YM^2N(
4\sin^2\gamma_i^+\sin^2\gamma_i^-
+(Q^{ii}_{\text{F}\,ii})^2-(1+\alpha)Q^{ii}_{\text{F}\,ii})\Kop[I_1]
\pnt
\end{aligned}
\end{equation}
It occurs in one of the remaining diagrams whose deformation-dependence is associated with the double-trace coupling $Q_{\text{F}\,ii}^{ii}$. The respective 1PI two-loop diagrams read\footnote{In the diagram involving the scalar self-energy in \eqref{subdiags} as a subdiagram, one must keep the finite contribution that is proportional to $\alpha-1$ and vanishes in Fermi-Feynman gauge. It contributes to the $\tfrac{1}{\varepsilon}$-pole of the two-loop diagram and is hence required for $\alpha$ to drop out of the final result.}

\begin{equation}
\begin{aligned}
\ltwoopren{
\fmfforce{(0.5w,0.721688h)}{vc1}
\fmf{plain_rar,left=0.75}{vc2,vc1}\fmf{plain_rar,right=0.75}{vc2,vc1}
\fmf{plain_rar,left=0.75}{vo,vc2}\fmf{plain_rar,right=0.75}{vo,vc2}
\fmf{plain_rar}{vc1,vl}\fmf{plain_rar}{vc1,vr}
\fmffreeze
\fmfposition
\fmfiv{label=$\scriptstyle Q_{\text{F}}$,l.a=-15,l.dist=6}{vloc(__vc1)}
}
&=\ltwoopren{
\fmfforce{(0.5w,0.721688h)}{vc1}
\fmf{plain_rar,left=0.75}{vc2,vc1}\fmf{plain_rar,right=0.75}{vc2,vc1}
\fmf{plain_rar,left=0.75}{vo,vc2}\fmf{plain_rar,right=0.75}{vo,vc2}
\fmf{plain_rar}{vc1,vl}\fmf{plain_rar}{vc1,vr}
\fmffreeze
\fmfposition
\fmfiv{label=$\scriptstyle Q_{\text{F}}$,l.a=0,l.dist=6}{vloc(__vc2)}
}
=-2g_\YM^4N^2 Q_{\text{F}\,ii}^{ii} I_1^2
\col\qquad
\ltwoopren{
\fmfforce{(0.5w,0.721688h)}{vc1}
\fmf{plain_rar,left=0.75}{vc2,vc1}\fmf{plain_rar,right=0.75}{vc2,vc1}
\fmf{plain_rar,left=0.75}{vo,vc2}\fmf{plain_rar,right=0.75}{vo,vc2}
\fmf{plain_rar}{vc1,vl}\fmf{plain_rar}{vc1,vr}
\fmffreeze
\fmfposition
\fmfiv{label=$\scriptstyle Q_{\text{F}}$,l.a=-15,l.dist=6}{vloc(__vc1)}
\fmfiv{label=$\scriptstyle Q_{\text{F}}$,l.a=0,l.dist=6}{vloc(__vc2)}
}
=4g_\YM^4N^2(Q_{\text{F}\,ii}^{ii})^2I_1^2
\col\\
\ltwoopren{
\fmfforce{(0.5w,0.5h)}{vc}
\fmf{phantom,left=0.75}{vo,vc}\fmf{phantom,left=0.75}{vc,vo}
\fmf{phantom}{vc,vl}\fmf{phantom}{vc,vr}
\fmffreeze\fmfposition
\fmfipath{p[]}
\fmfipair{vm[],vl[],vr[]}
\fmfiset{p1}{vpath(__vo,__vc)}
\fmfiset{p2}{reverse(vpath(__vc,__vo))}
\fmfiset{p3}{vpath(__vc,__vl)}
\fmfiset{p4}{vpath(__vc,__vr)}
\svertex{vm1}{p1}
\svertex{vm2}{p2}
\svertex{vm3}{p3}
\svertex{vm4}{p4}
\fmfi{plain_rar}{subpath (0,length(p3)/2) of p3}
\fmfi{plain_rar}{subpath (length(p3)/2,length(p3)) of p3}
\fmfi{plain_rar}{subpath (0,length(p4)/2) of p4}
\fmfi{plain_rar}{subpath (length(p4)/2,length(p2)) of p4}
\fmfi{plain_rar}{p1}
\fmfi{plain_rar}{p2}
\fmfi{photon}{vm3{dir 60}..{dir -60}vm4}
\fmfiv{label=$\scriptstyle Q_{\text{F}}$,l.a=0,l.dist=6}{vloc(__vc)}
}
&=-2g_\YM^4N^2Q_{\text{F}\,ii}^{ii}\alpha I_1^2
\col\\
\ltwoopren{
\fmfforce{(0.5w,0.5h)}{vc}
\fmf{phantom,left=0.75}{vo,vc}\fmf{phantom,left=0.75}{vc,vo}
\fmf{phantom}{vc,vl}\fmf{phantom}{vc,vr}
\fmffreeze\fmfposition
\fmfipath{p[]}
\fmfipair{vm[],vl[],vr[]}
\fmfiset{p1}{vpath(__vo,__vc)}
\fmfiset{p2}{reverse(vpath(__vc,__vo))}
\fmfiset{p3}{vpath(__vc,__vl)}
\fmfiset{p4}{vpath(__vc,__vr)}
\svertex{vm1}{p1}
\svertex{vm2}{p2}
\svertex{vm3}{p3}
\svertex{vm4}{p4}
\fmfi{plain_rar}{subpath (0,length(p1)/2) of p1}
\fmfi{plain_rar}{subpath (length(p1)/2,length(p1)) of p1}
\fmfi{plain_rar}{subpath (0,length(p2)/2) of p2}
\fmfi{plain_rar}{subpath (length(p2)/2,length(p2)) of p2}
\fmfi{plain_rar}{p3}
\fmfi{plain_rar}{p4}
\fmfi{photon}{vm1--vm2}
\fmfiv{label=$\scriptstyle Q_{\text{F}}$,l.a=0,l.dist=6}{vloc(__vc)}
}
&=-2g_\YM^4N^2Q_{\text{F}\,ii}^{ii}(2(3-\alpha)I_2-(3-2\alpha)I_1^2)
\col\\
\ltwoopren{
\fmfforce{(0.5w,0.5h)}{vc}
\fmf{phantom,left=0.75}{vo,vc}\fmf{phantom,left=0.75}{vc,vo}
\fmf{phantom}{vc,vl}\fmf{phantom}{vc,vr}
\fmffreeze\fmfposition
\fmfipath{p[]}
\fmfipair{vm[],vl[],vr[]}
\fmfiset{p1}{vpath(__vo,__vc)}
\fmfiset{p2}{reverse(vpath(__vc,__vo))}
\fmfiset{p3}{vpath(__vc,__vl)}
\fmfiset{p4}{vpath(__vc,__vr)}
\svertex{vm1}{p1}
\svertex{vm2}{p2}
\svertex{vm3}{p3}
\svertex{vm4}{p4}
\fmfi{plain_rar}{subpath (0,length(p1)*0.4) of p1}
\fmfi{plain_rar}{subpath (length(p1)*0.6,length(p1)) of p1}
\fmfi{plain_rar}{p2}
\fmfi{plain_rar}{p3}
\fmfi{plain_rar}{p4}
\vacpolp[0.25]{p1}
\fmfiv{label=$\scriptstyle Q_{\text{F}}$,l.a=0,l.dist=6}{vloc(__vc)}}
&=-4g_\YM^4N^2Q_{\text{F}\,ii}^{ii}(-(1+\alpha)I_2+(\alpha-1)(2I_2-I_1^2))
\pnt
\end{aligned}
\end{equation}
The 1PI one-loop diagrams involving one-loop counter terms read
\begin{equation}
\begin{aligned}
\ltwoopren{
\fmfforce{(0.5w,0.5h)}{vc}
\fmf{phantom,left=0.75}{vo,vc}\fmf{phantom,left=0.75}{vc,vo}
\fmf{phantom}{vc,vl}\fmf{phantom}{vc,vr}
\fmffreeze\fmfposition
\fmfipath{p[]}
\fmfipair{vm[],vl[],vr[]}
\fmfiset{p1}{vpath(__vo,__vc)}
\fmfiset{p2}{reverse(vpath(__vc,__vo))}
\fmfiset{p3}{vpath(__vc,__vl)}
\fmfiset{p4}{vpath(__vc,__vr)}
\svertex{vm1}{p1}
\svertex{vm2}{p2}
\svertex{vm3}{p3}
\svertex{vm4}{p4}
\fmfi{plain_rar}{subpath (0,length(p1)*0.5) of p1}
\fmfi{plain_rar}{subpath (length(p1)*0.5,length(p1)) of p1}
\fmfi{plain_rar}{p2}
\fmfi{plain_rar}{p3}
\fmfi{plain_rar}{p4}
\fmfi{phantom_cross}{p1}
\fmfiv{label=$\scriptstyle Q_{\text{F}}$,l.a=0,l.dist=6}{vloc(__vc)}}
&=4g_\YM^2N\delta^{(1)}_\phi Q_{\text{F}\,ii}^{ii}I_1
\col\qquad
&\ltwoopren{\fmfforce{(0.5w,0.5h)}{vc}
\fmf{plain_rar,left=0.75}{vo,vc}\fmf{plain_rar,right=0.75}{vo,vc}
\fmf{plain_rar}{vc,vl}\fmf{plain_rar}{vc,vr}
\fmffreeze
\fmfposition
\fmfcmd{draw (((polycross 4) scaled 8) rotated 45) shifted vloc(__vc);}
\fmfiv{label=$\scriptstyle Q_{\text{F}}$,l.a=0,l.dist=6}{vloc(__vc)}}
&=-2g_\YM^2N\delta Q_{\text{F}\,ii}^{ii}I_1
\col\\
\ltwoopren{\fmfforce{(0.5w,0.5h)}{vc}
\fmf{plain_rar,left=0.75}{vo,vc}\fmf{plain_rar,right=0.75}{vo,vc}
\fmf{plain_rar}{vc,vl}\fmf{plain_rar}{vc,vr}
\fmffreeze
\fmfposition
\fmfiv{d.sh=cross,d.size=8}{vloc(__vo)}
\fmfiv{label=$\scriptstyle Q_{\text{F}}$,l.a=0,l.dist=6}{vloc(__vc)}}
&=-2g_\YM^2N\delta J^{(1)}_{\mathcal{O}_2,\text{non-def}}Q_{\text{F}\,ii}^{ii}I_1
\col\qquad
&\ltwoopren{\fmfforce{(0.5w,0.5h)}{vc}
\fmf{plain_rar,left=0.75}{vo,vc}\fmf{plain_rar,right=0.75}{vo,vc}
\fmf{plain_rar}{vc,vl}\fmf{plain_rar}{vc,vr}
\fmffreeze
\fmfposition
\fmfiv{d.sh=cross,d.size=8}{vloc(__vo)}
\fmfiv{label=$\scriptstyle Q_{\text{F}}$,l.a=0,l.dist=6}{vloc(__vo)}}
&=g_\YM^2N\delta J^{(1)}_{\mathcal{O}_2,\text{def}}I_1
\col\\
\ltwoopren{\fmfforce{(0.5w,0.5h)}{vc}
\fmf{plain_rar,left=0.75}{vo,vc}\fmf{plain_rar,right=0.75}{vo,vc}
\fmf{plain_rar}{vc,vl}\fmf{plain_rar}{vc,vr}
\fmffreeze
\fmfposition
\fmfiv{d.sh=cross,d.size=8}{vloc(__vo)}
\fmfiv{label=$\scriptstyle Q_{\text{F}}$,l.a=0,l.dist=6}{vloc(__vc)}
\fmfiv{label=$\scriptstyle Q_{\text{F}}$,l.a=0,l.dist=6}{vloc(__vo)}}
&=-2g_\YM^2N\delta J^{(1)}_{\mathcal{O}_2,\text{def}}Q_{\text{F}\,ii}^{ii}I_1
\col\qquad
&\ltwoopren{
\fmf{plain_rar}{vo,vcl}\fmf{plain_rar}{vcl,vl}
\fmf{plain_rar}{vo,vcr}\fmf{plain_rar}{vcr,vr}
\fmffreeze
\fmf{photon}{vcl,vcr}
\fmfiv{d.sh=cross,d.size=8}{vloc(__vo)}
\fmfiv{label=$\scriptstyle Q_{\text{F}}$,l.a=0,l.dist=6}{vloc(__vo)}}
&=g_\YM^2N\delta J^{(1)}_{\mathcal{O}_2,\text{def}}\alpha I_1
\pnt
\end{aligned}
\end{equation}
The negative sum of the pole parts of the above diagrams is given by
\begin{equation}\label{deltaZO2dt1PI}
\begin{aligned}
\delta\mathcal{Z}^{(2)}_{\mathcal{O}_2,\text{dt},1\text{PI}}
&=g_\YM^4N^2
\big(16\sin^2\gamma_i^+\sin^2\gamma_i^-\Kop[\Kop[I_1]I_1]
+2Q_{\text{F}\,ii}^{ii}(\alpha+1
-2Q_{\text{F}\,ii}^{ii})\Kop\Rop[I_1^2]
\big)
\col
\end{aligned}
\end{equation}
where the contributions from terms in \eqref{deltaQ} depending on $Q_{\text{F}\,ii}^{ii}$ have combined with the remaining diagrams such that the results depends on the overall divergence $\Kop\Rop[I_1^2]=\Kop[I_1^2-2\Kop[I_1]I_1]=-\Kop[I_1]^2$ of the product $I_1^2$. Moreover, in contrast to the $L\ge 3 $ case, we have to consider also one-particle reducible (non-1PI) diagrams. They generate products of one-loop counter terms which contribute to $\delta\mathcal{Z}_{\mathcal{O}_2,\text{st}}^{(2)}$ in \eqref{Z2}, as follows from the loop expansion  \eqref{ZOLexpansion} of the renormalization constant. The only deformation-dependent non-1PI diagrams involve the one-loop counter term $\delta J^{(1)}_{\mathcal{O}_2,\text{def}}$, and they generate the contribution 
\begin{equation}
\begin{aligned}
\delta\mathcal{Z}^{(2)}_{\mathcal{O}_2,\text{dt},\text{non-$1$PI}}
&=
\frac{1}{2}\Bigg[
\ltwoopren{
\fmf{phantom}{vo,vl}
\fmf{plain_rar}{vo,vr}
\fmffreeze
\fmfposition
\fmfipath{p[]}
\fmfipair{vm[],vl[],vr[]}
\fmfiset{p1}{vpath(__vo,__vl)}
\fmfiset{p2}{vpath(__vo,__vr)}
\svertex{vm1}{p1}
\svertex{vm2}{p2}
\fmfi{plain_rar}{subpath (0,length(p1)*0.5) of p1}
\fmfi{plain_rar}{subpath (length(p1)*0.5,length(p1)) of p1}
\fmfi{phantom_cross}{p1}
\fmfiv{d.sh=cross,d.size=8}{vloc(__vo)}
\fmfiv{label=$\scriptstyle Q_{\text{F}}$,l.a=0,l.dist=6}{vloc(__vo)}}
+
\ltwoopren{
\fmf{plain_rar}{vo,vl}
\fmf{phantom}{vo,vr}
\fmffreeze
\fmfposition
\fmfipath{p[]}
\fmfipair{vm[],vl[],vr[]}
\fmfiset{p1}{vpath(__vo,__vl)}
\fmfiset{p2}{vpath(__vo,__vr)}
\svertex{vm1}{p1}
\svertex{vm2}{p2}
\fmfi{plain_rar}{subpath (0,length(p2)*0.5) of p2}
\fmfi{plain_rar}{subpath (length(p2)*0.5,length(p2)) of p2}
\fmfi{phantom_cross}{p2}
\fmfiv{d.sh=cross,d.size=8}{vloc(__vo)}
\fmfiv{label=$\scriptstyle Q_{\text{F}}$,l.a=0,l.dist=6}{vloc(__vo)}}
\Bigg]
=-\delta^{(1)}_\phi\delta J^{(1)}_{\mathcal{O}_2,\text{def}}
\pnt
\end{aligned}
\end{equation}
The complete two-loop contribution to the renormalization constant \eqref{Z2} is given by 
\begin{equation}\label{deltaZO2twoloop}
\begin{aligned}
\delta\mathcal{Z}^{(2)}_{\mathcal{O}_2}
&=\delta\mathcal{Z}^{(2)}_{\mathcal{O}_2,\text{st}}+\delta\mathcal{Z}^{(2)}_{\mathcal{O}_2,\text{dt},\text{$1$PI}}+\delta\mathcal{Z}^{(2)}_{\mathcal{O}_2,\text{dt},\text{non-$1$PI}}\\
&=
-g_\YM^4N^2(16\sin^2\gamma_i^+\sin^2\gamma_i^-\,\mathcal{I}_2
+4(Q_{\text{F}\,ii}^{ii})^2\Kop\Rop[I_1^2])
\pnt
\end{aligned}
\end{equation}
As discussed above, this result indeed contains the overall UV divergence $\mathcal{I}_2=\Kop\Rop[I_2]$ given in \eqref{KRI2}.

The two-loop renormalization constant is given by inserting the one- and two-loop corrections given respectively in \eqref{deltaZO2oneloop} and \eqref{deltaZO2twoloop} into \eqref{Z2}. Taking the logarithm and expanding it to second loop order, we obtain
\begin{equation}
\begin{aligned}\label{lnZO2}
\ln\mathcal{Z}_{\mathcal{O}_2}
&=\delta\mathcal{Z}^{(1)}_{\mathcal{O}_2}
+\delta\mathcal{Z}^{(2)}_{\mathcal{O}_2}
-\frac{1}{2}\big(\delta\mathcal{Z}^{(1)}_{\mathcal{O}_2}\big)^2+\mathcal{O}(g^6)\\
&=2g^2Q_{\text{F}\,ii}^{ii}\frac{1}{\varepsilon}
+2g^4\Big(
8\sin^2\gamma_i^+\sin^2\gamma_i^-\Big(\frac{1}{2\varepsilon^2}-\frac{1}{2\varepsilon}\Big)
+(Q_{\text{F}\,ii}^{ii})^2\frac{1}{\varepsilon^2}\Big)
+\mathcal{O}(g^6)
\pnt
\end{aligned}
\end{equation}
The coefficient of the $\frac{1}{\varepsilon^2}$-pole in \eqref{lnZO2} is $\frac12$ times the $\beta$-function $\beta_{Q_{\text{F}\,ii}^{ii}}$ given in \eqref{betaQ}. This guarantees that the anomalous dimension derived according to \eqref{gammaOLdef} is finite in the limit $\varepsilon\to0$. It reads
\begin{equation}\label{gammaO2}
\begin{aligned}
\gamma_{\mathcal{O}_2}
&=\Big(\varepsilon g\parderiv{g}
-\beta_{Q_{\text{F}\,ii}^{ii}}\parderiv{Q_{\text{F}\,ii}^{ii}}\Big)\ln\mathcal{Z}_{\mathcal{O}_2}
=4g^2Q_{\text{F}\,ii}^{ii}-32g^4\sin^2\gamma_i^+\sin^2\gamma_i^-
\end{aligned}
\end{equation}
in the dimensional reduction (DR) scheme which we used in the calculation.

The above result for $\gamma_{\mathcal{O}_2}$ depends on the chosen renormalization scheme, as discussed in Appendix \ref{app:schemedep}. At one loop, the coupling $Q^{ii}_{\text{F}\,ii}$ transforms under a scheme change as $Q^{\varrho\,ii}_{\text{F}\,ii}=Q^{ii}_{\text{F}\,ii}-\frac{\varrho}{2}\beta_{Q_{\text{F}\,ii}^{ii}}$, where the real parameter $\varrho$ labels the scheme and the $\beta$-function $\beta_{Q_{\text{F}\,ii}^{ii}}$ is given in \eqref{betaQ}.
In particular, $\varrho=0$ in the DR scheme. In a different scheme, the two-loop contribution in \eqref{gammaO2} acquires a contribution which is proportional to $\varrho$ and to  $\beta_{Q_{\text{F}\,ii}^{ii}}$. The result in the scheme $\varrho$ reads
\begin{equation}\label{gammaO2varrho}
\begin{aligned}
\gamma_{\mathcal{O}_2}^{\varrho}
&=4g^2Q_{\text{F}\,ii}^{ii}-32g^4\sin^2\gamma_i^+\sin^2\gamma_i^--2g^2\varrho\,\beta_{Q_{\text{F}\,ii}^{ii}}\\
&=4g^2Q_{\text{F}\,ii}^{ii}-8g^4\big(4(1+\varrho)\sin^2\gamma_i^+\sin^2\gamma_i^-+\varrho(Q^{ii}_{\text{F}\,ii})^2\big)
\col
\end{aligned}
\end{equation}
where in the second line we have inserted the explicit expression for $\beta_{Q_{\text{F}\,ii}^{ii}}$.

\section*{Acknowledgements}

We thank Burkhardt Eden and Stijn van Tongeren for useful discussions. 
J.F.\ and M.W.\ danken der Studienstiftung des Deutschen Volkes f\"ur ein 
Pro\-mo\-tions\-f\"or\-der\-ungs\-sti\-pen\-di\-um.

\appendix

\section{\texorpdfstring{The action of $\gamma_i$-deformed $\mathcal{N}=4$ SYM theory}{The action of gamma\_i-deformed N=4 SYM theory}}
\label{app:action}

In this appendix, we present the $\gamma_i$-deformation and our notation and 
conventions. For further details, we refer to our publication \cite{Fokken:2013aea}.

The gauge-fixed action of the $\gamma_i$-deformation with gauge group $SU(N)$ in Euclidean space can be written as
\begin{equation}
\begin{aligned}\label{S}
S
&=\int\de^4x\,\Big(\tr\Big[
-\frac{1}{4}F^{\mu\nu}F_{\mu\nu}-\frac{1}{2\alpha}(\partial^\mu A_\mu)^2
-(\D^\mu\bar\phi_i)\D_\mu\phi^i
+i\,\bar\psi^{\dot\alpha}_A\D_{\dot\alpha}{}^\alpha\psi_\alpha^A\\
&\hphantom{{}={}\int\de^4x\,\Big(\tr\Big[}
+g_\YM(\tilde\rho_i^{BA}\bar\psi^{\dot\alpha}_A\phi^i\bar\psi_{\dot\alpha\,B}
+(\tilde\rho^{\dagger\,i})_{BA}\psi^{\alpha\,A}\bar\phi_i\psi_\alpha^B)\\
&\phantom{{}={}\int\de^4x\,\Big(\tr\Big[}
+g_\YM(\rho_{i\,BA}\psi^{\alpha\,A}\phi^i\psi_\alpha^B
+(\rho^{\dagger\,i})^{BA}\bar\psi^{\dot\alpha}_A\bar\phi_i\bar\psi_{\dot\alpha\,B})
+\bar c\,\partial^\mu\D_\mu c\,\Big]
\\
&\phantom{{}={}\int\de^4x\,\Big(}
+g_\YM^2\Big(
\hat Q_{lk}^{ij}\tr[\bar\phi_i\bar\phi_j\phi^k\phi^l]
+\tilde Q_{kl}^{ij}\tr[\bar\phi_i\phi^k\bar\phi_j\phi^l]
-\frac{1}{N}Q_{lk}^{ij}\tr[\bar\phi_i\bar\phi_j]\tr[\phi^k\phi^l]\Big)
\Big)
\col
\end{aligned}
\end{equation}
where we have adopted the conventions of \cite{Gates:1983nr}, in particular the 
ones for raising, lowering and contractions of spinor indices.
Note that in this action doubled spacetime indices $\mu,\nu\in\{0,1,2,3\}$,
spinor indices $\alpha,\dot\alpha\in\{1,2\}$ and
flavor indices $i,j,k,l\in\{1,2,3\}$, $A,B\in\{1,2,3,4\}$ are summed.
This is the only exception to the rule that throughout this paper Einstein's summation convention never applies.
The deformation parameters $\gamma_i$ only enter the coupling tensors of the Yukawa-type scalar-fermion and F-term-type scalar couplings. 
Apart from the coupling \eqref{Qiiii}, the above action can be obtained by replacing the product of two fields by a non-commutative $\ast$-product  in the component expansion of the $\mathcal{N}=4$ SYM theory before the auxiliary fields are integrated out \cite{Fokken:2013aea}. This generates in particular the double-trace coupling with tensor $Q^{ij}_{lk}$. The coupling \eqref{Qiiii} can be introduced by redefining $Q^{ij}_{lk}$.

The $\ast$-products of two component fields $A$ and $B$ reads \cite{Beisert:2005if}
\begin{equation}\label{astprod}
A\ast B=\e^{\frac{i}{2}\mathbf{q}_A\wedge\mathbf{q}_B}\col
\end{equation}
where the antisymmetric product of the two $(q^1,q^2,q^3)$-charge vectors $\mathbf{q}_A$ and $\mathbf{q}_B$
is defined as
\begin{equation}
\mathbf{q}_A\wedge \mathbf{q}_B=(\mathbf{q}_A)^{\Top}\mathbf{C}\,\mathbf{q}_B
\col\qquad
\mathbf{C}=\begin{pmatrix}
0 & -\gamma_3 & \gamma_2 \\
\gamma_3 & 0 & -\gamma_1 \\
-\gamma_2 & \gamma_1 & 0 
\end{pmatrix}
\pnt
\end{equation} 
For the different fields, the components of the charge vectors are given by
\begin{equation}
\begin{array}{c|cccc|c|ccc}
B&\psi^1_{\alpha}&\psi^2_{\alpha}&\psi^3_{\alpha}&\psi^4_{\alpha}
&A_{\mu}&\phi^1&\phi^2&\phi^3\\
 & & & & & & & &\\[-0.4cm]
\hline
 & & & & & & & &\\[-0.4cm]
q^1_B & +\frac12 & -\frac12 & -\frac12 & +\frac12 & 0 & 1 & 0 & 0\\
 & & & & & & & &\\[-0.4cm]
q^2_B & -\frac12 & +\frac12 & -\frac12 & +\frac12 & 0 & 0 & 1 & 0\\
 & & & & & & & &\\[-0.4cm]
q^3_B & -\frac12 & -\frac12 & +\frac12 & +\frac12 & 0 & 0 & 0 & 1\\
\end{array}
\col
\end{equation}
and for the anti-fields their signs are reversed.

We define the antisymmetric phase tensors $\Gamma_{AB}$ and $\Gamma^+_{ij}$ via 
\begin{equation}
\begin{aligned}\label{Gammarel}
\Gamma_{i4}
=\mathbf{q}_{\psi^i}\wedge\mathbf{q}_{\psi^4}
=\gamma_i^-\col
\quad\Gamma_{i\,i+1}
=\mathbf{q}_{\psi^i}\wedge\mathbf{q}_{\psi^{i+1}}
=\gamma_{i+2}^+
\col\quad
\Gamma^+_{i\,i\pm1}
=\mathbf{q}_{\phi^i}\wedge\mathbf{q}_{\phi^{i\pm1}}
=\gamma_i^-\pm\gamma_i^+
\col
\end{aligned}
\end{equation}
where cyclic identification $i+3\sim i$ is understood. In terms of these phase tensors, the Yukawa type coupling tensors in the action \eqref{S} are given by
\begin{equation}\label{rhodef}
\rho_{i\,AB}=i\epsilon_{4iAB}\e^{\frac{i}{2}\Gamma_{AB}}
\col\qquad
\tilde\rho_i^{AB}=(\delta_4^A\delta_i^B-\delta_4^B\delta_i^A)\e^{\frac{i}{2}\Gamma_{AB}}
\col
\end{equation}
where we trust that the reader will not confuse the complex number $i$ with the index $i$. They obey the conjugation relations
\begin{equation}\label{rhorel}
\begin{aligned}
(\rho^{\dagger i})^{AB}
&=(\rho_{i\,BA})^\ast
=\rho_{i\,AB}
\col\qquad (\tilde\rho^{\dagger i})_{AB}
=(\tilde\rho_i^{BA})^\ast
=-\tilde\rho_i^{AB}\pnt
\end{aligned}
\end{equation}
The coupling tensors of the quartic scalar interactions read
\begin{equation}
\label{Qdef}
\hat Q_{lk}^{ij}
=\delta_k^i\delta_l^j\,{\e}^{i\Gamma^+_{ij}}-\frac{1}{2}\delta_l^i\delta_k^j\col\qquad
\tilde Q_{kl}^{ij}
=-\frac{1}{4}(\delta^i_k\delta^j_l+\delta^i_l\delta^j_k)\col\qquad
Q_{lk}^{ij}=\delta_k^i\delta_l^j\,{\e}^{i\Gamma^+_{ij}}-\delta_l^i\delta_k^j
+Q_{\text{F}\,lk}^{ij}
\col
\end{equation}
where $Q_{\text{F}\,lk}^{ij}$ is a tree-level coupling tensor with nontrivial components only for $i=j=k=l$ which have to vanish in the special case $\gamma_1=\gamma_2=\gamma_3$.

The Feynman rules for the $\gamma_i$-deformation can be found in Appendix B of our work \cite{Fokken:2013aea}. For the calculations in this paper, it is useful to alter the Feynman rules for the quartic scalar  interactions: in \cite{Fokken:2013aea} we have split these interactions into those originating from the F-term and D-term couplings in the supersymmetric special cases. Here, we split the interactions according to the two single-trace structures of four scalar fields in \eqref{S}. In this case, the entire F-term and parts of the D-term interactions contribute to the tensor $\hat Q_{lk}^{ij}$ in \eqref{Qdef}, while $\tilde Q_{kl}^{ij}$ is built from the remaining D-term interactions. Moreover, we have kept the double-trace couplings with tensor structure $Q_{\text{F}\,lk}^{ij}$ in a separate vertex.

\section{Tensor identities}
\label{app:tensorid}
In this appendix, we explicitly evaluate the combinations of the coupling tensors that are encountered in the Feynman diagram analysis in Section \ref{sec:Lge3}. Recall that Einstein's summation convention does not apply in the following expressions.

For the scalar diagrams, we need the expressions
\begin{equation}
\begin{aligned}\label{eq:Qidentity}
\sum_{j=1}^3 (\hat Q_{ij}^{ji})^L
&=\sum_{\substack{j=1\\ j\neq i}}^3(\hat Q_{ij}^{ji})^L
+(\hat Q_{ii}^{ii})^L
=\sum_{\substack{j=1\\ j\neq i}}^3\e^{iL\Gamma^+_{ij}}+\frac{1}{2^L}
=2\e^{iL\gamma_i^-}\cos{L\gamma_i^+}+\frac{1}{2^L}
\col\\
\sum_{j=1}^3 (\hat Q_{ji}^{ij})^L
&=\sum_{\substack{j=1\\ j\neq i}}^3(\hat Q_{ji}^{ij})^L
+(\hat Q_{ii}^{ii})^L
=\sum_{\substack{j=1\\ j\neq i}}^3\e^{-iL\Gamma^+_{ij}}+\frac{1}{2^L}
=2\e^{-iL\gamma_i^-}\cos{L\gamma_i^+}+\frac{1}{2^L}
\col
\end{aligned}
\end{equation}
where we have first used \eqref{Qdef} and then \eqref{Gammarel}.

For the diagrams with a fermionic wrapping loop, we first evaluate the contractions of two Yukawa-type coupling tensors. The resulting expressions read
\begin{equation}
\begin{aligned}
( \rho^{\dagger\,i})(\rho_{i})^{\T} {}^A{}_{B} &= \sum_{C=1}^4 (\rho^{\dagger\,i})^{AC} (\rho_{i})^{\T}_{CB}
 =-\delta^A_B\sum_{C=1}^4 (\epsilon_{ACi4})^2 \,{\e}^{i\Gamma_{AC}} \col \\
 (\tilde\rho^{\dagger\,i})(\tilde\rho_i)^{\T}{}^A{}_{B}&= \sum_{C=1}^4 (\tilde\rho^{\dagger\,i})^{AC}(\tilde\rho_i)^{\T}_{CB}
 =- \delta^A_4\delta^4_B\,{\e}^{i\Gamma_{4i}}-\delta^A_i\delta^i_B\,{\e}^{i\Gamma_{i4}} \col
\end{aligned}
\end{equation}
where we have used \eqref{rhodef} and \eqref{rhorel}. With these results as well as \eqref{Gammarel}, the required traces are determined as
\begin{equation}\label{eq:rhoidentity}
\begin{aligned}
 \tr[ (( \rho^{\dagger\,i})(\rho_{i})^{\T})^L] 
&= \sum_{A,C=1}^4 \left(- (\epsilon_{ACi4})^2 \e^{i\Gamma_{AC}} \right)^L 
= 2 (-1)^L \cos L\gamma_i^+ \col \\
 \tr[ ( (\tilde\rho^{\dagger\,i})(\tilde\rho_i)^{\T})^L] 
&= \sum_{A=1}^4 \left( - \delta_{A4}\e^{i\Gamma_{4i}}-\delta_{Ai}\e^{i\Gamma_{i4}} \right)^L
= 2 (-1)^L \cos L\gamma_i^- \pnt
\end{aligned}
\end{equation}

\section{Renormalization of composite operators}
\label{app:renormalization}

In this appendix, we review how composite operators are incorporated into the theory and how they are renormalized; see e.g.\ the textbooks \cite{ZinnJustin:2002ru,Collins:1984xc}.

Composite operators such as $\mathcal{O}_L$ in \eqref{groundstate} can be added to the action regularized in $D=4-2\varepsilon$ dimensions via a coupling to an external source $J_{\mathcal{O}_L}$. If $\mathcal{O}_L$ has scaling dimension $\Delta$, the source $J_{\mathcal{O}_L}$ has to have scaling dimension $D-\Delta$. The resulting term in the action then reads

\begin{equation}\label{SOL}
\delta S_{\mathcal{O}_L}
=\int\de^Dx\,J_{\mathcal{O}_L,0}\,\mathcal{O}_{L,0}(\phi_0^i)
=\int\de^Dx\,J_{\mathcal{O}_L}\big[\mathcal{O}_L(\phi^i)+\delta J_{\mathcal{O}_L}\mathcal{O}_L(\phi^i)\big]
\col
\end{equation}
where the explicit expression is given first in terms of the bare quantities and second in terms of renormalized quantities and a respective counter term. The renormalized and bare quantities are related via respective renormalization constants as
\begin{equation}
\begin{aligned}\label{phisourceren}
\phi^i=\mathcal{Z}_{\phi^i}^{-\frac{1}{2}}\phi_0^i
\col\qquad
J_{\mathcal{O}_L}=\mathcal{Z}_{J_{\mathcal{O}_L}}^{-1}J_{\mathcal{O}_L,0}
\col
\end{aligned}
\end{equation}
where $\mathcal{Z}_{\phi^i}$ and $\mathcal{Z}_{J_{\mathcal{O}_L}}$ are given in 
terms of the counter terms $\delta_{\phi^i}$ and $\delta J_{\mathcal{O}_L}$ 
as 
\begin{equation}
\begin{aligned}\label{Zsource}
 \mathcal{Z}_{\phi^i}=1+\delta_{\phi^i}
\col\qquad \mathcal{Z}_{J_{\mathcal{O}_L}}=\mathcal{Z}_{\mathcal{O}_L,1\text{PI}}\mathcal{Z}_{\phi^i}^{-\frac{L}{2}}
\col\qquad \mathcal{Z}_{\mathcal{O}_L,1\text{PI}}=1+\delta J_{\mathcal{O}_L}\pnt
\end{aligned}
\end{equation}
The counter term $\delta_{\phi^i}$ is $\frac{1}{p^2}$ times the sum of the divergences of the $1\text{PI}$ self-energy diagrams for the field $\phi^i$ with momentum $p_\nu$. The counter term $\delta J_{\mathcal{O}_L}$ is the negative sum of the divergences of the $1\text{PI}$ diagrams involving one operator $\mathcal{O}_L$.

Instead of renormalizing the sources, we can alternatively introduce a renormalization constant that expresses the renormalized operators in terms of the bare ones $\mathcal{O}_{L,0}$ as
\begin{equation}
\mathcal{O}_L(\phi^i)=\mathcal{Z}_{\mathcal{O}_L}\mathcal{O}_{L,0}(\phi_0^i)
\pnt
\end{equation}
We make contact with the source renormalization by demanding
\begin{equation}\label{JO}
J_{\mathcal{O}_L,0}\,\mathcal{O}_{L,0}(\phi_0^i)=J_{\mathcal{O}_L}\mathcal{O}_L(\phi^i)
\col
\end{equation}
which immediately yields 
\begin{equation}\label{ZOL}
\mathcal{Z}_{\mathcal{O}_L}=\mathcal{Z}_{J_{\mathcal{O}_L}}=\mathcal{Z}_{\mathcal{O}_L,1\text{PI}}\mathcal{Z}_{\phi^i}^{-\frac{L}{2}}\pnt
\end{equation}
Inserting the counter terms, the first two terms in the loop expansion of the above equation are given by
\begin{equation}\label{ZOLexpansion}
\begin{aligned}
\delta\mathcal{Z}^{(1)}_{\mathcal{O}_L}&=\delta J^{(1)}_{\mathcal{O}_L}-\frac{L}{2}\delta^{(1)}_{\phi^i}\col\\
\delta\mathcal{Z}^{(2)}_{\mathcal{O}_L}&=\delta J^{(2)}_{\mathcal{O}_L}
-\frac{L}{2}\delta^{(2)}_{\phi^i}-\frac{L}{2}\delta^{(1)}_{\phi^i}\Big(\delta J^{(1)}_{\mathcal{O}_L}-\frac{L+2}{4}\delta^{(1)}_{\phi^i}\Big)
\col
\end{aligned}
\end{equation}
where the superscript in parenthesis denotes the loop order of the respective contribution. The products of one-loop counter terms in the two-loop contribution can be interpreted in terms of non-1PI diagrams.

We consider Green functions that involve the operator $\mathcal{O}_{L}$ as well as $L$ anti-scalar fields $\bar\phi_i$. The bare connected Green function and its amputated counterpart are then given in terms of the  renormalized ones as
\begin{equation}\label{GFrel}
\begin{aligned}
\langle\mathcal{O}_{L,0}(x)\bar{\phi}_{i,0}(x_1)\dots\bar{\phi}_{i,0}(x_L)\rangle_{\text{c}}
&=\mathcal{Z}_{\mathcal{O}_L,1\text{PI}}^{-1}\mathcal{Z}_{\phi^i}^{L}\langle\mathcal{O}_{L}(x)\bar{\phi}_i(x_1)\dots\bar{\phi}_i(x_L)\rangle_{\text{c}}
\col\\
\langle\mathcal{O}_{L,0}(x)\bar{\phi}_{i,0}(x_1)\dots\bar{\phi}_{i,0}(x_L)\rangle_{\text{a}}
&=\mathcal{Z}_{\mathcal{O}_L,1\text{PI}}^{-1}\langle\mathcal{O}_{L}(x)\bar{\phi}_i(x_1)\dots\bar{\phi}_i(x_L)\rangle_{\text{a}}
\pnt
\end{aligned}
\end{equation}
The UV divergence of the connected Green function and the renormalization constant $\mathcal{Z}_{\mathcal{O}_L}$ in \eqref{ZOL} are determined by the same diagrams: these are the 1PI diagrams which renormalize the amputated Green function and the non-1PI diagrams which involve self-energy corrections of the non-amputated propagators. While the above Green functions are gauge dependent, the combination in \eqref{ZOL} is, however, gauge invariant and thus independent of the gauge-fixing parameter $\alpha$.

The renormalization constants and renormalized Green functions on the rhs.\ of \eqref{GFrel} depend on the renormalization scale given by the 't Hooft mass $\mu$. This scale is introduced in a relation for the bare Yang-Mills coupling constant $g_{\YM ,0}=\mu^\varepsilon g_\YM$. It guarantees that $g_\YM$ and hence the effective planar coupling constant $g=\frac{\sqrt{\lambda}}{4\pi}$, as well as $Q_{\text{F}\,ii}^{ii}$ of \eqref{Qiiii}, are dimensionless in $D=4-2\varepsilon$ dimensions. The bare Green functions have to be independent of $\mu$. This condition leads to renormalization group equations (RGEs) for the renormalized Green functions. They are given by
\begin{equation}
\Big(\mu\parderiv{\mu}+\beta_g\parderiv{g}+\beta_{Q_{\text{F}\,ii}^{ii}}\parderiv{{Q_{\text{F}\,ii}^{ii}}}+\delta\parderiv{\alpha}+\gamma_{\mathcal{O}_L}\pm L\gamma_{\phi^i}\Big)\langle\mathcal{O}_{L}(x)\bar{\phi}_i(x_1)\dots\bar{\phi}_i(x_L)\rangle_{\substack{\text{c} \\ \text{a}}}=0\col
\end{equation}
where the upper and lower sign holds for the connected and amputated Green function, respectively, and the renormalization group functions are defined as
\begin{equation}\label{definition of gamma}
\beta_g=\mu\deriv[g]{\mu}\col\quad
\beta_{Q_{\text{F}\,ii}^{ii}}=\mu\deriv[{Q_{\text{F}\,ii}^{ii}}]{\mu}\col\quad
\delta=\mu\deriv[\alpha]{\mu}\col\quad
\gamma_{\mathcal{O}_L}=-\mu\deriv{\mu}\ln\mathcal{Z}_{\mathcal{O}_L}\col\quad
\gamma_{\phi^i}=\frac{\mu}{2}\deriv{\mu}\ln\mathcal{Z}_{\phi^i}
\pnt
\end{equation}
Since $g_\YM$ and hence also $g$ is not renormalized in the theories we consider in this paper, $\beta_g$ can be determined exactly. Using that the bare coupling $g_0$ is independent of $\mu$ and that it obeys the relation $g_0=\mu^\varepsilon g$, one obtains
\begin{equation}
0=\mu\deriv[g_0]{\mu}=\Big(\mu\parderiv{\mu}+\beta_g\parderiv{g}\Big)\mu^\varepsilon g=\mu^\varepsilon(\varepsilon g+\beta_g)\col
\end{equation}
which yields $\beta_g=-\varepsilon g$.
Inserting this result into the definition of the anomalous dimension in \eqref{definition of gamma}, 
one finds
\begin{equation}\label{gammaOLdef}
\gamma_{\mathcal{O}_L}=\Big(\varepsilon g\parderiv{g}-\beta_{Q_{\text{F}\,ii}^{ii}}\parderiv{{Q_{\text{F}\,ii}^{ii}}}\Big)\ln\mathcal{Z}_{\mathcal{O}_L}
\pnt
\end{equation}
The above result must be finite in the limit $\varepsilon\to0$, which has to be taken at the end. If $\beta_{Q_{\text{F}\,ii}^{ii}}$ vanishes in the limit $\varepsilon\to0$, $\ln\mathcal{Z}_{\mathcal{O}_L}$ must not contain higher poles in $\varepsilon$. If $\beta_{Q_{\text{F}\,ii}^{ii}}$ does not vanish, however, $\ln\mathcal{Z}_{\mathcal{O}_L}$ has to contain also higher-order poles in $\varepsilon$ such that cancellations of all poles occur among both terms in \eqref{gammaOLdef}.

\section{Renormalization-scheme dependence}
\label{app:schemedep}

In this appendix, we discuss the renormalization-scheme dependence of the anomalous dimension $\gamma_{\mathcal{O}_2}$ given in \eqref{gammaO2}.

A renormalization scheme defines a prescription for the regularization of the UV divergences and their absorption into the counter terms. In particular, it specifies which finite contributions are absorbed into the counter terms together with the UV divergences and are hence subtracted from the regularized expressions.
In the dimensional reduction (DR) scheme \cite{Siegel:1979wq}, only the poles in $\varepsilon$ of the theory regularized in $D=4-2\varepsilon$ dimensions are absorbed into the counter terms, like in the famous minimal subtraction (MS) scheme \cite{'tHooft:1973mm}. 
In a modified dimensional reduction ($\overline{\text{DR}}$) scheme, the finite combination $\varrho=-\gamma_{\text{E}}+\ln4\pi$ is also absorbed, in analogy to the widely used modified minimal subtraction ($\overline{\text{MS}}$) scheme \cite{Bardeen:1978yd}.\footnote{See \cite{Mertig:1995ny,Vogelsang:1995vh} for a complete definition in the context of QCD, including also a description for handling $\gamma_5$ in $D=4-2\varepsilon$ dimensions.} Both, the DR and the $\overline{\text{DR}}$ scheme, are members of a family of schemes labeled by a free parameter $\varrho$. Two schemes of this family are related via a change of the 't Hooft mass $\mu$, which induces a change of the renormalized fields and coupling constants. In particular, the theory in the scheme $\varrho$ at 't Hooft mass $\mu$ is obtained by applying the subtractions of the DR scheme ($\varrho=0$) at 't Hooft mass $\mu_\varrho$, and then introducing $\mu$ via the relation $\mu_\varrho=\mu\e^{-\frac{\varrho}{2}}$. 

The effective planar coupling constant $g_{\varrho}$ and the running coupling $Q_{\text{F}\,ii}^{\varrho\,ii}$ of the $\gamma_i$-deformation in the scheme $\varrho$ can be expressed as expansions of those in the DR scheme. From the relation $\mu_\varrho^\varepsilon g_\varrho=\mu^\varepsilon g$, the coupling $g_\varrho$  is obtained as 
\begin{equation}\label{gvarrho}
g_\varrho=\e^{\frac{\varepsilon}{2}\varrho}g
\pnt
\end{equation}
This relation holds to all orders in planar perturbation theory, since $g$ is not renormalized. The one-loop renormalization of the running coupling $Q^{ii}_{\text{F}\,ii}$ was determined in the DR scheme in \cite{Fokken:2013aea}. Replacing the coupling $g$ by $g_\varrho$ in the respective counter terms yields the renormalization in the scheme $\varrho$. Then reexpressing the result in terms of the couplings $g$ and $Q^{ii}_{\text{F}\,ii}$ and neglecting terms which vanish in the limit $\varepsilon\to0$ yields the relation\footnote{The one-loop $\beta$-function $\beta_{Q_{\text{F}\,ii}^{ii}}$ given in \eqref{betaQ} itself is scheme independent, since the replacement only generates terms that vanish in the limit $\varepsilon\to0$ when no prefactor with poles in $\varepsilon$ is present.}
\begin{equation}
\begin{aligned}\label{Qredef}
Q^{\varrho\,ii}_{\text{F}\,ii}
=Q^{ii}_{\text{F}\,ii}-\frac{\varrho}{2}\beta_{Q_{\text{F}\,ii}^{ii}}+\mathcal{O}(g^4)
\pnt
\end{aligned}
\end{equation}

The anomalous dimension $\gamma_{\mathcal{O}_2}$ given in \eqref{gammaO2} depends on the renormalization scheme because of the finite redefinition of the coupling $Q^{\varrho\,ii}_{\text{F}\,ii}$ in \eqref{Qredef}. In the DR scheme, $\gamma_{\mathcal{O}_2}$ is determined from the logarithm of the renormalization constant \eqref{lnZO2}. The latter can be rewritten in terms of the individual one- and two-loop contributions $\gamma^{(1)}_{\mathcal{O}_2}=4g^2Q_{\text{F}\,ii}^{ii}$ and $\gamma^{(2)}_{\mathcal{O}_2}=-32g^4\sin^2\gamma_i^+\sin^2\gamma_i^-$ taken from \eqref{gammaO2} and the $\beta$-function $\beta_{Q_{\text{F}\,ii}^{ii}}$ given in \eqref{betaQ} as
\begin{equation}
\begin{aligned}
\ln\mathcal{Z}_{\mathcal{O}_2}
&=\frac{1}{2\varepsilon}\gamma^{(1)}_{\mathcal{O}_2}
+\frac{g^2}{2\varepsilon^2}\beta_{Q_{\text{F}\,ii}^{ii}}
+\frac{1}{4\varepsilon}\gamma^{(2)}_{\mathcal{O}_2}
+\mathcal{O}(g^6)
\pnt
\end{aligned}
\end{equation}
In the scheme $\varrho$,  the result for $\ln\mathcal{Z}^\varrho_{\mathcal{O}_2}$ is obtained by inserting the couplings $g_\varrho$ and $Q_{\text{F}\,ii}^{\varrho\,ii}$ of this new scheme into the above expression. Expressing the result in terms of the couplings in the DR scheme via \eqref{gvarrho} and \eqref{Qredef}, one finds that the difference of the logarithms in the two schemes is finite and given by 
\begin{equation}
\begin{aligned}\label{lnZO2rho}
\ln\mathcal{Z}^\varrho_{\mathcal{O}_2}
-\ln\mathcal{Z}_{\mathcal{O}_2}&=
\frac{\varrho}{2}\gamma^{(1)}_{\mathcal{O}_2}
+\mathcal{O}(g^4)
\pnt
\end{aligned}
\end{equation}
Using this result to determine the anomalous dimension in the scheme $\varrho$ via \eqref{gammaOLdef}, one obtains
\begin{equation}
\gamma_{\mathcal{O}_2}^{\varrho}=\Big(\varepsilon g\parderiv{g}-\beta_{Q_{\text{F}\,ii}^{ii}}\parderiv{{Q_{\text{F}\,ii}^{ii}}}\Big)\ln\mathcal{Z}^\varrho_{\mathcal{O}_2}
=\gamma_{\mathcal{O}_2}-2g^2\varrho\,\beta_{Q_{\text{F}\,ii}^{ii}}+\mathcal{O}(g^6)
\col
\end{equation}
where the term which is added to the anomalous dimension of the DR scheme originates from the partial derivative of the one-loop term in \eqref{lnZO2rho} with respect to $Q_{\text{F}\,ii}^{ii}$. Since this derivative is multiplied by the one-loop $\beta$-function $\beta_{Q_{\text{F}\,ii}^{ii}}$, the resulting term is a two-loop contribution. Accordingly, it was possible to discard higher-loop contributions to \eqref{lnZO2rho}, since they contribute to the anomalous dimension only beyond two loops. The individual one- and two-loop contributions to the anomalous dimension in the two schemes are hence related as\footnote{Note that a direct identification of $\gamma^\varrho_{\mathcal{O}_2}(g_\varrho,Q_{\text{F}\,ii}^{\varrho\,ii})=\gamma_{\mathcal{O}_2}(g,Q_{\text{F}\,ii}^{ii})$ leads to the same result, which immediately shows that the scheme dependence of the two-loop term originates from the finite redefinition \eqref{Qredef} of the coupling constant $Q_{\text{F}\,ii}^{ii}$.}
\begin{equation}\label{gamma12varrho}
\gamma^{\varrho\,(1)}_{\mathcal{O}_2}=\gamma^{(1)}_{\mathcal{O}_2}\col\qquad
\gamma^{\varrho\,(2)}_{\mathcal{O}_2}=\gamma^{(2)}_{\mathcal{O}_2}-2g^2\varrho\,\beta_{Q_{\text{F}\,ii}^{ii}}\pnt
\end{equation}

\footnotesize
\bibliographystyle{JHEP}
\bibliography{references}

\providecommand{\href}[2]{#2}\begingroup\raggedright\begin{thebibliography}{10}

\bibitem{Frolov:2005dj}
S.~Frolov, {\it {Lax pair for strings in Lunin-Maldacena background}},  {\em
  JHEP} {\bf 05} (2005) 069,
  [\href{http://xxx.lanl.gov/abs/hep-th/0503201}{{\tt hep-th/0503201}}].

\bibitem{Maldacena:1997re}
J.~M. Maldacena, {\it {The large $N$ limit of superconformal field theories and
  supergravity}},  {\em Adv. Theor. Math. Phys.} {\bf 2} (1998) 231--252,
  [\href{http://xxx.lanl.gov/abs/hep-th/9711200}{{\tt hep-th/9711200}}].

\bibitem{Gubser:1998bc}
S.~S. Gubser, I.~R. Klebanov, and A.~M. Polyakov, {\it {Gauge theory
  correlators from non-critical string theory}},  {\em Phys. Lett.} {\bf B428}
  (1998) 105--114, [\href{http://xxx.lanl.gov/abs/hep-th/9802109}{{\tt
  hep-th/9802109}}].

\bibitem{Witten:1998qj}
E.~Witten, {\it {Anti-de Sitter space and holography}},  {\em Adv. Theor. Math.
  Phys.} {\bf 2} (1998) 253--291,
  [\href{http://xxx.lanl.gov/abs/hep-th/9802150}{{\tt hep-th/9802150}}].

\bibitem{Lunin:2005jy}
O.~Lunin and J.~M. Maldacena, {\it {Deforming field theories with $U(1)\times
  U(1)$ global symmetry and their gravity duals}},  {\em JHEP} {\bf 05} (2005)
  033, [\href{http://xxx.lanl.gov/abs/hep-th/0502086}{{\tt hep-th/0502086}}].

\bibitem{Leigh:1995ep}
R.~G. Leigh and M.~J. Strassler, {\it {Exactly marginal operators and duality
  in four-dimensional $\mathcal{N}=1$ supersymmetric gauge theory}},  {\em
  Nucl. Phys.} {\bf B447} (1995) 95--136,
  [\href{http://xxx.lanl.gov/abs/hep-th/9503121}{{\tt hep-th/9503121}}].

\bibitem{'tHooft:1973jz}
G.~'t~Hooft, {\it {A planar diagram theory for strong interactions}},  {\em
  Nucl. Phys.} {\bf B72} (1974) 461.

\bibitem{Sieg:2005kd}
C.~Sieg and A.~Torrielli, {\it {Wrapping interactions and the genus expansion
  of the $2$-point function of composite operators}},  {\em Nucl.Phys.} {\bf
  B723} (2005) 3--32, [\href{http://xxx.lanl.gov/abs/hep-th/0505071}{{\tt
  hep-th/0505071}}].

\bibitem{Filk:1996dm}
T.~Filk, {\it {Divergencies in a field theory on quantum space}},  {\em
  Phys.Lett.} {\bf B376} (1996) 53--58.

\bibitem{Beisert:2005if}
N.~Beisert and R.~Roiban, {\it {Beauty and the twist: The Bethe ansatz for
  twisted $\mathcal{N}=4$ SYM}},  {\em JHEP} {\bf 08} (2005) 039,
  [\href{http://xxx.lanl.gov/abs/hep-th/0505187}{{\tt hep-th/0505187}}].

\bibitem{Beisert:2003jj}
N.~Beisert, {\it {The complete one loop dilatation operator of $\mathcal{N}=4$
  superYang-Mills theory}},  {\em Nucl.Phys.} {\bf B676} (2004) 3--42,
  [\href{http://xxx.lanl.gov/abs/hep-th/0307015}{{\tt hep-th/0307015}}].

\bibitem{Roiban:2003dw}
R.~Roiban, {\it {On spin chains and field theories}},  {\em JHEP} {\bf 09}
  (2004) 023, [\href{http://xxx.lanl.gov/abs/hep-th/0312218}{{\tt
  hep-th/0312218}}].

\bibitem{Beisert:2010jr}
N.~Beisert, C.~Ahn, L.~F. Alday, Z.~Bajnok, J.~M. Drummond, et~al., {\it
  {Review of AdS/CFT Integrability: An Overview}},  {\em Lett.Math.Phys.} {\bf
  99} (2012) 3--32, [\href{http://xxx.lanl.gov/abs/1012.3982}{{\tt
  arXiv:1012.3982}}].

\bibitem{Zoubos:2010kh}
K.~Zoubos, {\it {Review of AdS/CFT Integrability, Chapter IV.2: Deformations,
  Orbifolds and Open Boundaries}},  {\em Lett.Math.Phys.} {\bf 99} (2012)
  375--400, [\href{http://xxx.lanl.gov/abs/1012.3998}{{\tt arXiv:1012.3998}}].

\bibitem{Gromov:2010dy}
N.~Gromov and F.~Levkovich-Maslyuk, {\it {Y-system and $\beta$-deformed
  $\mathcal{N}=4$ Super-Yang-Mills}},  {\em J. Phys.} {\bf A44} (2011) 015402,
  [\href{http://xxx.lanl.gov/abs/1006.5438}{{\tt arXiv:1006.5438}}].

\bibitem{Arutyunov:2010gu}
G.~Arutyunov, M.~de~Leeuw, and S.~J. van Tongeren, {\it {Twisting the Mirror
  TBA}},  {\em JHEP} {\bf 02} (2011) 025,
  [\href{http://xxx.lanl.gov/abs/1009.4118}{{\tt arXiv:1009.4118}}].

\bibitem{Ahn:2010ws}
C.~Ahn, Z.~Bajnok, D.~Bombardelli, and R.~I. Nepomechie, {\it {Twisted Bethe
  equations from a twisted S-matrix}},  {\em JHEP} {\bf 02} (2011) 027,
  [\href{http://xxx.lanl.gov/abs/1010.3229}{{\tt arXiv:1010.3229}}].

\bibitem{Fiamberti:2008sm}
F.~Fiamberti, A.~Santambrogio, C.~Sieg, and D.~Zanon, {\it {Finite-size effects
  in the superconformal $\beta$-deformed $\mathcal{N}=4$ SYM}},  {\em JHEP}
  {\bf 0808} (2008) 057, [\href{http://xxx.lanl.gov/abs/0806.2103}{{\tt
  arXiv:0806.2103}}].

\bibitem{Sieg:2010tz}
C.~Sieg, {\it {Superspace calculation of the three-loop dilatation operator of
  $\mathcal{N}=4$ SYM theory}},  {\em Phys.Rev.} {\bf D84} (2011) 045014,
  [\href{http://xxx.lanl.gov/abs/1008.3351}{{\tt arXiv:1008.3351}}].

\bibitem{Gross:2002su}
D.~J. Gross, A.~Mikhailov, and R.~Roiban, {\it {Operators with large R charge
  in $\mathcal{N}=4$ Yang-Mills theory}},  {\em Annals Phys.} {\bf 301} (2002)
  31--52, [\href{http://xxx.lanl.gov/abs/hep-th/0205066}{{\tt
  hep-th/0205066}}].

\bibitem{Ambjorn:2005wa}
J.~Ambjorn, R.~A. Janik, and C.~Kristjansen, {\it {Wrapping interactions and a
  new source of corrections to the spin-chain / string duality}},  {\em Nucl.
  Phys.} {\bf B736} (2006) 288--301,
  [\href{http://xxx.lanl.gov/abs/hep-th/0510171}{{\tt hep-th/0510171}}].

\bibitem{Fiamberti:2008sn}
F.~Fiamberti, A.~Santambrogio, C.~Sieg, and D.~Zanon, {\it {Single impurity
  operators at critical wrapping order in the $\beta$-deformed $\mathcal{N}=4$
  SYM}},  {\em JHEP} {\bf 0908} (2009) 034,
  [\href{http://xxx.lanl.gov/abs/0811.4594}{{\tt arXiv:0811.4594}}].

\bibitem{Gunnesson:2009nn}
J.~Gunnesson, {\it {Wrapping in maximally supersymmetric and marginally
  deformed $\mathcal{N}=4$ Yang-Mills}},  {\em JHEP} {\bf 0904} (2009) 130,
  [\href{http://xxx.lanl.gov/abs/0902.1427}{{\tt arXiv:0902.1427}}].

\bibitem{Freedman:2005cg}
D.~Z. Freedman and U.~Gursoy, {\it {Comments on the $\beta$-deformed
  $\mathcal{N}=4$ SYM theory}},  {\em JHEP} {\bf 0511} (2005) 042,
  [\href{http://xxx.lanl.gov/abs/hep-th/0506128}{{\tt hep-th/0506128}}].

\bibitem{Hollowood:2004ek}
T.~J. Hollowood and S.~P. Kumar, {\it {An N=1 duality cascade from a
  deformation of $\mathcal{N}=4$ SUSY Yang-Mills theory}},  {\em JHEP} {\bf
  0412} (2004) 034, [\href{http://xxx.lanl.gov/abs/hep-th/0407029}{{\tt
  hep-th/0407029}}].

\bibitem{Fokken:2013aea}
J.~Fokken, C.~Sieg, and M.~Wilhelm, {\it {Non-conformality of
  $\gamma_i$-deformed $\mathcal{N}=4$ SYM theory}},
  \href{http://xxx.lanl.gov/abs/1308.4420}{{\tt arXiv:1308.4420}}.

\bibitem{Fokken:2013mza}
J.~Fokken, C.~Sieg, and M.~Wilhelm, {\it {The complete one-loop dilatation
  operator of planar real $\beta$-deformed $\mathcal{N}=4$ SYM theory}},
  \href{http://xxx.lanl.gov/abs/1312.2959}{{\tt arXiv:1312.2959}}.

\bibitem{Penati:2005hp}
S.~Penati, A.~Santambrogio, and D.~Zanon, {\it {Two-point correlators in the
  $\beta$-deformed $\mathcal{N}=4$ SYM at the next-to-leading order}},  {\em
  JHEP} {\bf 0510} (2005) 023,
  [\href{http://xxx.lanl.gov/abs/hep-th/0506150}{{\tt hep-th/0506150}}].

\bibitem{Frolov:2009in}
S.~Frolov and R.~Suzuki, {\it {Temperature quantization from the TBA
  equations}},  {\em Phys.Lett.} {\bf B679} (2009) 60--64,
  [\href{http://xxx.lanl.gov/abs/0906.0499}{{\tt arXiv:0906.0499}}].

\bibitem{deLeeuw:2012hp}
M.~de~Leeuw and S.~J. van Tongeren, {\it {The spectral problem for strings on
  twisted $\AdS_5\times\text{S}^5$}},  {\em Nucl.Phys.} {\bf B860} (2012)
  339--376, [\href{http://xxx.lanl.gov/abs/1201.1451}{{\tt arXiv:1201.1451}}].

\bibitem{FrolovPC}
S.~Frolov.
\newblock private communication.

\bibitem{Jin:2013baa}
Q.~Jin, {\it {The Emergence of Supersymmetry in $\gamma_i$-deformed
  $\mathcal{N}=4$ super-Yang-Mills theory}},
  \href{http://xxx.lanl.gov/abs/1311.7391}{{\tt arXiv:1311.7391}}.

\bibitem{Ahn:2011xq}
C.~Ahn, Z.~Bajnok, D.~Bombardelli, and R.~I. Nepomechie, {\it {TBA, NLO Luscher
  correction, and double wrapping in twisted AdS/CFT}},  {\em JHEP} {\bf 1112}
  (2011) 059, [\href{http://xxx.lanl.gov/abs/1108.4914}{{\tt
  arXiv:1108.4914}}].

\bibitem{Broadhurst:1985vq}
D.~J. Broadhurst, {\it {Evaluation of a class of Feynman diagrams for all
  numbers of loops and dimensions}},  {\em Phys. Lett.} {\bf B164} (1985) 356.

\bibitem{'tHooft:1973mm}
G.~'t~Hooft, {\it {Dimensional regularization and the renormalization group}},
  {\em Nucl.Phys.} {\bf B61} (1973) 455--468.

\bibitem{Bardeen:1978yd}
W.~A. Bardeen, A.~Buras, D.~Duke, and T.~Muta, {\it {Deep Inelastic Scattering
  Beyond the Leading Order in Asymptotically Free Gauge Theories}},  {\em
  Phys.Rev.} {\bf D18} (1978) 3998.

\bibitem{Khoze:2005nd}
V.~V. Khoze, {\it {Amplitudes in the $\beta$-deformed conformal Yang-Mills}},
  {\em JHEP} {\bf 02} (2006) 040,
  [\href{http://xxx.lanl.gov/abs/hep-th/0512194}{{\tt hep-th/0512194}}].

\bibitem{Chetyrkin:1980pr}
K.~G. Chetyrkin, A.~L. Kataev, and F.~V. Tkachov, {\it {New Approach to
  Evaluation of Multiloop Feynman Integrals: The Gegenbauer Polynomial $x$
  Space Technique}},  {\em Nucl. Phys.} {\bf B174} (1980) 345--377.

\bibitem{Vladimirov:1979zm}
A.~Vladimirov, {\it {Method for Computing Renormalization Group Functions in
  Dimensional Renormalization Scheme}},  {\em Theor.Math.Phys.} {\bf 43} (1980)
  417.

\bibitem{Gates:1983nr}
S.~J. Gates, M.~T. Grisaru, M.~Rocek, and W.~Siegel, {\it {Superspace, or one
  thousand and one lessons in supersymmetry}},  {\em Front. Phys.} {\bf 58}
  (1983) 1--548, [\href{http://xxx.lanl.gov/abs/hep-th/0108200}{{\tt
  hep-th/0108200}}].

\bibitem{ZinnJustin:2002ru}
J.~Zinn-Justin, {\em {Quantum field theory and critical phenomena}}, vol.~113
  of {\em Int.Ser.Monogr.Phys.}
\newblock Clarendon Press, Oxford, 1996.

\bibitem{Collins:1984xc}
J.~C. Collins, {\em {Renormalization. An introduction to renormalization, the
  renormalization group, and the operator product expansion}}.
\newblock Cambridge University Press, 1984.

\bibitem{Siegel:1979wq}
W.~Siegel, {\it {Supersymmetric Dimensional Regularization via Dimensional
  Reduction}},  {\em Phys. Lett.} {\bf B84} (1979) 193.

\bibitem{Mertig:1995ny}
R.~Mertig and W.~van Neerven, {\it {The Calculation of the two loop spin
  splitting functions P(ij)(1)(x)}},  {\em Z.Phys.} {\bf C70} (1996) 637--654,
  [\href{http://xxx.lanl.gov/abs/hep-ph/9506451}{{\tt hep-ph/9506451}}].

\bibitem{Vogelsang:1995vh}
W.~Vogelsang, {\it {A Rederivation of the spin dependent next-to-leading order
  splitting functions}},  {\em Phys.Rev.} {\bf D54} (1996) 2023--2029,
  [\href{http://xxx.lanl.gov/abs/hep-ph/9512218}{{\tt hep-ph/9512218}}].

\end{thebibliography}\endgroup

\end{fmffile}
\end{document}